\documentclass[sigchi]{acmart}

\AtBeginDocument{%
  \providecommand\BibTeX{{%
    \normalfont B\kern-0.5em{\scshape i\kern-0.25em b}\kern-0.8em\TeX}}}

 \newcommand{\edits}[1]{{#1}}

\copyrightyear{2023}
\acmYear{2023}
\setcopyright{rightsretained}
\acmConference[DIS '23]{Designing Interactive Systems Conference}{July 10--14, 2023}{Pittsburgh, PA, USA}
\acmBooktitle{Designing Interactive Systems Conference (DIS '23), July 10--14, 2023, Pittsburgh, PA, USA}
\acmDOI{10.1145/3563657.3596012}
\acmISBN{978-1-4503-9893-0/23/07}

\begin{document}

\title[Broadening Privacy and Surveillance]{Broadening Privacy and Surveillance: Eliciting Interconnected Values with a Scenarios Workbook on Smart Home Cameras}
\author{Richmond Y. Wong}
\email{rwong34@gatech.edu}
\orcid{0000-0001-8613-0380}
\affiliation{%
  \institution{Georgia Institute of Technology}
  \city{Atlanta}
  \state{Georgia}
  \country{USA}
  \postcode{30308}
}

\author{Jason Caleb Valdez}
\email{jvalde3@uw.edu}
\orcid{0009-0008-4019-570X}
\affiliation{%
  \institution{University of Washington}
  \city{Seattle}
  \state{Washington}
  \country{USA}
  \postcode{98195}
}

\author{Ashten Alexander}
\email{ashtendesign@gmail.com}
\orcid{0009-0001-1608-6043}
\affiliation{%
  \institution{University of Washington}
  \city{Seattle}
  \state{Washington}
  \country{USA}
  \postcode{98195}
}

\author{Ariel Chiang}
\email{ariel62212@gmail.com}
\orcid{0009-0003-4091-6515}
\affiliation{%
  \institution{University of Washington}
  \city{Seattle}
  \state{Washington}
  \country{USA}
  \postcode{98195}
}

\author{Olivia Quesada}
\email{oliviagq@uw.edu}
\orcid{0009-0003-7760-9324}
\affiliation{%
  \institution{University of Washington}
  \city{Seattle}
  \state{Washington}
  \country{USA}
  \postcode{98195}
}

\author{James Pierce}
\email{jjpierce@uw.edu}
\orcid{0000-0002-2192-1728}
\affiliation{%
  \institution{University of Washington}
  \city{Seattle}
  \state{Washington}
  \country{USA}
  \postcode{98195}
}

\renewcommand{\shortauthors}{Wong, et al.}

\begin{abstract}
 We use a design workbook of speculative scenarios as a values elicitation activity with 14 participants. The workbook depicts use case scenarios with smart home camera technologies that involve surveillance and uneven power relations. The scenarios were initially designed by the researchers to explore scenarios of privacy and surveillance within three social relationships involving “primary” and “non-primary” users: Parents-Children, Landlords-Tenants, and Residents-Domestic Workers. When the scenarios were utilized as part of a values elicitation activity with participants, we found that they reflected on a broader set of interconnected social values beyond privacy and surveillance, including autonomy and agency, physical safety, property rights, trust and accountability, and fairness. The paper suggests that future research about ethical issues in smart homes should conceptualize privacy as interconnected with a broader set of social values (which can align or be in tension with privacy), and reflects on considerations for doing research with non-primary users. 
\end{abstract}



\begin{teaserfigure}
    \centering
  \includegraphics[width=0.9\textwidth]{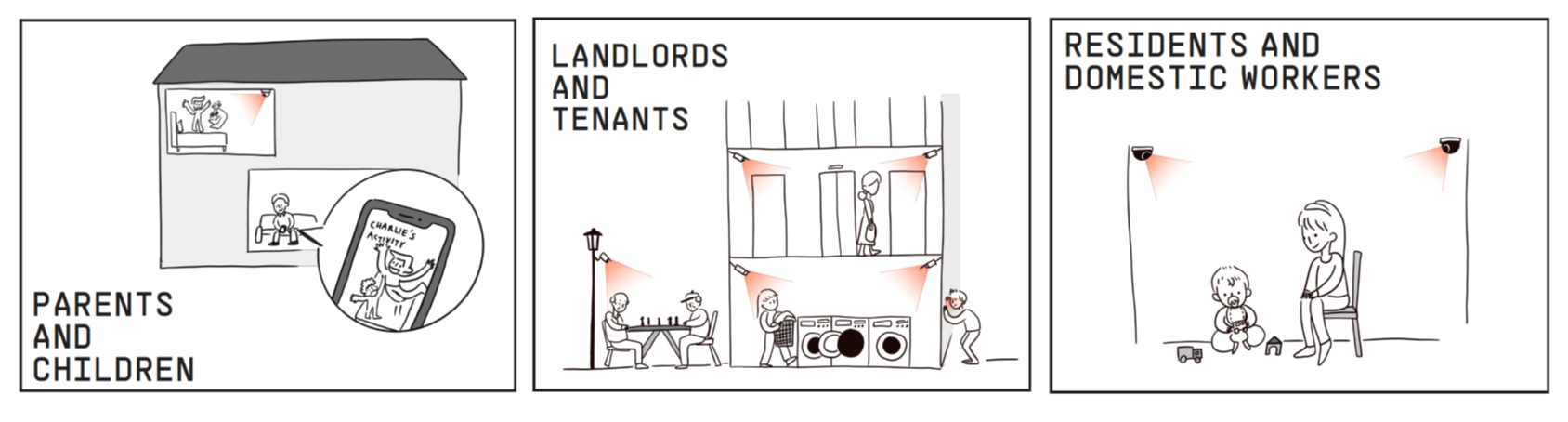}
  \caption{Section “title pages” from our Scenario Workbook (images of the scenarios are in Section \ref{sec:scenariosInWorkbook} and Appendix \ref{sec:appendix}.)}
  \Description{3 screenshots of "title pages" from the workbook. Title page saying "parents and children" with an illustration of a house. "Landlords and tenants" title, with an illustration of an apartment building with smart sensors. "Residents and domestic workers" title page with an image of cameras watching a child and babysitter.}
  \label{fig:teaser}
\end{teaserfigure}


\maketitle

\section{Introduction}
Consumers are increasingly adopting and deploying “smart” products or Internet of Things (IoT) technologies, including speakers, doorbells, locks, cameras, appliances, and others. While these may provide new ways to live and interact with others, they also present concerns related to privacy and surveillance. Furthermore, these technologies may affect additional social values, such as mediating social interactions and changing how trust or autonomy are considered within relationships. This paper focuses on \textit{smart home cameras} because they are  one of the most popular, growing \cite{Tan2022Monitoring}, and controversial applications of smart home products \cite{Pierce2019Smart,Pierce2022Addressing,Tan2022Monitoring}. Values in design scholars have sought methods to proactively identify and discuss the potential social values impacts and ethical harms related to new technologies \cite{Shilton2018Values,Wong2020Beyond}. Forward-looking and open ended design methods may assist in these efforts \cite{Shilton2020Role-Playing,Wong2019Bringing}. 

This paper presents a case study using a design workbook of conceptual designs proposals and hypothetical use case scenarios to elicit discussion about social values and ethical issues with users and stakeholders of smart home camera technologies (Figure \ref{fig:teaser}). We created a workbook to initially explore how smart camera technologies might continue to amplify and exacerbate issues of privacy and surveillance \cite{Pierce2019Smart,Tan2022Monitoring}, especially in social relationships between “primary” and “non-primary” users. \textit{Primary users} have greater forms of control and access to smart camera devices, while \textit{non-primary users} \cite{Tan2022Monitoring} have less control and access. Specifically, the workbook explores how future smart home camera systems might affect relationships among Parents \& Children, Landlords \& Tenants, and Residents \& Domestic Workers. 

We shared the workbook with 14 participants who were users of smart home technologies and experienced at least one of the social relationship categories described in the scenarios (e.g., parent, landlord). We conducted interviews to understand: \textbf{When using these scenarios as elicitation tools, how do participants assess values and ethics? What perspectives, values, opinions, preferences, and judgments do participants reveal? }While participants directly mentioned privacy and surveillance in their reactions to the scenarios, we also found that they contextualized these among a broader (and sometimes conflicting) set of interconnected social values. 

This paper makes two empirical contributions. \edits{First, it builds on prior research broadening conceptions of privacy by finding that participants employed multiple values and concepts in their ethical assessments of the scenarios. Contributing to HCI research on privacy, ethics and values, this suggests that smart home research should similarly broaden how it conceptualizes problems of privacy and surveillance.}
Second, it provides a case study of doing research that engages both primary and non-primary users. We reflect on our design process for constructing scenarios that can engage in discussion from multiple points of view.   

In the following sections, we discuss related work, our design process, and methods. We then present our findings of how participants identified multiple and sometimes conflicting social values at stake in the scenarios including: privacy and surveillance, autonomy and agency, physical safety, property rights, trust and accountability, and fairness. We end with reflections on (1) expanding our conceptions of privacy to be entangled with other values, (2) the fluidity of the categories of “primary” and “non-primary” users, (3) the modes of engagement participants used to reflect on values, and (4) future implications for research and design.

\section{Related Work}

\subsection{Values \& Ethics, Primary \& Non-Primary Users, and Smart Home Technologies}

Prior research has studied privacy in smart home settings. Traditionally, digital privacy research has focused on the relationships between device owners and the companies that collect the data, such as providing privacy notices in smart home settings \cite{Naeini2017Privacy}, providing users control over personal data \cite{Bahirat2021Overlooking,Chalhoub2021itdidnot,Feng2021Design}, or creating systems that make privacy decisions for users \cite{Lee2020Confident}. These focus on the so-called “vertical” relationships between users and companies. However in practice, not everyone is equally able to exercise these rights or make these decisions. Furthermore, many consumer IoT devices involve spatial sensors such as cameras and microphones which affect the privacy of nearby actors \cite{Pierce2022Addressing,Tan2022Monitoring}. 

A growing body of HCI research has begun studying how privacy and ethical concerns arise in “lateral” \cite{Andrejevic2005work,Mulligan2011Bridging} social relationships between different types of users. In particular we consider \textbf{primary users} as users who have greater forms of control and access to smart home devices (often by owning them), while \textbf{non-primary users} have less control and access \cite{Tan2022Monitoring}.  \edits{Koshy et al. use the terms “pilot” and “passenger” users to similarly describe these relationships \cite{Koshy2021WeJustUse}, though we build on Pierce et al.’s conception of  non-primary users \cite{Pierce2022Addressing} to consider a range of relationships (such as secondary and tertiary users who may have some but restricted access to devices, or indirect/incidental users like delivery workers interacting with cameras). Pierce et al combine prior work (e.g., \cite{Baumer2015Usees,Bernd2022Balancing,Yao2019Privacy,Zeng2017End}) to outline a range of terminology to grapple with various types of non-primary users. They refer to people who meaningfully interact with a digital system but experience low levels of control and benefits as “adjacent users” \cite[p.37]{Pierce2022Addressing}.}

\edits{Regardless of the specific terminology used to describe non-primary users,} the introduction of smart home devices has the potential to cause tension in social relationships in the home \cite{Choe2012Investigating,Yao2019Privacy}. Prior research suggests that non-primary users have limited understanding of how smart devices collect their data \cite{Marky2020You,Marky2021Roles}, and limited ability to access or control device settings \cite{Zeng2017End}. In some contexts, primary users may wish to protect non-primary users' privacy but often lack features to help do this \cite{Wu2023doStreamers,ahmad2020tangible}. Given this, researchers have explored how to design to increase non-primary users’ awareness, understandings, and ability to control how their data is collected and used \cite{Cobb2021Iwould,Lau2018Alexa,Thakkar2022It}.

Importantly, non-primary users often have less social power than primary users, potentially exacerbating their privacy violations. For instance, Ur et al. study a smart lock system, and find that parents would configure the system in ways that surveils their teenage children \cite{Ur2014Intruders}. Participatory design workshops with smart home sensor systems have suggested that some participants might actively seek to create applications to surveil or control non-primary users’ behaviors \cite{Kurze2020Guess,Tabassum2020Smart}. 

Several projects study these dynamics specifically with smart cameras. Bernd et al. study privacy practices of in-home nannies in relation to smart cameras installed by home residents (their employers), finding that nannies may have the right to data protection, but in practice do not have the ability to request control over the data because of the power imbalance in their employer-employee relationship \cite{Bernd2022Balancing,Bernd2020Bystanders'}. Tan et al. find that many primary users of smart cameras have limited concern for non-primary users’ privacy, and that some non-primary users feel concerned about privacy but are uncomfortable confronting primary users about it \cite{Tan2022Monitoring}.  \edits{While there are a few examples of design interventions specifically intended to improve privacy and social tensions with smart sensing devices (e.g., \cite{Chen2020Wearable,Cheng2019Peekaboo,david2021let}), there remains a need to better understand and foreground primary and non-primary user dynamics.}

Prior research has also investigated additional values and ethical implications of smart home environments. Smart home devices have implications for trust in social relationships \cite{Bernd2022Balancing,Choe2012Investigating}, and for autonomy and agency \cite{Denefleh2019Sensorstation:,Ehrenberg2021Technology,Garg2022Social}. Others have questioned the values and politics of smart homes more broadly by asking who gets considered as a smart home user and what counts as a smart home \cite{Jenkins2017Living,Kozubaev2019Spaces,Odom2019Diversifying,Oogjes2018Designing}, and highlighting power imbalances in gender roles in the smart home \cite{Strengers2019Protection}.

In our design explorations, we consider three particular relationships of primary and non-primary users: Parents \& Children, Landlords \& Tenants, and Residents \& Domestic Workers. 

\subsection{Conceptualizing Privacy}

Privacy scholars have attempted to define privacy in many ways, such as privacy as: control over personal information, as protections from government searches, as the protection of physical spaces and bodies, as freedom of thought, and more. However, no single definition applies well to every situation. Reacting to this, privacy scholar Dan Solove has written that privacy as “a concept in disarray. Nobody can articulate what it means” \cite{Solove2008Privacy:}.

Shifting away from universal theories of privacy, recent work has acknowledged multifaceted conceptions of privacy, such as Solove’s taxonomy of data privacy harms \cite{Solove2003Taxonomy}, and his articulation of multiple conceptions of privacy \cite{Solove2002Conceptualizing}. Nissenbaum’s contextual integrity formulates how privacy depends on contextual and situational norms \cite{Nissenbaum2009Privacy}. Mulligan et al. present a set of dimensions that can be used to describe privacy in specific contexts, asking why is privacy important, who and what is protected by privacy, what actions violate privacy, what protects privacy, and how broadly privacy applies \cite{Mulligan2016Privacy}. 

Notably, recent privacy scholarship has begun to discuss how the harms of privacy violations are multifaceted and include social values related to many aspects of human experience. Citron and Solove’s typology of privacy harms describes how violations of privacy can lead to physical harms (such as bodily injury and death), economic and monetary losses, harms to individuals’ reputations, psychological and emotional harms, harms to people’s autonomy and ability to make choices, discrimination harms, and harms to personal and professional social relationships \cite{Citron2022Privacy}. They write that each of these types of harms implicates different values, legal theories, and histories. \edits{Citron and Solove articulate this typology using legal case studies. Our research surfaces perspectives and experiences from actual users of smart technology who suggest a similarly wide range of interconnected privacy harms.}

\edits{While these theories broadening the conceptualization of privacy are acknowledged and cited within HCI, privacy is still often operationalized in narrower ways within HCI research.} McDonald and Forte describe \cite{McDonald2020Politics} that HCI privacy work tends to either conceptualize privacy as individual control (e.g., individual privacy boundaries \cite{Altman1975environment}), or as social community norms (e.g., contextual integrity \cite{Nissenbaum2009Privacy}). Wong and Mulligan describe how much of HCI research frames privacy as a technical problem that can be solved through design solutions such as encryption or data architecture; or frames privacy as an informational problem which requires the re-design of notices and other informational content to assist users’ decision-making \cite{Wong2019Bringing}. McDonald and Forte suggest that HCI would benefit by expanding its conceptions of privacy to focus on vulnerabilities from feminist and queer perspectives \cite{McDonald2020Politics}, building on other research identifying how people’s experiences of privacy can vary with their social power positions \cite{McDonald2020Privacy,Freed2018"A}. This paper similarly suggests that HCI can usefully expand its conceptions of privacy, but by considering a broader network of interconnected social values. 

\subsection{Design Futuring Methods}
Imagining different sociotechnical configurations of the world with design futuring provides opportunities to consider matters of public concern, such as how social values, ethical outcomes, and arrangements of power might change \cite{DiSalvo2014Making,Lockton2019New,Wong2020Infrastructural}. “Ethical speculation” practices can be useful for pedagogical and reflective discussion of technology’s ethical impacts \cite{Fiesler2019Ethical,Fiesler2021Ethical}. Yet at times, these practices have been critiqued for not engaging adequately with people and perspectives beyond designers or not providing empirical knowledge \cite{Elsden2017On,Kozubaev2020Expanding}.

In response, researchers and designers have involved participants in design futuring by co-designing futures \cite{Kozubaev2019Spaces,TranOLeary2019Who}, creating activities or experiences for participants to engage in \cite{Ballard2019Judgment,Elsden2017On,Kozubaev2020Expanding,Merrill2020Security}, or using design futuring artifacts as probes to elicit participants’ reflections in values-centered research \cite{Briggs2015Inclusive,Harrington2021Eliciting,Noortman2019HawkEye,Wong2017Eliciting}. Several design futuring projects consider smart home cameras and themes related to privacy and surveillance, including Pierce et al.’s speculative exploration of smart camera surveillance metaphors \cite{Pierce2019Lamps,Pierce2019Smart,Pierce2020Sensor,Pierce2021eccentric}, Tan et al.’s in-home speculative probes with participants to imagine new relationships with their own cameras \cite{Tan2022Critical-Playful}, and Cheng et al.’s design research camera platform that allows for home participants to actively or passively share their recording data \cite{Cheng2019Peekaboo}. 

\edits{With varying degrees of participation, HCI and design researchers have further explored a variety of forms for presenting design scenarios to various audiences and stakeholders, including concept videos \cite{Rogers2019OurFriends,Lindley2020DesignResearch}, textual narratives \cite{Baumer2018WhatWouldYou},  interface elements such as icons \cite{Disalvo2016Designing,Lindley2020Researching}, GIFs \cite{Biggs2021Moving}, and brief design proposals involving photographs, collage, illustration, diagrams, and annotations \cite{Aipperspach2008Heterogenous,Pierce2018AddressingNetwork,Gaver2000Alternatives,Boucher2012Power}.}

\subsubsection{Design Workbooks and Values Elicitation}
Design workbooks, while often used as an internal tool for design teams to explore a design space \cite{Gaver2011Making}, have more recently been used as research probes to elicit participant reflection and engagement. Wong et al. use workbooks of imagined IoT surveillance devices to understand how technologists might think about different aspects of privacy, finding that participants engaged with the workbooks in multiple ways when discussing privacy, such as seeing themselves as both users and designers, imagining the designs as real, and comparing the designs to real-world examples and experiences \cite{Wong2017Eliciting}. Based on this, the authors suggest that design workbooks could be useful in a values in design process. Wyche utilizes workbooks as part of a visual elicitation study in an ICTD context, finding that design concepts afforded rich participant responses that gave insight into participants’ memories and experiences, helped participants note practical problems and question the designs’ assumptions, and showed how local community members conceptualized values (such as surveillance) differently than the researchers \cite{Wyche2021Benefits}. Similar to these elicitation activities, Chen et al. utilize workbooks of visual design concepts to broach topics with participants that might be difficult to discuss in an interview, such as the role of data after death \cite{Chen2021What}. Harrington and Dillahunt take design workbooks in a more participatory direction, creating an interactive design workbook with explicit prompts and spaces for participants to brainstorm, allowing for co-speculating with Black youth \cite{Harrington2021Eliciting}. \edits{Other researchers have creatively used design probes, participatory engagements, and design scenarios in other ways to engage people with the topics privacy, security, and surveillance \cite{Pierce2018Interface,Knowles2019Scenario-Based,Nagele2018PDFi,Rogers2019OurFriends}.}

This paper uses workbooks as a form of visual elicitation and values elicitation, showing participants scenarios created by the researchers to understand their reactions and reflections on social values.

\subsubsection{Scenario Literature that Motivates Our Approach}

Scenarios, broadly defined as strategically produced visions of possible futures \cite{Buehring2018Embracing}, vary in their form and purpose across fields. Design scenarios in HCI are typically brief textual and/or visual descriptions (e.g., use cases) used to creatively imagine design solutions \cite{1995Scenario-based}, or to communicate, validate and endorse design decisions about user actions \cite{Evans2003Trend,Martin2009Design}. 

A scenario lineage from outside HCI comes from strategic planning and foresight \cite{Heijden2005Scenarios:}—often used in public policy, crisis management, and business strategy—to help organizations deal with uncertainty by envisioning medium to long-term futures (5-15 years) to influence an organization’s strategic decisions \cite{Vaara2012Strategy-as-Practice:}. Scenarios are not predictive but rather anticipatory tools \cite{Weber1996Counterfactuals} for imagining, discussing, shaping, planning, and preparing for possible futures. We found this lineage’s use of trend analysis to create scenarios useful for our design explorations. 

While largely separate today, HCI design scenarios and strategic planning scenarios do have common origins in the 1960s \cite{Wong2018Speculative,maze2019politics}. Recent work has sought to re-integrate the two approaches of design scenarios and strategic planning scenarios, highlighting opportunities to: improve strategic scenarios using design expertise \cite{Buehring2018Embracing}, improve the “worldmaking” aspects of scenarios \cite{Vervoort2015Scenarios}, involve diverse stakeholders in the planning process \cite{Heijden2005Scenarios:,Vaara2012Strategy-as-Practice:}, combine both traditions to explore ethical issues related to emerging technologies \cite{Epp2022Reinventing,Wong2021Timelines:} including privacy and security \cite{CLTC2016Cybersecurity,Knowles2019Scenario-Based}. This paper uses scenarios as a probe \cite{Hutchinson2003Technology,Noortman2019HawkEye} in a values elicitation activity. We draw inspiration from value sensitive design approaches that use scenarios as a way to elicit stakeholder understandings of values \cite{Briggs2015Inclusive,Friedman2017Survey,Nathan2008Envisioning}. 

\section{Developing the Scenarios Workbook}
The scenarios we present in this study form part of a larger, ongoing project that aims to (a) conceptually explore trends of smart home devices, (b) methodologically combine the strengths of design scenarios and strategic planning scenarios, and (c) adapt this approach to engage broader participation from diverse stakeholders. To develop scenarios for the design workbook, we identified current trends in smart home technologies synthesized from prior research and news reporting. In parallel, we generated a wide range of hypothetical scenarios that explored a range of themes related to smart home devices and datafication of daily life. (We present more details about this process in Appendix \ref{sec:appendix}). 

\textbf{Scenarios Design Process.} Developing the scenarios was an iterative and emergent process. We began generating scenarios using short textual descriptions. We ideated by creating lists of scenarios that consisted of a short and often catchy title followed by a few sentences describing the scenario. From here, we selected some to iterate by adding rough sketches. We then continued to select and refine.  This process allowed us to explore many scenarios before committing further efforts toward refining them through visual illustration and polished copywriting. The next stage was splitting the scenarios into text and illustration. One group of authors created rough illustrations while another focused on refining the text. To achieve consistency, we then had one author finalize all illustrations and one author finalize all captions. 

\edits{We initially explored at least 50 possible scenario ideas covering multiple themes including (1) the sensorification of daily life, (2) perceptually powerful devices, (3) behind-the-scenes actors and data misuse, (4) beyond safety and security: social, reflective, and aesthetic applications, (5) environmental sensor pollution, and (6) social tensions and asymmetric power relations. Ultimately, our study demanded that we narrow down and select fewer scenarios that we could discuss with participants in a 60-90 minute interview. We decided to focus on the theme of smart camera surveillance within everyday asymmetric power relations as it captured important social and ethical issues at stake with smart devices, and this theme elicited much discussion among the authors. Scenario ideas on other themes  were excluded, leaving us with approximately a dozen scenarios. Of those, most  fit into one of three categories of social relationships: Parents \& Children, Landlords \& Tenants, and Homeowners \& Domestic Workers, so we focused on developing scenarios in these categories.} Other relationships with fewer scenario ideas (such as one scenario between homeowners and delivery drivers) were not selected for further exploration. We ended up developing text and visuals for 3 scenarios from those 3 categories which we felt would explore different facets of these relationships as they relate to privacy and surveillance. 

\begin{figure}[h]
  \centering
  \includegraphics[width=\linewidth]{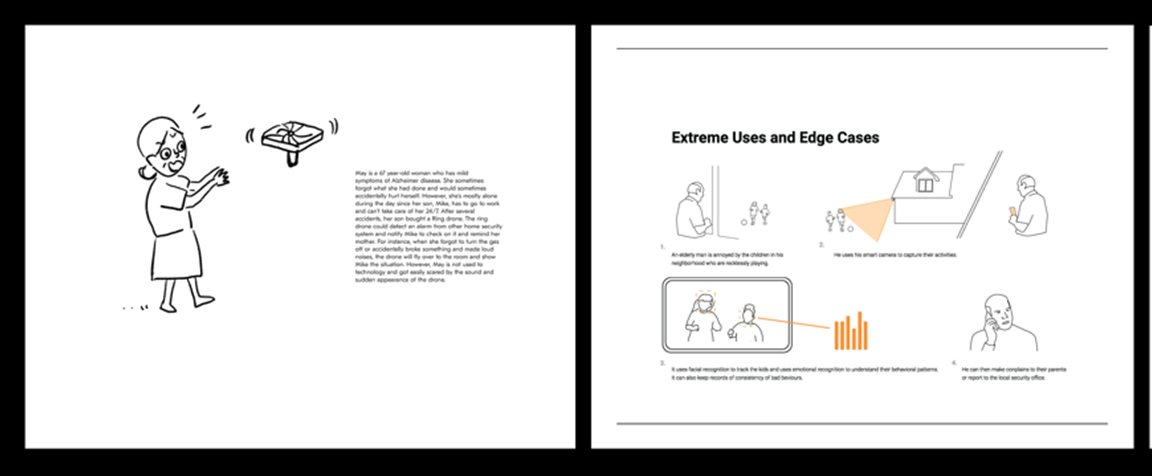}
  \caption{We explored different levels of detail, for instance, single-image snapshots (left) versus multi-frame storyboards (right). Ultimately we decided to create single-image snapshot scenarios. }
  \Description{Screenshot on the left of a scenario with 1 image and some text; screenshot on the right of a storyboard scenario, with 5 visual scenes and text descriptions}
   \label{fig:2process}
\end{figure}

\textbf{Design Composition of Scenarios.} We wanted the scenarios to be legible, comprehensible, and relatable to participants so that it might lead to discussion, reflection, criticism, and imagination. We formalized a design framework of 4 core considerations to help us achieve these goals:

\textit{1. Ethical tone: }Do the scenarios suggest a use case that is positive, negative, or ambiguous (from our perspective or the anticipated perspective of our audience)? We intended each scenario to be ethically ambiguous in order to avoid being immediately interpreted as completely utopian or dystopian, by incorporating positive and negative aspects. \edits{We note that “ambiguous” is different than neutral; rather than avoiding being positive or negative (neutral), we find that ambiguity allows us to acknowledge how many scenarios can have both positive and negative elements, depending on one’s perspective or viewpoint.}

\textit{2. Timescale:} Does the scenario take place in the near-term (soon?) or long-term (someday?). We decided to place each scenario closer to the present to ensure they were more relatable and comprehensible.

\textit{3. Levels of detail:} Does a scenario focus on breadth or depth? Are they conveyed through a single snapshot or a longer storyboard (e.g., a single image or a multi-panel storyboard)? We decided to make each scenario simple, easy to digest, and presentable within a single page to ensure we cover approximately 10 scenarios within the relatively short interview time. We constrained each scenario to a single-image with short captions, rather than a longer storyboard narrative showing multiple scenes (Figure \ref{fig:2process}). We referred to these as a ``snapshot'' scenario format.

\textit{4. Mundanity:} Does the scenario depict an ordinary versus extraordinary scene within the fictional future world? We tended to select scenarios that, within the speculative world, appeared to be fairly regular, commonplace events in order to prompt normative judgements about future “new normals.” The question of “mundane for whom?” was one we grappled with, given that what might appear as “normal” depends on one’s life experiences. We tried to include a mix of experiences, such as an overprotective parent, a tenant living in a large apartment building, and nanny reporting for work. However, we note that the research team lived in close proximity to large cities on the U.S. west coast, all of whom lived in rental housing. We recognize that there are broader housing situations than our own lived experiences \cite{Kozubaev2019Spaces,Odom2019Diversifying}. We note that one limitation of this approach is that by building scenarios focused on more mundane situations (from our perspective) we missed an opportunity to engage with more marginal, idiosyncratic, or subcultural scenes—something we intend to explore in future work.

\subsection{Scenarios in the Workbook Study}
\label{sec:scenariosInWorkbook}
The scenarios in the final version of the workbook (Table \ref{tab:1summary}) are grouped in sections featuring different relationships: (1) Parents \& Children, where parents monitor or take care of children; (2) Landlords \& Tenants, where apartment landlords and property managers manage the tenants who rent from them; and (3) Residents \& Domestic Workers, where home residents hire people to work within their home such as nannies or babysitters.

\begin{table}
\renewcommand*{\arraystretch}{1.25}
\caption{Summary of Scenarios in the Workbook }
\label{tab:1summary}
\begin{tabular}{p{0.15\linewidth} p{0.76\linewidth}}

\toprule
Scenario Name & Scenario Description \\
\midrule

\multicolumn{2}{c}{\textit{Parents \& Children}}    \\
\midrule

Out After Curfew Detector   & Smart cameras designed to keep intruders out might be also used to help keep kids in. (Fig \ref{fig:A1curfew}) \\
Remote Chaperone           & Smart cameras might be used to help parents remotely monitor kids (Fig \ref{fig:A2chaperone}).   \\
Drone Parents              & A helicopter parent is one who “hovers” over their child to closely monitor their experiences and problems. Smart   cameras might enable a whole new level of helicopter parenting (Fig \ref{fig:3drone})  \\

\midrule
\multicolumn{2}{c}{\textit{Landlords \& Tenants}}  \\
\midrule

Extreme Lease Enforcer        & Smart cameras and microphones might be used to precisely and strictly enforce rental agreements by helping landlords and   property managers send warnings, fines, or eviction notices. (Fig \ref{fig:A3extremelease})  \\
Dealing with a Problem Tenant & Smart devices might enable tenants to report dangerous or annoying behavior so that property managers can deal with it (Fig \ref{fig:A4problemtenant}) \\
Approved Access Only          & Cameras and other smart devices might be used to increase security for tenants. The same technology may be used to   limit access for non-approved guests and tenants not in good standing. (Fig \ref{fig:4approvedaccess})  \\

\midrule
\multicolumn{2}{c}{\textit{Residents \& Domestic Workers}} \\
\midrule                                                                                                                                                                                                            
Dropping in, Keeping Tabs     & Smart cameras might allow people to easily check in on nannies, housekeepers, babysitters, plumbers, painters, and other hired   help.  (Fig \ref{fig:A5droppingin})   \\
Incident Report               & An indoor smart camera system might   allow an adult to monitor their parent’s caregiver and their child’s nanny. The system sends incident reports whenever a notable event occurs such as a   grandparent falling, a child having a meltdown, or a caregiver yelling or neglecting their patient. (Fig \ref{fig:A6incident}) \\
Mood Check                    & Smart camera systems might also monitor   the internal emotional states of people. This could be used to assess whether   a nanny, caregiver, or pet sitter enjoys their job or not, and whether they have   the right social skills and temperament. (Fig \ref{fig:5moodcheck}) \\

\end{tabular}
\end{table}

\begin{figure}[h]
  \centering
  \includegraphics[width=0.95\linewidth]{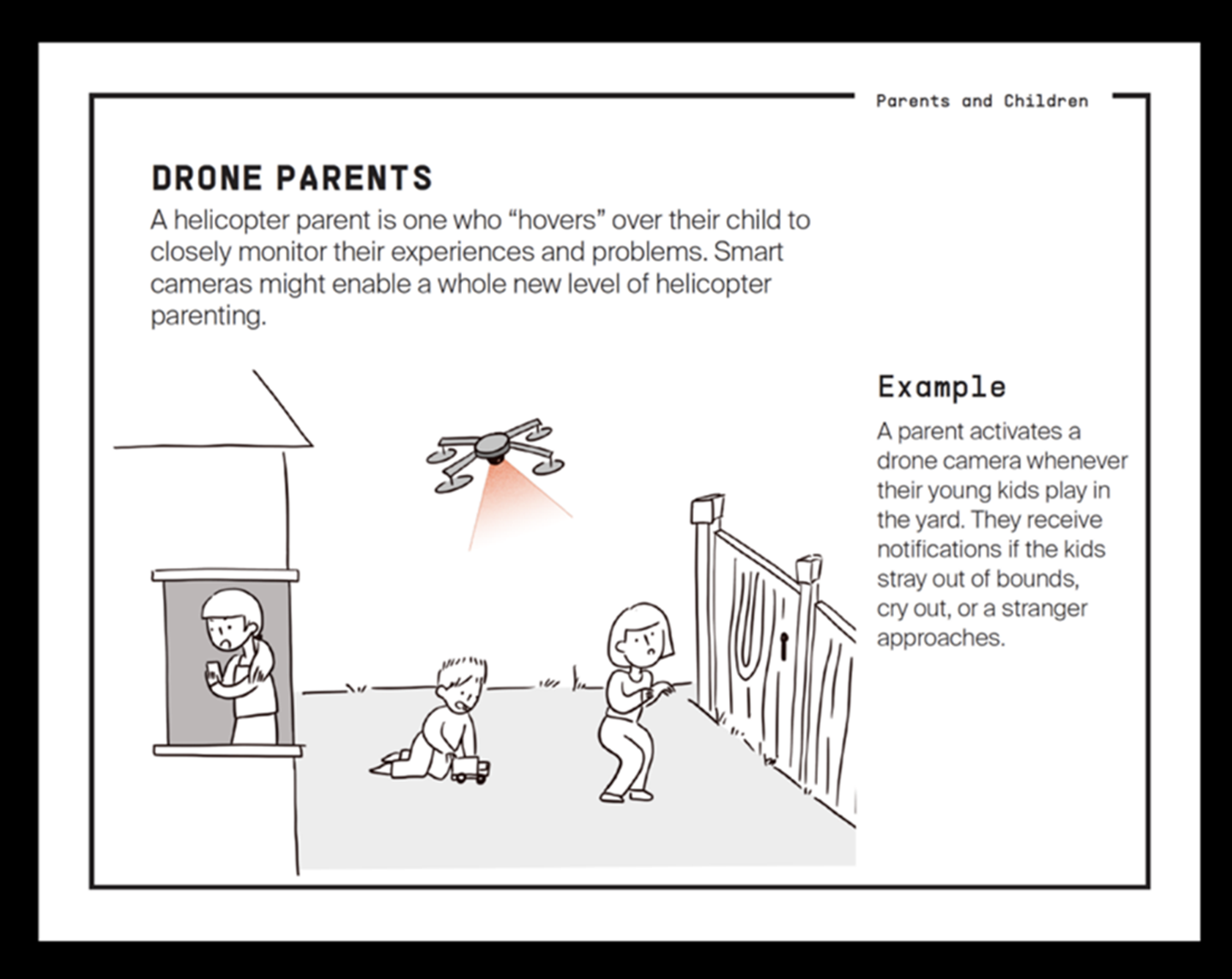}
  \caption{Detailed view of the \textit{Drone Parents} scenario shown to participants. }
  \Description{Drawing of "drone parents" scenario - Drone Parents. A helicopter parent is one who “hovers” over their child to closely monitor their experiences and problems. Smart cameras might enable a whole new level of helicopter parenting}
   \label{fig:3drone}
\end{figure}

\begin{figure}[h]
  \centering
  \includegraphics[width=0.95\linewidth]{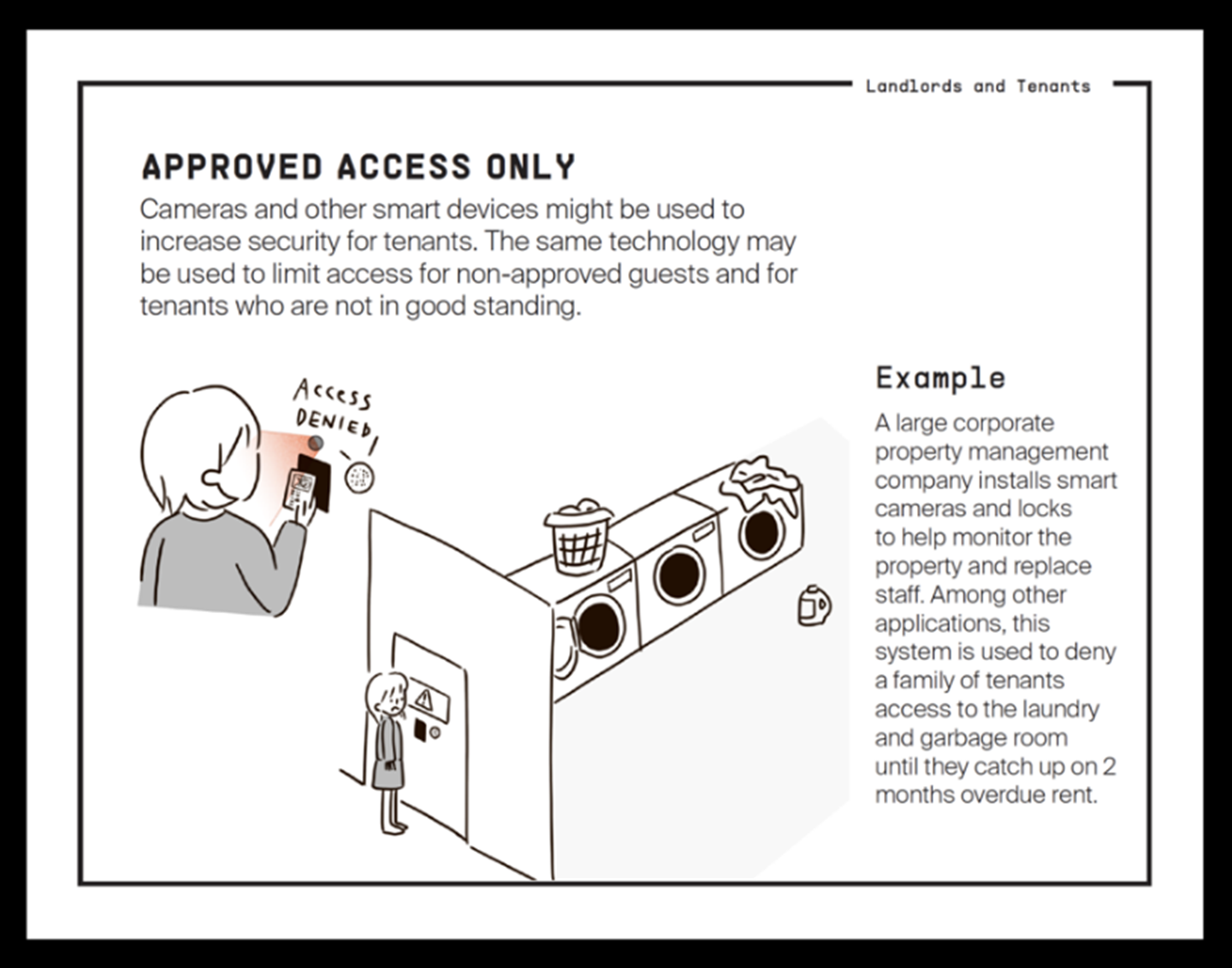}
  \caption{Detailed view of the \textit{Approved Access Only} scenario shown to participants. }
  \Description{Drawing of approved access only scenario -   Cameras and other smart devices might be used to increase security for tenants. The same technology may be used to limit access for non-approved guests and tenants not in good standing.}
   \label{fig:4approvedaccess}
\end{figure}

\begin{figure}[h]
  \centering
  \includegraphics[width=0.95\linewidth]{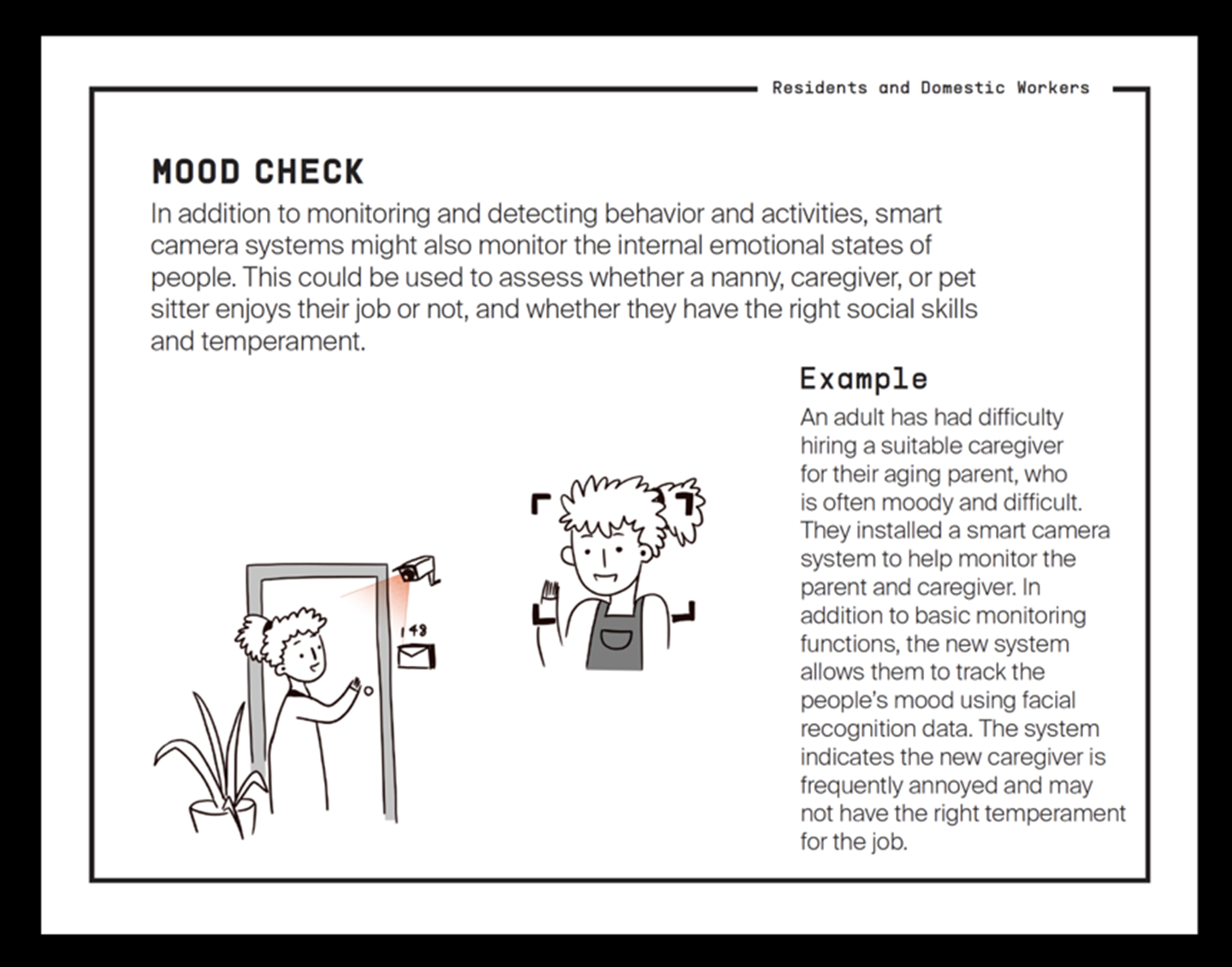}
  \caption{Detailed view of the \textit{Mood Check} scenario shown to participants }
  \Description{Drawing of mood check scenario - Mood Check. Smart camera systems might also monitor the internal emotional states of people. This could be used to assess whether a nanny, caregiver, or pet sitter enjoys their job or not, and whether they have the right social skills and temperament.}
   \label{fig:5moodcheck}
\end{figure}

\section{Study Methods}
In 2021, we recruited participants through advertisements using Craigslist in the San Francisco Bay area and Seattle metro area, and social media posts. We initially recruited near the researchers’ institutions, planning to meet participants in-person. Due to COVID-19 restrictions, we conducted all interviews using Zoom. As our social media posts were re-shared online, we ended up recruiting from across the United States. The recruitment messages described that we were looking for people who either “own a smart home device” or “interact with smart home devices that you don't own.” We did not define “smart home device.” Interested participants filled out a screener survey asking them to list smart home devices that they own or interact with, describe a previous experience with a smart device, and basic demographic data. The study was approved by the Institutional Review Boards at the authors’ institutions.

All participants we reached out to for interviews reported having some prior experience with smart home devices, either as a primary or non-primary user. All participants also fit into one of the social relationship categories in the scenarios, through their current or past experiences (e.g., currently a parent, or previously a domestic worker).   Some met more than one of these qualifications. All participants resided in the United States. A limitation of our recruitment is that we did not have IRB permission to do research with people under 18 years old. As a proxy, we interviewed several younger adults who currently live with their parents (P8, P12) both to understand their current home dynamics and their recent experiences from when they were younger. Table \ref{tab:2participants} provides key information about the participants, while Table \ref{tab:3demographics} provides aggregated demographic information.\footnote{Demographic information is provided in aggregate form in order to help preserve participants’ anonymity. Classifications of current and previous experiences were determined by coding participants’ explicit discussion of their experiences, so this table likely undercounts people’s experiences. For instance, most people probably lived with their parents previously at some point, but we only indicate participants who explicitly discussed it. }

\begin{table*}
\caption{Summary of Participants and their Relevant Experiences (C = currently does this, P = previously did this) }
\label{tab:2participants}
\begin{tabular}{p{0.02\linewidth} p{0.05\linewidth} p{0.02\linewidth} p{0.12\linewidth} p{0.05\linewidth} p{0.09\linewidth} p{0.07\linewidth} p{0.07\linewidth} p{0.05\linewidth} p{0.05\linewidth} p{0.07\linewidth} p{0.07\linewidth}}
\toprule
\#  & Gender & Age & Occupation                 & House-hold Size & Other Household Members            & Lives w/ Parent(s) & Lives w/ Child(ren) & Is Landlord & Is Tenant/ Renter & Hires Domestic Workers & Is Domestic Worker \\
\midrule
P1  & Female & 35  & Grad Student                      & 1              & n/a                               & P                  &                     &               & C                   &                        & P                    \\
P2  & Male   & 35  & Project Manager                   & 4              & Spouse and children                &                    & C                   &               &                     &                        &                      \\
P3  & Male   & 28  & Electrical Engineer               & 2              & Roommate                           &                    &                     &               & C                   &                        &                      \\
P4  & Male   & 24  & Graduate Student                  & 4              & Roommates                          & P                  &                     &               & C                   &                        &                      \\
P5  & Female & 37  & Finance Manager                   & 2              & Spouse                             &                    &                     & C             & P                   & C                      &                      \\
P6  & Female & 36  & Administrative Assistant          & 4              & Spouse and children                &                    & C                   &               & P                   & C                      &                      \\
P7  & Male   & 37  & Sales Executive                   & 7              & Spouse, children, mother, brother  &                    & C                   & C             & P                   &                        &                      \\
P8  & Male   & 22  & Student, Administrative Assistant & 5              & Parents, siblings                  & C                  &                     &               &                     &                        &                      \\
P9  & Male   & 50  & Banquet server \& Food Delivery   & 2              & Spouse                             &                    &                     & P             & C                   & C                      &                      \\
P10 & Female & 47  & Administrative Assistant          & 3              & Spouse, child                      &                    & C                   &               &                     &                        &                      \\
P11 & Female & 41  & Administrative                    & 4              & Spouse, children                   &                    & C                   &               &                     &                        &                      \\
P12 & Male   & 28  & Student \& Pet Sitter/Dog Walker  & 3              & Parent and roommate                & C                  &                     &               & C                   &                        & C                    \\
P13 & Female & 60  & Operations Manager                & 2              & Spouse                             &                    &                     & C             & C                   & C                      &                      \\
P14 & Male   & 40  & Social Worker                     & 2              & Spouse                             &                    &                     &               & C                   &                        &                     
\end{tabular}
\end{table*}

\begin{table*}
\caption{Aggregated demographics of participants }
\label{tab:3demographics}
\begin{tabular}{llll}
\toprule
Ethnicity           & Area Lived In      & Highest Education      & Annual Household Income \\
\midrule
White/Caucasian (7) & Big City/Urban (7) & Some college (1)       & \$10k-50k (2)           \\
Asian (6)           & Suburban (5)       & Associate’s Degree (2) & \$50k-100k (5)          \\
Hispanic (1)        & Mid-Sized City (1) & Bachelor’s Degree (6)  & \$100k-150k (5)         \\
                    & Small Town (1)     & Master’s Degree (5)    & \$150k+ (1)             \\
                    &                    &                        & Prefer not to share (1)
\end{tabular}
\end{table*}

We interviewed 14 participants for between 60-90 minutes, conducted by the first or second authors in a semi-structured fashion. Participants were compensated with a \$30 Amazon.com gift card. Interviews consisted of two parts. First, participants were asked background questions about their prior experiences with smart home technologies to get them comfortable talking. However, this part of the interview is not analyzed, as we focus instead on the participants’ interactions with the scenarios.

The second part focused on the scenario workbook. The interviewer would screen share the workbook, explaining that these are fictional scenarios depicting smart home cameras. We decided the researcher would share their screen to show participants the scenarios, rather than having interviewees share their screens to avoid potential technical difficulties. The workbook was presented in a linear fashion, similar to a slideshow. It was shown in the order listed in Table \ref{tab:1summary} to all participants, since the workbook was grouped into sections and had summary pages listing the designs in order, to help communicate the three groups of relationships. The interviewer would present the scenarios in one section, asking participants to “think aloud” and provide any initial thoughts or reactions. For instance, with Parents \& Children, the interviewer would go through the Out After Curfew Detector, Remote Chaperone, and Drone Parent. After viewing a section, the interviewer would present a section summary page. While on this page, participants were asked a range of questions about the scenarios, such as if they have had similar experiences, what they think is positive or negative, how they think the people in the scenarios might feel, and if there were any new ideas or scenarios that came to mind. This was repeated for the Landlords \& Tenants and Residents \& Domestic Worker sections. While we asked participants if they had any concerns about the scenarios, we never explicitly prompted them to speak about privacy, surveillance, or other social values. 

Recordings were transcribed using Zoom’s captioning features. The first author conducted the first round of analysis. While reading transcripts he inductively used action coding to identify participants’ interactions with the workbooks \cite[pg.96]{Saldana2013Coding}, and values coding to identify statements related to social values, attitudes, beliefs, and ethics \cite[pg.110]{Saldana2013Coding}. Text excerpts with these codes were extracted from the transcripts and grouped into themes. The first, second, and last authors met periodically to discuss and refine the codes and themes of the analysis. 

\subsection{Cultural Context of the Study}
This research was conducted by researchers living in large cities on the west coast of the United States and with participants living in the U.S. This cultural and legal context affected participants’ responses and the researchers’ analysis. At the time of research. the U.S. does not have a national data privacy law, meaning that many expectations of privacy depend on people’s beliefs about proper norms and practices rather than on a legal definition. Culturally, ownership of private property is encouraged and often seen as a way to increase generational wealth (although the ability to purchase property is not evenly distributed among social, class, and racial differences). Furthermore, in most states and cities in the U.S., there are relatively limited protections legal protections or recourse for renters and tenants (such as if tenants face rate increases or evictions by landlords) or employees (in many cases it is very easy to be fired by an employer). Knowledge of these dynamics affected how the researchers interpreted participants’ statements about their beliefs and experiences in these domains. 

\section{Findings: Eliciting Participants' Values Reflections on the Scenarios}
In our findings, we report on prominent values that surfaced across participant reactions: (1) privacy and surveillance, (2) autonomy and agency, (3) physical safety and care, (4) private property interests, (5) trust and accountability, (6) fairness, and (7) how values differ across social identities. We note that participants introduced these values; \textit{we did not explicitly name} any of the values when asking participants questions. One takeaway is that participants rarely thought in terms of a single value or concern, such as privacy, but rather suggested a network of interrelated values and concerns. 

Before discussing those findings, we briefly note how participants expressed a range of mixed reactions to the scenarios \textit{both within and across subjects}. Participants varied widely on their feelings of positivity or negativity of each scenario. No scenario was universally liked or disliked by participants. Some disliked almost all the scenarios, some liked almost all the scenarios, but most liked some and disliked others. Some participants also expressed mixed feelings about the same scenario when considering it from different points of view. For instance, P1 immediately reacted to the designs by stating “I can appreciate why the parent might \textit{want} [that] and the child might \textit{not}” when presented with the first scenario, Curfew Detector. Across all the scenarios, participants were able to consider the situations from multiple stakeholder perspectives. 

\subsection{Privacy and Surveillance}
Participants explicitly brought up privacy in multiple ways, often describing how primary users (Parents, Landlords, and Residents) were in positions to surveil or violate the privacy of non-primary users (Children, Tenants, and Domestic Workers) in the scenarios. For instance, in response to the Landlords \& Tenants scenarios P4 (current  renter) explained “I think a problem on one side would be abuse of power by the landlord and again the invasion of privacy.”

Participants described how their expectations of privacy related to practices of surveillance and data collection. Several participants observed that smart cameras might allow an owner to constantly collect data. P3 (current renter), reflecting on the Problem Tenant scenario said “I wouldn’t want them [landlords] to record everything all the time. But for a specific time when you receive a complaint or something from others, and they can just keep that, it’s fine.” P3 suggests that landlords keeping specific recordings related to tenant complaints might be acceptable, but that recording all the time violates a tenant’s expectations of privacy. 

Another concern with privacy expectations emerged with the Mood Check scenario, which depicts a smart home camera used for new, speculative types of emotional surveillance.

\begin{quote}
P5 (current landlord, previous renter): It strikes me as a little strange to have something that was never available to me. I don't know if it's necessary. We've all gone through life without mood checks. [...] I feel like this is a little invasive, like you're trying to read someone's mind, basically and boil it down to one simple description.  
\end{quote}

P5’s reflection suggests that the scenario reconfigures Resident \& Domestic Worker relationships by introducing a new form of intimate data that was not available before this speculative technology, violating current social expectations of privacy. 

Several participants drew connections to other experiences of privacy and surveillance, including surveillance of workers at stores and offices, or tracking others’ location on smartphones. P1 (current renter) noted how the Landlord \& Tenant scenarios fit into a longer (pre-IoT) history of landlords surveilling and taking advantage of tenants, stating that “these aren’t really new problems; this is a new way to achieve an old technique.” 

Overall, participants reflected on multiple aspects of privacy and surveillance, including the role of social power differences between primary and non-primary users, current social expectations of privacy, and connecting the scenarios to issues of privacy and surveillance in other domains and time periods.  Next, we discuss how participants typically other social values  (though often still in relation to privacy and surveillance). 

\subsection{Autonomy and Agency}
Many participants who responded negatively to the scenarios described threats to individual autonomy and agency. These values arose most prominently in the Parents \& Children scenarios. For instance, P9 (does not live with children) responded to the Remote Chaperone scenario saying:

\begin{quote}
 P9: Kids need to be able to bad talk their parents in front of friends and stuff like that and without worrying about their dad or mom listening or hearing that. They need to be themselves. Use cuss words or whatever with their friends. […]And it [the scenario] just takes away from kids’ freedom and the ability to be it to be themselves.   
\end{quote}

Here, autonomy away from a parent’s watchful eye---which is amplified to an unnatural degree by technology---is seen as an important part of children’s development. P1,5,6,8,13,14 expressed similar concerns (this includes participants who currently live with parents, currently live with children, and some who live with neither). P4 (used to live with parents) expressed a desire for children to have more autonomy by comparing the Parent \& Children scenarios to how his parents used to keep in touch with him using a landline phone when they were away from the home—which allowed them to check in, but not monitor him at all times, granting him more autonomy compared to the scenarios’ smart cameras.

Several participants described agency through the concept of informed consent, particularly for domestic workers. P13 (hires domestic workers) reacted to the childcare scenarios in Residents and Domestic Workers by saying “the camera for a small child’s carer would be great but there needs to be agreement, I wouldn’t want to keep it hidden. In other words, I want the babysitter to know, just in fairness to them.” While many researchers discuss consent in relation to data privacy, P13 frames consent as an issue of dignity in the power relationship between residents and domestic workers. P4,9,11 (one of whom discussed currently hiring domestic workers) also discussed how to notify domestic workers about the use of smart home cameras and obtain their consent. These reflections suggest home residents' recognition of the autonomy and agency of domestic workers. 

Others considered how scenarios could be re-designed to give non-primary users more agency. P1 (previous babysitter) imagined if babysitters could use or access smart home recordings to understand if there is potential parental abuse in the household. 

\begin{quote}
 P1 (previous domestic worker): As a babysitter you see a very intimate picture of a parent and child relationship […] But what if you're there to, like, secretly assess child abuse? Is this mother giving her teenage daughter an eating disorder? What is the child saying to you when the parents are not there? Yes, there's a predominant power relationship [...] but what would the reverse be?     
\end{quote}

Similarly, P6 (previous renter) imagined that monitoring systems in the scenarios could be re-designed to help tenants protect themselves from landlords when ending a lease by having video evidence that can help tenants dispute cases when landlords erroneously claim that tenants have caused property damage and owe extra money. 

Overall, participants noted how the scenarios presented challenges to different conceptions of autonomy and agency. Some framed autonomy as allowing for self-development (particularly during childhood), others discussed autonomy and agency in terms of consent or making informed choices, and others connected issues of autonomy and agency to the power differences between primary and non-primary users.

\subsection{Physical Safety \& Care}
Many participants who found the scenarios positive viewed scenarios as protecting people’s physical safety or expressing care. These themes were most prominent with the Parents \& Children and Residents \& Domestic Worker scenarios (and often in tension with autonomy and agency). Several parents described the scenarios as providing physical safety for their children. P13 (did not disclose if living with children) described how Curfew Detector would help “keep them safe” by tracking a child’s comings and goings from their room and home.

Some acknowledged that the scenarios might violate children’s privacy but that was outweighed by the parents’ interest in safety and care. In a mixed reaction, P6 viewed the Remote Chaperone as useful even if her kids resisted.

\begin{quote}
 P6 (currently lives with children): I can see where kids or teenagers might really hate that. But as a parent myself, I really like that, especially now. Like you want to give them some semblance of privacy, but if you learn or hear that things are going wrong you can check in and make sure. It's just nice to be able to have an eye on them.    
\end{quote}

P6 recognizes the potential privacy invasions, but she would still value the system as a form of parental care. At the same time, she considered ways to provide her kids with more privacy by suggesting that she would only check the system if something “goes wrong,” rather than continuously monitoring it.

Several participants discussed protecting the physical safety of home residents and private property from domestic workers. While not explicitly identifying as a “domestic worker,” P8 had experience as a behavioral interventionist working with children in their homes, and related that to the Domestic Workers scenarios:

\begin{quote}
 P8: This past year I was a behavior interventionist […] You're trusting your kid with disabilities, with a stranger who is supposed to help you with them. As someone who's providing that care, I wouldn't feel invaded if my client's family happened to record what I was doing to make sure that their child is safe. […] They're looking for the best interest of their kid, not necessarily trying to pinpoint something that I do wrong.     
\end{quote}

These participants felt that surveilling children’s activities and people working with children in the home was justified due to physical safety and security concerns, discussing it as an expression of care. In these cases, participants often viewed themselves as responsible for the care of others. Echoing tensions expressed by some participants, researchers have noted that even if socially accepted, surveillance as care can still reinforce problematic power dynamics and obscure other forms of care \cite{Jacobs2016Attending,Stark2018surveillant}. 

\subsection{Property Rights}
Participants discussed values related to private property ownership in the U.S. in both positive and negative reactions to the scenarios. \textbf{For those who responded more positively, beliefs about private property owners' rights were used to justify the acceptability of owners' control} over devices and spaces. P10 (did not disclose property ownership), reacting to the Problem Tenant scenario, said “Remember you're renting somebody’s space. You have to be respectful. And those [cameras] could be some of the rules.” P6 juxtaposed tenants’ privacy concerns with a landlord’s private property rights with the Extreme Lease Enforcer:

\begin{quote}
 P6 (previously rented): If it's in the rental agreement, I don't see an issue with having something like this. A lot of people probably wouldn't like it just because they feel like it's invading their privacy. But it is the landlord's property. I think that if they wanted to really enforce how many guests are staying, this would be a good option for landlords.    
\end{quote}

This suggests a belief that a landlord’s private property interests outweigh the privacy interests of tenants, even though P6’s reaction acknowledges tenants might react negatively. P5 similarly justified parental surveillance of children by explaining that “the house is my domain” and parents should be able to set the rules within the home because they own it. 

\textbf{Those who responded negatively noted that property owners do not have absolute control over their private property in practice}. P9 drew on his landlord and property management experience to describe how the Extreme Lease Enforcer changes the landlord-tenant relationship in significant ways, even though it purportedly just enforces terms of a lease: 

\begin{quote}
 P9 (previous landlord, current renter): I’ve been a landlord and this crosses the line. Most rental agreements—I’ve been an apartment manager for many years—they’re loosely enforced. If someone has a guest for an extra week or so it’s not that big a deal. If they’re causing a problem, then that’s one thing. But to me this is really micromanaging your tenants.     
\end{quote}

P9 points out that in practice, many provisions of a landlord’s lease are loosely enforced as long as tenants are not causing problems, even if technically prohibited. The Extreme Lease Enforcer would change these norms and practices by strictly enforcing a lease’s terms, providing less leeway than what is socially expected. 

Several participants who were tenants described feeling that the physical space of a home should still afford them privacy or autonomy without being constantly controlled by a landlord.

\begin{quote}
    P13 (current landlord, current renter): So right now I’m a tenant, I have an apartment in the Bay Area. I wouldn’t want any of these. I wouldn’t welcome them because I consider myself a good tenant. Why would I want something that was potentially tracking my behavior, my daily activities when I want some privacy? And that’s part of the deal when you’re renting from someone. You’re signing a lease with clear expectations. 
\end{quote}

Moreover, for P13 these expectations about a tenant’s rights are communicated through the leasing agreement with the landlord. 

After looking at the workbook, P5 expressed mixed feelings about her own use of smart security cameras for property protection. 

\begin{quote}
 P5 (current landlord, previous renter): In San Francisco we see a ton of theft. I bought a sign that says “secured by Ring, Ring enabled security.” […] But maybe in that very act of trying to deter theft, it leaves people feeling like this is not a welcoming home or not a welcoming building. So it's a little bit of a double edged sword. What are these little things that we're doing around the home and how is it impacting the community?    
\end{quote}

While P5 initially bought smart cameras to secure her property and felt positive towards them, the scenarios led her to consider that others in the neighborhood might interpret her cameras as being unwelcoming to the community. Overall, participants’ varying opinions highlight tensions in considering how to balance rights, social power, and control between privacy property owners and other stakeholders. 

\subsection{Trust, Accountability, and Social Relationships}
Participants described how the scenarios’ smart cameras might mediate social relationships and affect trust and accountability. A few participants suggested that some scenarios might \textbf{increase trust and improve social relationships. }P6 imagined that Approved Access might improve landlord-tenant relationships. 

\begin{quote}
 P6 (previous renter): I think this is a good way to hold people accountable. If they were having an issue where they couldn’t pay rent or they needed something else approved, they would actually have to communicate with their landlord, which I think is always a good thing.    
\end{quote}

P6 felt that the penalty of automatically revoking tenants’ access to facilities after a lease violation would motivate tenants to proactively communicate potential issues with landlords and strengthen tenants’ accountability to landlords. P6 (who also hires domestic workers) also thought that the scenarios might improve communication between residents and domestic workers such as plumbers or electricians by reducing awkward conversations, helping the worker communicate their progress, or remotely communicating if they need to leave the home temporarily to purchase materials and return later.

However, many other participants expressed concern that these might \textbf{negatively affect social relationships and decrease trust}. Viewing the Residents \& Domestic Workers scenarios skeptically, P1 discussed her experiences as a babysitter and how these scenarios alter social dynamics.

\begin{quote}
 P1 (previous domestic worker): When I babysat, [...] I loved all the kids. I appreciated their [parents] caring about who I was as a person, as it related to their kid. Taking out that relationship, would I still do the job for the money? Yeah, but it would change how I feel about it and I feel like that would change over time. […] I'm just checking the boxes.     
\end{quote}

P1 felt that these smart camera applications would make her job feel narrowly task-focused rather than being about holistically building deeper social relationships with the kids and their parents. 

P13, a landlord, similarly critiqued the Landlord \& Tenant scenarios:

\begin{quote}
 P13 (current landlord, current renter): I prefer to do things more directly with people. The [scenario] seems like the tenant-landlord relationship is much more removed. I would like to have a stronger communication plan or process with my tenants so that I can deal with these things proactively.    
\end{quote}

P13 notes that in-person she can “try to determine if this is just a complainer, or is this someone who’s got a legitimate issue.” She feels these judgements gained through social interactions would be more difficult to make if mediated by smart cameras. 

Relatedly, many participants (P1,4,5,6,3,9,11) felt that the Parents \& Children scenarios suggested that parents mistrust their children. P7 (currently lives with children), drew comparisons to his childhood which lacked the technology-enforced rules of the scenarios, saying “I grew up without limits. […] [this scenario] is like taking the parenting out of being a parent.” Across these examples, participants expressed concerns that the responsibilities and trust built through in-person social interactions would suffer if mediated by a smart camera or its analytical capabilities. 

Participants’ varying opinions in relation to the scenarios’ effects on trust and accountability highlight how the scenarios suggest changes to existing social relationships, and may affect these relationships over long term use. 

\subsection{Fairness}
Participants also discussed fairness, particularly noting how the power imbalance between primary and non-primary users might lead to unfair outcomes. Some participants imagined new social rules, behaviors, or norms that could help make the scenarios more fair. P9 viewed the Out After Curfew Detector as a form of punishment for misbehavior, and argued that fair parental surveillance of children should have a time limit and its use should be collectively understood, rather than being used indiscriminately.

\begin{quote}
 P9: It has to be an issue the parent and child decide. “You know you’re being punished, I’m going to have this up for six months, as a form of being grounded.” I think there needs to be a time limitation. “If you don’t try to disable the camera, we’ll take it down or shut it off at a certain time within six months,” something like that.    
\end{quote}

Participants also imagined alternative scenarios of their own that might be fairer by giving more discretion to non-primary users. For instance, some participants imagined using alternate technologies that might achieve similar ends but limit visual surveillance. P3 suggested that parents might monitor children’s activities with smartwatches and a geofence tagging system. P9 suggested that to check if a pet sitter or domestic worker came to the home, a sensor detecting the door opening would be enough. P8 suggested that if landlords are concerned about tenant noise, a decibel monitor would avoid collecting private audio. 

P9 (previously landlord, current renter) also described how housing laws in U.S. cities and states promote fairness by providing some basic protections for renters. Drawing on his former experiences as a property manager, P9 discussed how Approved Access might be illegal, or at least goes against the spirit of the law. 

\begin{quote}
 P9 (previous landlord, current renter): If you’re behind on your rent and not allowing a tenant to do laundry or throw away their garbage, that is cruel. […] To me that's kind of akin to turning off the lights. It's illegal for landlord to turn off power in your house, or shut the water off to try to get someone out.     
\end{quote}

P9 later reflected that stronger tenant laws may be needed to preserve fairness and to protect tenants from landlord surveillance overreach. Scholars have noted that fairness, like other values, has multiple conceptions and definitions in different contexts \cite{Mulligan2019This}; participants tended to discuss fairness as a sense of empathy and respect for others, and as a belief in equal process and treatment (rather than being about fair distributions of resources).  

\subsection{Reflecting on Values Across Social Identities}

Some participants noted how values implications may vary across identity and cultural differences, which were not explicitly depicted in the scenarios. P1,8,12 expressed concerns that women and girls may be disproportionately and unfairly surveilled in some of the scenarios. P4 described how the Approved Access scenario’s determination of whether a tenant is in “good standing” might embed and promote stereotypes, particularly for residents who are “Black, Indigenous, or people of color […] and there’s not a good justification behind it.” P1,4,12 also discussed how scenarios make assumptions about defining “good” neighbors and tenants that may exacerbate gentrification tensions in U.S. cities.

\begin{quote}
 P1 (current renter): Noise complaints are one of the early signs of gentrification. People moving into neighborhoods and then saying, “This neighborhood is too loud” and calling the police. That leads to escalations and can be a tool to evict lower income people in a neighborhood where the prices are getting higher. The problem tenant is not the loud tenant. It's the new tenant who comes in and expect things to be quiet. […] I think there are class and racial elements of what is considered loud, what music. Are you blasting Tchaikovsky or hip hop? You know it can be the same decibel but I think it would have a different cultural read and have different impacts.     
\end{quote}

Participants not only considered different values at play, but some participants discussed how the experience of these values varied among groups, and can be entangled with unequal distributions of social power. This aligns with contemporary research on values that view them as situated, lived experiences rather than a set of static norms and principles \cite{JafariNaimi2015Values,LeDantec2009Values}.

\section{Discussion}
We discuss how participants considered values beyond privacy and surveillance, how the concepts of primary and non-primary users are fluid categories, and the modes of engagement participants had with the workbooks when reflecting on values. We then suggest implications for future research and design, and reflect on our limitations and future work. 

\subsection{Expanding Conceptions of Privacy: How Participants Reflected on Multiple Interconnected Values }

We find that participants engaged in complex reflections about the values “at play” in the scenarios \cite{Flanagan2014Groundwork}. Prior HCI research tends to focus on privacy as being about individual management or local norms, allowing privacy problems to be addressed by technical and informational solutions \cite{McDonald2020Politics,Wong2019Bringing}. However, participants’ conceptions of privacy in relation to other values reflects the wide-ranging conceptualizations described by privacy law and social science scholars \cite{Citron2022Privacy,Mulligan2016Privacy,Solove2002Conceptualizing}. We note several ways how participants discussed ethical issues that encompassed privacy and other values. 

Participants highlighted how \textit{surveillance and invasions of privacy can violate individuals’ autonomy and agency}, and how this may have an outsized effect on individuals with less social power (such as non-primary users). Autonomy and agency are complex concepts as well. Privacy scholars Citron and Solove discuss many types of autonomy harms, including: coercion, manipulation, failing to be given sufficient information to make decisions, thwarted expectations, lack of control, or chilling effects inhibiting people from conducting lawful activities \cite{Citron2022Privacy}. Participants discussed several of these conceptions, such as wanting to make cameras visible to provide sufficient information to non-primary users, or raising concerns that parental surveillance might lead to chilling effects on the activities children choose to engage in. 

Some participants suggested that \textit{physical safety and care was important enough to warrant surveillance of others}, even if it raised privacy concerns. Sometimes this was to make sure physical harm did not occur to the watched subject, such as a parent watching their child outdoors. Other times this was to make sure that the watched subject did not cause harm, such as a parent watching a babysitter’s behaviors, or a homeowner trying to protect themselves against property damage. While “surveillance as care” may be useful in some situations, privacy researchers have noted that this can still reinforce problematic power dynamics, and obscure other forms of care that are less surveillant \cite{Stark2018surveillant}.

Participants also discussed tensions in \textit{how much power a private property owner has to surveil others.} Property owners and landlords have interests in knowing what occurs on their property, but home dwellers (such as renters, tenants, children, or domestic workers) also have a reasonable expectation of privacy. Some local housing laws in the U.S. attempt to create balances between stakeholders, such as requiring landlords to give tenants at least 24 hours of notice before physically entering their residences, recognizing both tenants’ privacy expectations and a landlord’s private property rights. Recording data in homes with multiple stakeholders and competing interests presents new challenges, which a few design researchers have begun to explore, such as the implications of using smart technology in public housing \cite{Kozubaev2019Spaces}. Legal approaches to balancing competing rights in these situations may serve as models or inspiration for designing systems that similarly need to balance competing stakeholder interests.

Participants also discussed how \textit{surveillance might decrease or increase social trust and accountability}. Those who felt that surveillance might increase social trust and accountability tended to suggest that it would create information parity or symmetry between people, which might improve communication. Those who felt it might decrease social trust and accountability raised concerns about how mediating human-to-human relationships through smart home cameras suggested a lack of trust or care in a social relationship. 

Participants noted how \textit{surveillance and violations of privacy can threaten fairness}. Prior researchers have noted how the values of privacy and fairness can be in tension with each other, in the context of collecting sensitive personal data in order to understand if a marginalized community is being fairly treated \cite{Dwork2013It's}. However, our participants focused more on power imbalances between primary and non-primary users and how surveillance could harm people’s dignity and respect in unequal and unfair ways. 

Participants’ reflections show the ripple effects that social values can have, sometimes supporting each other and sometimes being in tension. For instance, values such as fairness might be used to address violations of privacy (for instance, finding ways to try to fairly notify non-primary users that they are being recorded). In some situations, values like privacy and trust may align, while in other situations the same values may be in tension with one another. And these effects are not experienced evenly. Stakeholders with different amounts of social power or different social identities may be affected in dissimilar ways. 

We highlight how privacy is a wide-ranging concept, as that was our entry point as researchers. However, we note that the values discussed by participants are themselves entangled too (for instance, participants’ discussion of fairness often implicated the values of autonomy and agency). This suggests that it may be difficult to study privacy without consideration of other social values, which we return to in the Implications section. 

\edits{Finally, we note that understanding privacy as interconnected to other ethical concerns aligns with an “ethics of care” perspective articulated by feminist scholars (e.g., \cite{delaBellacasa2010Matters}). A care ethics perspective points to how the ethical positions and responses of people are, in practice, often highly contextual, relational, and complex. We find empirical evidence of a care ethics perspective across participants’ responses, particularly when they described nuanced consideration of multiple actors, perspectives, and details of the particular situation (e.g., relating to both the competing perspectives of a child and a parent). Future work should continue to place discourses of privacy and surveillance in conversation with burgeoning work within design and HCI advocating for a care ethics perspectives \cite{Bardzell2010FeministHCI,Key2021Proceed}, which includes both human and non-human actors \cite{wakkary2021things,Akama2020Expanding,Liu2018Design,Giaccardi2016ThingEthnography}.}

\subsection{Primary \& Non-Primary Stakeholders are Fluid Categories}
While we initially recruited participants based on our ideas of primary and non-primary users from the scenarios, we find that in practice, these categories are fluid and dynamic and can overlap or change. Put simply, a person can be a primary user at one time and place, and a non-primary user in another time and place. 

Sometimes this change occurs over time. Several participants reflected on their previous experiences as children (non-primary users) as compared to their current experiences as adults or parents (primary users). Similarly, other participants described having previously worked as domestic workers and no longer doing so, or having previously rented a home to now becoming a homeowner. In another example, a primary user may be later surveilled by a camera that they initially acquired and set up. 

Other times, people may shift between primary and non-primary use as they move between contexts. Some participants described being monitored as a non-primary user at their workplaces by an employer, while being in control of the smart devices in their own home as a primary user. 

Others may play multiple roles at the same time. One participant discussed how they both manage rental properties but are a renter themselves. Others described how in public areas of an apartment building they are a non-primary user (because the landlord installed and monitors cameras in those areas), but simultaneously they are a primary user of smart camera devices used within their own apartment.

This suggests a need not only to consider non-primary users and their privacy when designing smart camera systems, but to also design systems in ways that acknowledge that these roles can vary with time and context. Designing a system that assigns someone a static “primary” or “non-primary” user role with different permissions or access controls may not fully account for the dynamic experience of these roles.

\subsection{Reflecting on Participants’ Engagements with Workbooks }
Building on prior research using workbooks as a values elicitation activity with participants [18,98,100], we reflect on how participants engaged with, responded to, and thought through our scenarios when discussing values or ethical issues. These engagement types may be of use to researchers seeking to engage participants in reflective discussions using future-oriented design scenarios. We identified five types of engagement that build on and extend prior work. 

\textit{1. Participants viewed designs from different stakeholder positions. }Wong et al.’s workbooks elicitation study found that expert participants viewed designs through the lens of being a technologist and being a user \cite{Wong2017Eliciting}. We find that non-technologist participants are also able to view designs from user and stakeholder perspectives beyond their own. This suggests that workbooks may be useful for sensitizing people to \textit{multiple} user and stakeholder perspectives, particularly with primary/non-primary user relationships, rather than just a single type of user.

\textit{2. Participants related the scenarios to their historical and lived experiences.} Discussing scenarios elicited memories from participants (such as comparisons to their own experiences as children) and discussions of current lived experiences (such interacting with tenants and landlords). This aligns with Wyche’s finding that workbooks can help elicit memories and understandings of local context \cite{Wyche2021Benefits}. We found that participants discussed their lived experiences in service of two goals: (a) describing how new technologies might change existing and historical social relationships, such as introducing new forms of technical mediation into a social relationship; (b) comparing the scenarios’ technology to a previous or existing technology, such as parents’ use of landline phones to check in on children. Prior research has described how using new technologies to achieve the same function can have ethical consequences \cite{Mulligan2020Concept}. For instance, the function of parents keeping track of their children is the same when using a landline phone and smart camera, but these technologies have different ethical implications, and suggest different expectations of privacy. Participants’ discussion of lived and historical experiences helps elicit details about how and why technical choices can have ethical consequences. 

\textit{3. Participants directly contested and critiqued the scenarios.} Some participants suggested the scenarios granted primary users too much power, or discussed how the technologies in the scenarios might exacerbate issues of social and economic inequality. This aligns with Wyche’s finding that people find practical problems with the proposed designs \cite{Wyche2021Benefits} and Wong et al.’s finding that participants may question workbook designs’ framings and motivations \cite{Wong2017Eliciting}. Moreover, it suggests how scenarios can be used for more than just values elicitation, but also to open up a space to engage in more critical conversations about power and justice. 

\textit{4. Participants extended the scenarios} by imagining ripple effects, and imagining what else would need to change in order for the scenario’s fictional world to come true. Participants speculated on new social rules, behaviors, norms, and laws that could change or would be required if the scenarios’ imagined futures became true. They then reflected on the values embedded in those norms and rules. This aligns with Wong et al.’s finding that technologists discussed implications of workbook designs as if they were real \cite{Wong2017Eliciting}, and extends it by showing that a non-technologist audience can engage with workbooks in a similar way. 

\textit{5. Participants imagined alternate scenarios.} Even though we did not explicitly create our design workbooks as a tool for co-speculation (unlike Harrington and Dillahunt who used workbooks containing interactive activities explicitly for co-speculation \cite{Harrington2021Eliciting}), we find that participants still engaged in practices of speculation while responding to a workbook of visual designs and scenarios. Participants’ alternatives included systems that might achieve similar ends but limit the amount of data surveillance—such as using geofencing to lessen the need for visual surveillance, or using a decibel monitor instead of recording private audio to measure noise levels. Others suggested new scenarios that could invert the power relationships and give the non-primary users more power, such as helping tenants protect themselves from bad landlords or helping babysitters rather than parents. While not all participants engaged in this way, it nevertheless shows the possibility for using visual-based scenarios in participatory speculative design activities. 

Several scenario design decisions may have helped encourage participants’ reflections through these forms of engagement.  As this paper is exploratory, future research can explore the causal relationship between these design dimensions and the types of reflections they elicit: 
\begin{itemize}
    \item \textit{Ethical tone}: We attempted to make the ethical tone ambiguous—neither utopian nor dystopian, nor neutral. This may have been reflected in participants’ mixed reactions to the scenarios, considering them useful in some circumstances but ethically problematic in others. 
    \item \textit{Timescale}: We designed the scenarios to appear closer to the present, which may have helped participants relate the scenarios to lived experiences. 
    \item \textit{Levels of detail}: We provided short “single snapshot” scenarios. This allowed participants to view multiple scenarios and compare them. It also may have invited them to extend the scenarios and speculate about “missing details.”
    \item \textit{Mundanity}: We depicted “everyday” scenarios, which potentially helped participants relate the scenarios to everyday real-world experiences. Future work depicting more diverse situations in the scenarios may help elicit additional forms of values reflections. 

\end{itemize}

This study adds to prior research using design workbooks, suggesting that visual and textual scenarios can be an effective method of values elicitation with participants. Beyond design research, this also suggests opportunities for participatory engagement in policy and business uses of scenarios. In these contexts, scenarios are often created by and for decision-making experts \cite{Vervoort2015Scenarios,Wack1985Scenarios:}. Actively soliciting engagement from a broader range of stakeholders may help surface discussion of values tensions within the scenarios, and generate further ideas to explore. 

\subsection{Implications}
We highlight several implications for design and research related to smart home cameras, privacy, and primary and non-primary users. 

\subsubsection{Defining Privacy in Smart Home Systems Differs Based on Whose Privacy is Considered}

While “privacy” is often discussed as a social value that needs to be considered in the design of smart home camera systems, the actual problem of privacy and how to design for it may differ based on the answer to “privacy for whom?” \cite{Mulligan2016Privacy}. Prior research notes how privacy is not experienced equally due to people’s different amounts of social power \cite{McDonald2020Privacy,Ur2014Intruders}, historical inequities that persist in the present \cite{Browne2015Dark}, and different groups’ threat models \cite{Pierce2018Differential}. 

Our results demonstrate how participants conceptualized privacy differently when considering different stakeholders. When discussing children, participants discussed children’s privacy and their autonomy to develop into adults, versus parents’ interests in the physical safety of their children. With tenants, participants considered tenants’ expectation of privacy in their homes, versus a landlord’s interests in protecting their private property. With domestic workers, participants weighed a domestic worker’s agency versus the private property interests of the home residents or cultural norms about workplace surveillance. 

While all the scenarios contained smart home cameras, the problem of privacy is quite different in each. Answers to questions such as “why is privacy important?”, “who should have access to the camera data?”, and “for what purposes should the data be used?” differs based on who the stakeholders are. \textit{This suggests that the design of privacy systems in smart home cameras and similar sensing technologies need to be flexible to adapt to different conceptions of privacy when used in different contexts by different stakeholders.} 

\subsubsection{Researchers and Designers Can Shift from Studying Privacy Per Se to Entangled Values}

This study also shows how values related to smart home camera use and deployment are entangled. While privacy and surveillance were our entry point into this project, we found that “privacy” rarely emerged as the single or paramount concern for participants. \edits{Rather, participants conceptualized privacy in multiple ways, often in relation to other social values, aligning with a body of theoretical research that conceptualizes privacy in multifaceted ways, and related to values such as agency, autonomy, safety, community \cite{Citron2022Privacy,Mulligan2016Privacy,Solove2002Conceptualizing}. }

However, much prior smart home research frames privacy as a distinct problem or concern. Formulating research questions and conducting empirical studies exclusively focused on “privacy” may lead researchers to miss out on understanding a broader range of interconnected ethical and moral issues related to smart homes. While some prior research considers tradeoffs of privacy versus utility \cite{Butler2015Privacy-Utility,Zeng2019Understanding}, “utility” may conflate multiple specific goals and values, and oversimplify the relationship between privacy and other values. 

Values can be entangled in a network of interrelated ethical concerns.  While we described values independently, privacy is tied to a range of other values (in this case, autonomy and agency, safety and care, private property expectations, trust, and fairness). Sometimes promoting another value violates privacy (or vice versa), at other times protecting privacy also protects other values. Promoting a value for one stakeholder may simultaneously violate a value of another stakeholder.

It is important—in fact, necessary—for designers and researchers to define a scope of the problems or issues they are addressing. However, our findings suggest that in some cases, it may be a mistake for designers and researchers to assume that “privacy” represents the overarching concern from the perspective of users and stakeholders. This is not a call for smart home designers and researchers to abandon privacy, but rather to \edits{build on research that} considers a broader and more multifaceted range of values that may be desirable to promote, protect, and negotiate—and to consider the interplay among those values.

We consider how our results might have changed if we explicitly focused our discussion around questions of “privacy” and refrained from probing other issues such as property rights, autonomy, and safety. If our primary research questions had strictly focused on privacy and subtopics such as information disclosure, we may have missed the wider range of concerns that our participants surfaced. While future empirical research would be needed to test this counterfactual hypothesis, our study results show that the scenarios we initially created to probe issues of privacy also surfaced a broader range of interconnected values.

Practically, we suggest that \textit{even when smart home research is conducted on a specific value (like privacy), designers and researchers should remain open to additional values that participants may surface, and allow participants opportunities to bring these into the conversation. }

\subsubsection{Designers and Researchers Should Consider Both Primary and Non-Primary User Roles and Interactions}
Designers and researchers of smart home systems should consider both primary and non-primary user perspectives, and how those categories might shift. A primary user of a technology will likely also find themselves as a non-primary user of the same device (e.g., someone else in their household watches them on the smart camera). We found that participants discussed their experiences as both primary and non-primary users of technologies, and were able to imagine the scenarios from multiple user perspectives. \textit{Because people shift between being primary and non-primary users across times and contexts, considering non-primary users can help primary users if they later interact with system as a non-primary user. }

Furthermore, smart home devices with spatial sensors—such as cameras, microphones, location tracking, and lidar—will invariably affect non-primary users and subjects. In a general moral sense, it is desirable to consider and respect the perspectives of non-primary users. We also found that many primary users may want to better respect and consider non-primary users. Many participants considered if there were ways that they might make camera surveillance systems “fair” to people being monitored. \textit{Considerations of non-primary users should be included throughout the design process, to ensure that features designed for non-primary users are integrated from the start. }

\subsection{Limitations and Future Work}
We consider several limitations and opportunities for future work. We shared the workbooks with participants via the interviewer’s screen as a slide show, providing greater ease of use for participants. However, this may have limited their ability to make comparisons across scenarios as they might have done if all scenarios were laid out on a table, as shown in prior research \cite{Wong2017Eliciting}. We also showed the scenarios in the same order listed in Table \ref{tab:1summary} to each participant, since the workbook was organized into 3 “sections” and each section had a summary page listing the designs in order. While this was done to communicate the types of social relationships for each section, these ordering effects means that participants’ reactions to earlier scenarios may have affected reactions to later ones. Future research using virtual design workbooks may consider using alternate platforms or forms that might help participants compare multiple scenarios, such as by creating scenarios in the form of concise virtual cards. 

\edits{Our participants were not evenly distributed across the relationship categories (for instance, we had fewer participants who identified as domestic workers). While we do not claim generalizability from this project, future work could recruit more people from the specific populations and relationships we identified, and from more diverse populations to further investigate the question “privacy for whom?”. }

Our design choices may affect how participants engaged with the material. \edits{The development of our scenarios was foregrounded ideas of surveillance, which may have encouraged participants to discuss privacy. In part, we were less interested in understanding if participants raised privacy issues, and more interested in \textit{how} participants would discuss privacy and other ethical issues when confronted with them. However, it is possible that a different set of scenarios less focused on surveillance or with different ethical tones might elicit a different set of ethical issues.} We are mindful of our own biases and predispositions as researchers and designers working on this project, and how these may come through in our scenarios and discussions. Future iterations of these scenarios might \edits{depict additional scenarios and} increase the visual diversity of the people depicted in the scenarios, such as: incorporating a broader range of skin tones, less explicitly gendered characters, and a range of different abilities. These may help participants consider additional ethical issues and a more diverse range of stakeholders. 


\section{Conclusion}
 This paper presents a case study utilizing a design workbook of speculative scenarios that ground privacy and surveillance concerns related to smart cameras in three types of primary/non-primary user relationships: Parents \& Children, Landlords \& Tenants, and Residents \& Domestic Workers. We found that our scenarios elicited discussion from participants about social values and ethical issues beyond (and sometimes in tension with) privacy and surveillance, including: autonomy and agency, safety, property, trust and accountability, and fairness. We found that participants rarely thought in terms of a single value or concern, but rather suggested a network of interrelated values and concerns, \edits{conceptualizing privacy in multiple ways. This builds on prior theoretical work that views privacy as inherently multifaceted, entangled and connected with other values and concerns. This suggests that smart home researchers would benefit by being open to multiple conceptions of privacy, and considering values together rather than separating out “privacy” from other values and concerns.} We further found that many participants also identified both as primary and non-primary users at different times and contexts. This suggests a need for smart home research to both consider primary and non-primary users as fluid categories, and to consider a larger constellation of ethical concerns that represent how stakeholders think about social tensions and values with smart home technology.

\begin{acks}
Thank you to Noura Howell, Ji Su Yoo, Jon Gillick, Elizabeth Resor, and the anonymous reviewers for their feedback. This research was supported by National Science Foundation Grants 1910218, 2142795, 2230825, and the UC Berkeley Center for Long-Term Cybersecurity.
\end{acks}

\bibliographystyle{ACM-Reference-Format}
\bibliography{main}


\begin{thebibliography}{126}


\ifx \showCODEN    \undefined \def \showCODEN     #1{\unskip}     \fi
\ifx \showDOI      \undefined \def \showDOI       #1{#1}\fi
\ifx \showISBNx    \undefined \def \showISBNx     #1{\unskip}     \fi
\ifx \showISBNxiii \undefined \def \showISBNxiii  #1{\unskip}     \fi
\ifx \showISSN     \undefined \def \showISSN      #1{\unskip}     \fi
\ifx \showLCCN     \undefined \def \showLCCN      #1{\unskip}     \fi
\ifx \shownote     \undefined \def \shownote      #1{#1}          \fi
\ifx \showarticletitle \undefined \def \showarticletitle #1{#1}   \fi
\ifx \showURL      \undefined \def \showURL       {\relax}        \fi
\providecommand\bibfield[2]{#2}
\providecommand\bibinfo[2]{#2}
\providecommand\natexlab[1]{#1}
\providecommand\showeprint[2][]{arXiv:#2}

\bibitem[Ahmad et~al\mbox{.}(2020)]%
        {ahmad2020tangible}
\bibfield{author}{\bibinfo{person}{Imtiaz Ahmad}, \bibinfo{person}{Rosta Farzan}, \bibinfo{person}{Apu Kapadia}, {and} \bibinfo{person}{Adam~J. Lee}.} \bibinfo{year}{2020}\natexlab{}.
\newblock \showarticletitle{Tangible Privacy: Towards User-Centric Sensor Designs for Bystander Privacy}.
\newblock \bibinfo{journal}{\emph{Proc. ACM Hum.-Comput. Interact.}} \bibinfo{volume}{4}, \bibinfo{number}{CSCW2}, Article \bibinfo{articleno}{116} (\bibinfo{date}{oct} \bibinfo{year}{2020}), \bibinfo{numpages}{28}~pages.
\newblock
\urldef\tempurl%
\url{https://doi.org/10.1145/3415187}
\showDOI{\tempurl}


\bibitem[Aipperspach et~al\mbox{.}(2008)]%
        {Aipperspach2008Heterogenous}
\bibfield{author}{\bibinfo{person}{Ryan Aipperspach}, \bibinfo{person}{Ben Hooker}, {and} \bibinfo{person}{Allison Woodruff}.} \bibinfo{year}{2008}\natexlab{}.
\newblock \showarticletitle{The Heterogeneous Home}. In \bibinfo{booktitle}{\emph{Proceedings of the 10th International Conference on Ubiquitous Computing}} (Seoul, Korea) \emph{(\bibinfo{series}{UbiComp '08})}. \bibinfo{publisher}{Association for Computing Machinery}, \bibinfo{address}{New York, NY, USA}, \bibinfo{pages}{222–231}.
\newblock
\showISBNx{9781605581361}
\urldef\tempurl%
\url{https://doi.org/10.1145/1409635.1409666}
\showDOI{\tempurl}


\bibitem[Akama et~al\mbox{.}(2020)]%
        {Akama2020Expanding}
\bibfield{author}{\bibinfo{person}{Yoko Akama}, \bibinfo{person}{Ann Light}, {and} \bibinfo{person}{Takahito Kamihira}.} \bibinfo{year}{2020}\natexlab{}.
\newblock \showarticletitle{Expanding Participation to Design with More-Than-Human Concerns}. In \bibinfo{booktitle}{\emph{Proceedings of the 16th Participatory Design Conference 2020 - Participation(s) Otherwise - Volume 1}} (Manizales, Colombia) \emph{(\bibinfo{series}{PDC '20})}. \bibinfo{publisher}{Association for Computing Machinery}, \bibinfo{address}{New York, NY, USA}, \bibinfo{pages}{1–11}.
\newblock
\showISBNx{9781450377003}
\urldef\tempurl%
\url{https://doi.org/10.1145/3385010.3385016}
\showDOI{\tempurl}


\bibitem[Altman(1975)]%
        {Altman1975environment}
\bibfield{author}{\bibinfo{person}{Irwin Altman}.} \bibinfo{year}{1975}\natexlab{}.
\newblock \bibinfo{booktitle}{\emph{The environment and social behavior: privacy, personal space, territory, crowding}}.
\newblock \bibinfo{publisher}{Brooks/Cole Pub. Co.}, \bibinfo{address}{Monterey, California}.
\newblock


\bibitem[Andrejevic(2005)]%
        {Andrejevic2005work}
\bibfield{author}{\bibinfo{person}{Mark Andrejevic}.} \bibinfo{year}{2005}\natexlab{}.
\newblock \showarticletitle{The work of watching one another: Lateral surveillance, risk, and governance}.
\newblock \bibinfo{journal}{\emph{Surveillance and Society}} \bibinfo{volume}{2}, \bibinfo{number}{4} (\bibinfo{year}{2005}), \bibinfo{pages}{479--497}.
\newblock
\showISSN{14777487}
\urldef\tempurl%
\url{https://doi.org/10.24908/ss.v2i4.3359}
\showDOI{\tempurl}


\bibitem[Bahirat et~al\mbox{.}(2021)]%
        {Bahirat2021Overlooking}
\bibfield{author}{\bibinfo{person}{Paritosh Bahirat}, \bibinfo{person}{Martijn~C. Willemsen}, \bibinfo{person}{Yangyang He}, \bibinfo{person}{Qizhang Sun}, {and} \bibinfo{person}{Bart~P. Knijnenburg}.} \bibinfo{year}{2021}\natexlab{}.
\newblock \showarticletitle{Overlooking Context : How do Defaults and Framing Reduce Deliberation in Smart Home Privacy Decision-Making?}. In \bibinfo{booktitle}{\emph{Proceedings of the 2021 CHI Conference on Human Factors in Computing Systems}} (2021). \bibinfo{publisher}{ACM}.
\newblock
\urldef\tempurl%
\url{https://doi.org/10.1145/3411764.3445672}
\showDOI{\tempurl}


\bibitem[Ballard et~al\mbox{.}(2019)]%
        {Ballard2019Judgment}
\bibfield{author}{\bibinfo{person}{Stephanie Ballard}, \bibinfo{person}{Karen~M. Chappell}, {and} \bibinfo{person}{Kristen Kennedy}.} \bibinfo{year}{2019}\natexlab{}.
\newblock \showarticletitle{Judgment Call the Game: Using value sensitive design and design fiction to surface ethical concerns related to technology}. In \bibinfo{booktitle}{\emph{Proceedings of the 2019 on Designing Interactive Systems Conference - DIS '19}} (2019). \bibinfo{publisher}{ACM Press}, \bibinfo{address}{New York, New York, USA}, \bibinfo{pages}{421--433}.
\newblock
\showISBNx{978-1-4503-5850-7}
\urldef\tempurl%
\url{https://doi.org/10.1145/3322276.3323697}
\showDOI{\tempurl}


\bibitem[Bardzell(2010)]%
        {Bardzell2010FeministHCI}
\bibfield{author}{\bibinfo{person}{Shaowen Bardzell}.} \bibinfo{year}{2010}\natexlab{}.
\newblock \showarticletitle{Feminist HCI: Taking Stock and Outlining an Agenda for Design}. In \bibinfo{booktitle}{\emph{Proceedings of the SIGCHI Conference on Human Factors in Computing Systems}} (Atlanta, Georgia, USA) \emph{(\bibinfo{series}{CHI '10})}. \bibinfo{publisher}{Association for Computing Machinery}, \bibinfo{address}{New York, NY, USA}, \bibinfo{pages}{1301–1310}.
\newblock
\showISBNx{9781605589299}
\urldef\tempurl%
\url{https://doi.org/10.1145/1753326.1753521}
\showDOI{\tempurl}


\bibitem[Barrett(2015)]%
        {Barrett2015Fuhu's}
\bibfield{author}{\bibinfo{person}{Brian Barrett}.} \bibinfo{year}{2015}\natexlab{}.
\newblock \showarticletitle{Fuhu's Kid-Centric Smart Home Is a Helicopter Parent's Dream}.
\newblock \bibinfo{journal}{\emph{Wired}} (\bibinfo{date}{8 7} \bibinfo{year}{2015}).
\newblock
\urldef\tempurl%
\url{https://www.wired.com/2015/07/fuhu-smart-home/}
\showURL{%
\tempurl}


\bibitem[Baumer(2015)]%
        {Baumer2015Usees}
\bibfield{author}{\bibinfo{person}{Eric~P.S. Baumer}.} \bibinfo{year}{2015}\natexlab{}.
\newblock \showarticletitle{Usees}. In \bibinfo{booktitle}{\emph{Proceedings of the 33rd Annual ACM Conference on Human Factors in Computing Systems}} (Seoul, Republic of Korea) \emph{(\bibinfo{series}{CHI '15})}. \bibinfo{publisher}{Association for Computing Machinery}, \bibinfo{address}{New York, NY, USA}, \bibinfo{pages}{3295–3298}.
\newblock
\showISBNx{9781450331456}
\urldef\tempurl%
\url{https://doi.org/10.1145/2702123.2702147}
\showDOI{\tempurl}


\bibitem[Baumer et~al\mbox{.}(2018)]%
        {Baumer2018WhatWouldYou}
\bibfield{author}{\bibinfo{person}{Eric~P.S. Baumer}, \bibinfo{person}{Timothy Berrill}, \bibinfo{person}{Sarah~C. Botwinick}, \bibinfo{person}{Jonathan~L. Gonzales}, \bibinfo{person}{Kevin Ho}, \bibinfo{person}{Allison Kundrik}, \bibinfo{person}{Luke Kwon}, \bibinfo{person}{Tim LaRowe}, \bibinfo{person}{Chanh~P. Nguyen}, \bibinfo{person}{Fredy Ramirez}, \bibinfo{person}{Peter Schaedler}, \bibinfo{person}{William Ulrich}, \bibinfo{person}{Amber Wallace}, \bibinfo{person}{Yuchen Wan}, {and} \bibinfo{person}{Benjamin Weinfeld}.} \bibinfo{year}{2018}\natexlab{}.
\newblock \showarticletitle{What Would You Do? Design Fiction and Ethics}. In \bibinfo{booktitle}{\emph{Proceedings of the 2018 ACM International Conference on Supporting Group Work}} (Sanibel Island, Florida, USA) \emph{(\bibinfo{series}{GROUP '18})}. \bibinfo{publisher}{Association for Computing Machinery}, \bibinfo{address}{New York, NY, USA}, \bibinfo{pages}{244–256}.
\newblock
\showISBNx{9781450355629}
\urldef\tempurl%
\url{https://doi.org/10.1145/3148330.3149405}
\showDOI{\tempurl}


\bibitem[Bernd et~al\mbox{.}(2022)]%
        {Bernd2022Balancing}
\bibfield{author}{\bibinfo{person}{Julia Bernd}, \bibinfo{person}{Ruba Abu-Salma}, \bibinfo{person}{Junghyun Choy}, {and} \bibinfo{person}{Alisa Frik}.} \bibinfo{year}{2022}\natexlab{}.
\newblock \showarticletitle{Balancing Power Dynamics in Smart Homes: Nannies' Perspectives on How Cameras Reflect and Affect Relationships}. In \bibinfo{booktitle}{\emph{Eighteenth Symposium on Usable Privacy and Security (SOUPS 2022)}} (2022). \bibinfo{publisher}{USENIX Association}, \bibinfo{pages}{687---706}.
\newblock
\urldef\tempurl%
\url{https://www.usenix.org/conference/soups2022/presentation/bernd}
\showURL{%
\tempurl}


\bibitem[Bernd et~al\mbox{.}(2020)]%
        {Bernd2020Bystanders'}
\bibfield{author}{\bibinfo{person}{Julia Bernd}, \bibinfo{person}{Ruba Abu-Salma}, {and} \bibinfo{person}{Alisa Frik}.} \bibinfo{year}{2020}\natexlab{}.
\newblock \showarticletitle{Bystanders' privacy: The perspectives of nannies on smart home surveillance}. In \bibinfo{booktitle}{\emph{FOCI 2020 - 10th USENIX Workshop on Free and Open Communications on the Internet}} (2020).
\newblock


\bibitem[Biggs et~al\mbox{.}(2021)]%
        {Biggs2021Moving}
\bibfield{author}{\bibinfo{person}{Heidi Biggs}, \bibinfo{person}{Cayla Key}, \bibinfo{person}{Audrey Desjardins}, {and} \bibinfo{person}{Afroditi Psarra}.} \bibinfo{year}{2021}\natexlab{}.
\newblock \showarticletitle{Moving Design Research: GIFs as Research Tools}. In \bibinfo{booktitle}{\emph{Designing Interactive Systems Conference 2021}} (Virtual Event, USA) \emph{(\bibinfo{series}{DIS '21})}. \bibinfo{publisher}{Association for Computing Machinery}, \bibinfo{address}{New York, NY, USA}, \bibinfo{pages}{1927–1940}.
\newblock
\showISBNx{9781450384766}
\urldef\tempurl%
\url{https://doi.org/10.1145/3461778.3462144}
\showDOI{\tempurl}


\bibitem[Bohn(2020)]%
        {Bohn2020Ring}
\bibfield{author}{\bibinfo{person}{Dieter Bohn}.} \bibinfo{year}{2020}\natexlab{}.
\newblock \showarticletitle{The Ring drone is just the latest Amazon privacy puzzle box}.
\newblock \bibinfo{journal}{\emph{The Verge}} (\bibinfo{year}{2020}).
\newblock
\urldef\tempurl%
\url{https://www.theverge.com/2020/9/25/21455197/amazon-ring-drone-home-security-surveillance-sidewalk-halo-privacy}
\showURL{%
\tempurl}
\newblock
\shownote{[Online; accessed 2021-09-09]}.


\bibitem[Boucher et~al\mbox{.}(2012)]%
        {Boucher2012Power}
\bibfield{author}{\bibinfo{person}{Andy Boucher}, \bibinfo{person}{David Cameron}, {and} \bibinfo{person}{Nadine Jarvis}.} \bibinfo{year}{2012}\natexlab{}.
\newblock \showarticletitle{Power to the People: Dynamic Energy Management through Communal Cooperation}. In \bibinfo{booktitle}{\emph{Proceedings of the Designing Interactive Systems Conference}} (Newcastle Upon Tyne, United Kingdom) \emph{(\bibinfo{series}{DIS '12})}. \bibinfo{publisher}{Association for Computing Machinery}, \bibinfo{address}{New York, NY, USA}, \bibinfo{pages}{612–620}.
\newblock
\showISBNx{9781450312103}
\urldef\tempurl%
\url{https://doi.org/10.1145/2317956.2318048}
\showDOI{\tempurl}


\bibitem[Briggs and Thomas(2015)]%
        {Briggs2015Inclusive}
\bibfield{author}{\bibinfo{person}{Pam Briggs} {and} \bibinfo{person}{Lisa Thomas}.} \bibinfo{year}{2015}\natexlab{}.
\newblock \showarticletitle{An Inclusive, Value Sensitive Design Perspective on Future Identity Technologies}.
\newblock \bibinfo{journal}{\emph{ACM Transactions on Computer-Human Interaction}} \bibinfo{volume}{22}, \bibinfo{number}{5} (\bibinfo{date}{10 8} \bibinfo{year}{2015}), \bibinfo{pages}{1--28}.
\newblock
\showISSN{10730516}
\urldef\tempurl%
\url{https://doi.org/10.1145/2778972}
\showDOI{\tempurl}


\bibitem[Browne(2015)]%
        {Browne2015Dark}
\bibfield{author}{\bibinfo{person}{Simone Browne}.} \bibinfo{year}{2015}\natexlab{}.
\newblock \bibinfo{booktitle}{\emph{Dark Matters: On the Surveillance of Blackness}}.
\newblock \bibinfo{publisher}{Duke University Press}, \bibinfo{address}{Durham}.
\newblock


\bibitem[Buehring and Liedtka(2018)]%
        {Buehring2018Embracing}
\bibfield{author}{\bibinfo{person}{Joern~Henning Buehring} {and} \bibinfo{person}{Jeanne Liedtka}.} \bibinfo{year}{2018}\natexlab{}.
\newblock \showarticletitle{Embracing systematic futures thinking at the intersection of Strategic Planning, Foresight and Design}.
\newblock \bibinfo{journal}{\emph{Journal of Innovation Management}} \bibinfo{volume}{6}, \bibinfo{number}{3} (\bibinfo{date}{23 11} \bibinfo{year}{2018}), \bibinfo{pages}{134}.
\newblock
\showISSN{2183-0606}
\urldef\tempurl%
\url{https://doi.org/10.24840/2183-0606_006.003_0006}
\showDOI{\tempurl}


\bibitem[Butler et~al\mbox{.}(2015)]%
        {Butler2015Privacy-Utility}
\bibfield{author}{\bibinfo{person}{Daniel~J. Butler}, \bibinfo{person}{Justin Huang}, \bibinfo{person}{Franziska Roesner}, {and} \bibinfo{person}{Maya Cakmak}.} \bibinfo{year}{2015}\natexlab{}.
\newblock \showarticletitle{The Privacy-Utility Tradeoff for Remotely Teleoperated Robots}. In \bibinfo{booktitle}{\emph{Proceedings of the Tenth Annual ACM/IEEE International Conference on Human-Robot Interaction}} (2015-03-02). \bibinfo{publisher}{ACM}, \bibinfo{address}{Portland Oregon USA}, \bibinfo{pages}{27--34}.
\newblock
\showISBNx{978-1-4503-2883-8}
\urldef\tempurl%
\url{https://doi.org/10.1145/2696454.2696484}
\showDOI{\tempurl}


\bibitem[Carroll(1995)]%
        {1995Scenario-based}
\bibfield{editor}{\bibinfo{person}{John~M Carroll}} (Ed.). \bibinfo{year}{1995}\natexlab{}.
\newblock \bibinfo{booktitle}{\emph{Scenario-based design: Envisioning work and technology in system development}}.
\newblock \bibinfo{publisher}{John Wiley \& Sons, Inc.}, \bibinfo{address}{New York}.
\newblock


\bibitem[Chalhoub et~al\mbox{.}(2021)]%
        {Chalhoub2021itdidnot}
\bibfield{author}{\bibinfo{person}{George Chalhoub}, \bibinfo{person}{Martin~J Kraemer}, \bibinfo{person}{Norbert Nthala}, {and} \bibinfo{person}{Ivan Flechais}.} \bibinfo{year}{2021}\natexlab{}.
\newblock \showarticletitle{“It Did Not Give Me an Option to Decline”: A Longitudinal Analysis of the User Experience of Security and Privacy in Smart Home Products}. In \bibinfo{booktitle}{\emph{Proceedings of the 2021 CHI Conference on Human Factors in Computing Systems}} (Yokohama, Japan) \emph{(\bibinfo{series}{CHI '21})}. \bibinfo{publisher}{Association for Computing Machinery}, \bibinfo{address}{New York, NY, USA}, Article \bibinfo{articleno}{555}, \bibinfo{numpages}{16}~pages.
\newblock
\showISBNx{9781450380966}
\urldef\tempurl%
\url{https://doi.org/10.1145/3411764.3445691}
\showDOI{\tempurl}


\bibitem[Chen et~al\mbox{.}(2021)]%
        {Chen2021What}
\bibfield{author}{\bibinfo{person}{Janet~X. Chen}, \bibinfo{person}{Francesco Vitale}, {and} \bibinfo{person}{Joanna McGrenere}.} \bibinfo{year}{2021}\natexlab{}.
\newblock \showarticletitle{What Happens After Death? Using a Design Workbook to Understand User Expectations for Preparing their Data}. In \bibinfo{booktitle}{\emph{Proceedings of the 2021 CHI Conference on Human Factors in Computing Systems}} (2021-05-06). \bibinfo{publisher}{ACM}, \bibinfo{address}{New York, NY, USA}, \bibinfo{pages}{1--13}.
\newblock
\showISBNx{978-1-4503-8096-6}
\urldef\tempurl%
\url{https://doi.org/10.1145/3411764.3445359}
\showDOI{\tempurl}


\bibitem[Chen et~al\mbox{.}(2020)]%
        {Chen2020Wearable}
\bibfield{author}{\bibinfo{person}{Yuxin Chen}, \bibinfo{person}{Huiying Li}, \bibinfo{person}{Shan-Yuan Teng}, \bibinfo{person}{Steven Nagels}, \bibinfo{person}{Zhijing Li}, \bibinfo{person}{Pedro Lopes}, \bibinfo{person}{Ben~Y. Zhao}, {and} \bibinfo{person}{Haitao Zheng}.} \bibinfo{year}{2020}\natexlab{}.
\newblock \showarticletitle{Wearable Microphone Jamming}. In \bibinfo{booktitle}{\emph{Proceedings of the 2020 CHI Conference on Human Factors in Computing Systems}} (Honolulu, HI, USA) \emph{(\bibinfo{series}{CHI '20})}. \bibinfo{publisher}{Association for Computing Machinery}, \bibinfo{address}{New York, NY, USA}, \bibinfo{pages}{1–12}.
\newblock
\showISBNx{9781450367080}
\urldef\tempurl%
\url{https://doi.org/10.1145/3313831.3376304}
\showDOI{\tempurl}


\bibitem[Cheng et~al\mbox{.}(2019)]%
        {Cheng2019Peekaboo}
\bibfield{author}{\bibinfo{person}{Yu-Ting Cheng}, \bibinfo{person}{Mathias Funk}, \bibinfo{person}{Wenn-Chieh Tsai}, {and} \bibinfo{person}{Lin-Lin Chen}.} \bibinfo{year}{2019}\natexlab{}.
\newblock \showarticletitle{Peekaboo Cam: Designing an observational camera for home ecologies concerning privacy}. In \bibinfo{booktitle}{\emph{Proceedings of the 2019 on Designing Interactive Systems Conference}} (2019-06-18). \bibinfo{publisher}{ACM}, \bibinfo{address}{New York, NY, USA}, \bibinfo{pages}{823--836}.
\newblock
\showISBNx{978-1-4503-5850-7}
\urldef\tempurl%
\url{https://doi.org/10.1145/3322276.3323699}
\showDOI{\tempurl}


\bibitem[Choe et~al\mbox{.}(2012)]%
        {Choe2012Investigating}
\bibfield{author}{\bibinfo{person}{Eun~Kyoung Choe}, \bibinfo{person}{Sunny Consolvo}, \bibinfo{person}{Jaeyeon Jung}, \bibinfo{person}{Beverly Harrison}, \bibinfo{person}{Shwetak~N. Patel}, {and} \bibinfo{person}{Julie~A. Kientz}.} \bibinfo{year}{2012}\natexlab{}.
\newblock \showarticletitle{Investigating receptiveness to sensing and inference in the home using sensor proxies}. In \bibinfo{booktitle}{\emph{Proceedings of the 2012 ACM Conference on Ubiquitous Computing - UbiComp '12}} (2012). \bibinfo{publisher}{ACM Press}, \bibinfo{address}{Pittsburgh, Pennsylvania}, \bibinfo{pages}{61}.
\newblock
\showISBNx{978-1-4503-1224-0}
\urldef\tempurl%
\url{https://doi.org/10.1145/2370216.2370226}
\showDOI{\tempurl}


\bibitem[Citron and Solove(2022)]%
        {Citron2022Privacy}
\bibfield{author}{\bibinfo{person}{Danielle~Keats Citron} {and} \bibinfo{person}{Daniel~J. Solove}.} \bibinfo{year}{2022}\natexlab{}.
\newblock \showarticletitle{Privacy Harms}.
\newblock \bibinfo{journal}{\emph{Boston University Law Review}} \bibinfo{volume}{102}, \bibinfo{number}{3} (\bibinfo{year}{2022}), \bibinfo{pages}{793--864}.
\newblock
\showISSN{1556-5068}
\urldef\tempurl%
\url{https://doi.org/10.2139/ssrn.3782222}
\showDOI{\tempurl}


\bibitem[Cobb et~al\mbox{.}(2021)]%
        {Cobb2021Iwould}
\bibfield{author}{\bibinfo{person}{Camille Cobb}, \bibinfo{person}{Sruti Bhagavatula}, \bibinfo{person}{Kalil~Anderson Garrett}, \bibinfo{person}{Alison Hoffman}, \bibinfo{person}{Varun Rao}, {and} \bibinfo{person}{Lujo Bauer}.} \bibinfo{year}{2021}\natexlab{}.
\newblock \showarticletitle{“I would have to evaluate their objections”: Privacy tensions between smart home device owners and incidental users}.
\newblock \bibinfo{journal}{\emph{Proceedings on Privacy Enhancing Technologies}} \bibinfo{volume}{2021}, \bibinfo{number}{4} (\bibinfo{date}{1 10} \bibinfo{year}{2021}), \bibinfo{pages}{54--75}.
\newblock
\showISSN{2299-0984}
\urldef\tempurl%
\url{https://doi.org/10.2478/popets-2021-0060}
\showDOI{\tempurl}


\bibitem[David-John et~al\mbox{.}(2021)]%
        {david2021let}
\bibfield{author}{\bibinfo{person}{Brendan David-John}, \bibinfo{person}{Diane Hosfelt}, \bibinfo{person}{Kevin Butler}, {and} \bibinfo{person}{Eakta Jain}.} \bibinfo{year}{2021}\natexlab{}.
\newblock \showarticletitle{Let’s SOUP up XR: Collected thoughts from an IEEE VR workshop on privacy in mixed reality}. In \bibinfo{booktitle}{\emph{VR4Sec: Security for VR and VR for Security, SOUPS 2021 Workshop}}.
\newblock


\bibitem[de~la Bellacasa(2010)]%
        {delaBellacasa2010Matters}
\bibfield{author}{\bibinfo{person}{Maria~Puig de~la Bellacasa}.} \bibinfo{year}{2010}\natexlab{}.
\newblock \showarticletitle{Matters of care in technoscience: Assembling neglected things}.
\newblock \bibinfo{journal}{\emph{Social Studies of Science}} \bibinfo{volume}{41}, \bibinfo{number}{1} (\bibinfo{date}{Dec.} \bibinfo{year}{2010}), \bibinfo{pages}{85--106}.
\newblock
\urldef\tempurl%
\url{https://doi.org/10.1177/0306312710380301}
\showDOI{\tempurl}


\bibitem[Denefleh et~al\mbox{.}(2019)]%
        {Denefleh2019Sensorstation:}
\bibfield{author}{\bibinfo{person}{Teresa Denefleh}, \bibinfo{person}{Arne Berger}, \bibinfo{person}{Albrecht Kurze}, {and} \bibinfo{person}{Chris Frauenberger}.} \bibinfo{year}{2019}\natexlab{}.
\newblock \showarticletitle{Sensorstation: Exploring Simple Sensor Data in the Context of a Shared Apartment}. In \bibinfo{booktitle}{\emph{Proceedings of the 2019 Designing Interactive Systems Conference (DIS)}} (2019).
\newblock
\urldef\tempurl%
\url{https://doi.org/10.1145/3322276.3322309}
\showDOI{\tempurl}


\bibitem[DiSalvo et~al\mbox{.}(2016)]%
        {Disalvo2016Designing}
\bibfield{author}{\bibinfo{person}{Carl DiSalvo}, \bibinfo{person}{Tom Jenkins}, {and} \bibinfo{person}{Thomas Lodato}.} \bibinfo{year}{2016}\natexlab{}.
\newblock \showarticletitle{Designing Speculative Civics}. In \bibinfo{booktitle}{\emph{Proceedings of the 2016 CHI Conference on Human Factors in Computing Systems}} (San Jose, California, USA) \emph{(\bibinfo{series}{CHI '16})}. \bibinfo{publisher}{Association for Computing Machinery}, \bibinfo{address}{New York, NY, USA}, \bibinfo{pages}{4979–4990}.
\newblock
\showISBNx{9781450333627}
\urldef\tempurl%
\url{https://doi.org/10.1145/2858036.2858505}
\showDOI{\tempurl}


\bibitem[DiSalvo et~al\mbox{.}(2014)]%
        {DiSalvo2014Making}
\bibfield{author}{\bibinfo{person}{Carl DiSalvo}, \bibinfo{person}{Jonathan Lukens}, \bibinfo{person}{Thomas Lodato}, \bibinfo{person}{Tom Jenkins}, {and} \bibinfo{person}{Tanyoung Kim}.} \bibinfo{year}{2014}\natexlab{}.
\newblock \showarticletitle{Making public things: How HCI Design Can Express Matters of Concern}. In \bibinfo{booktitle}{\emph{Proceedings of the SIGCHI Conference on Human Factors in Computing Systems}} (2014-04-26). \bibinfo{publisher}{ACM}, \bibinfo{address}{New York, NY, USA}, \bibinfo{pages}{2397--2406}.
\newblock
\showISBNx{978-1-4503-2473-1}
\urldef\tempurl%
\url{https://doi.org/10.1145/2556288.2557359}
\showDOI{\tempurl}


\bibitem[Dwork and Mulligan(2013)]%
        {Dwork2013It's}
\bibfield{author}{\bibinfo{person}{Cynthia Dwork} {and} \bibinfo{person}{Mulligan}.} \bibinfo{year}{2013}\natexlab{}.
\newblock \showarticletitle{It's not privacy, and it's not fair}.
\newblock \bibinfo{journal}{\emph{Stanford Law Review Online}}  \bibinfo{volume}{66} (\bibinfo{year}{2013}), \bibinfo{pages}{35--40}.
\newblock
\showISSN{03014215}


\bibitem[Ehrenberg and Keinonen(2021)]%
        {Ehrenberg2021Technology}
\bibfield{author}{\bibinfo{person}{Nils Ehrenberg} {and} \bibinfo{person}{Turkka Keinonen}.} \bibinfo{year}{2021}\natexlab{}.
\newblock \showarticletitle{The Technology Is Enemy for Me at the Moment: How Smart Home Technologies Assert Control Beyond Intent}. In \bibinfo{booktitle}{\emph{Proceedings of the 2021 CHI Conference on Human Factors in Computing Systems}} (2021-05-06). \bibinfo{publisher}{ACM}, \bibinfo{address}{New York, NY, USA}, \bibinfo{pages}{1--11}.
\newblock
\showISBNx{978-1-4503-8096-6}
\urldef\tempurl%
\url{https://doi.org/10.1145/3411764.3445058}
\showDOI{\tempurl}


\bibitem[Elsden et~al\mbox{.}(2017)]%
        {Elsden2017On}
\bibfield{author}{\bibinfo{person}{Chris Elsden}, \bibinfo{person}{David Chatting}, \bibinfo{person}{Abigail~C. Durrant}, \bibinfo{person}{Andrew Garbett}, \bibinfo{person}{Bettina Nissen}, \bibinfo{person}{John Vines}, {and} \bibinfo{person}{David~S. Kirk}.} \bibinfo{year}{2017}\natexlab{}.
\newblock \showarticletitle{On Speculative Enactments}. In \bibinfo{booktitle}{\emph{Proceedings of the 2017 CHI Conference on Human Factors in Computing Systems}} (2017-05-02). \bibinfo{publisher}{ACM}, \bibinfo{address}{New York, NY, USA}, \bibinfo{pages}{5386--5399}.
\newblock
\showISBNx{978-1-4503-4655-9}
\urldef\tempurl%
\url{https://doi.org/10.1145/3025453.3025503}
\showDOI{\tempurl}


\bibitem[Epp et~al\mbox{.}(2022)]%
        {Epp2022Reinventing}
\bibfield{author}{\bibinfo{person}{Felix~Anand Epp}, \bibinfo{person}{Tim Moesgen}, \bibinfo{person}{Antti Salovaara}, \bibinfo{person}{Emmi Pouta}, {and} \bibinfo{person}{İdil Gaziulusoy}.} \bibinfo{year}{2022}\natexlab{}.
\newblock \showarticletitle{Reinventing the Wheel: The Future Ripples Method for Activating Anticipatory Capacities in Innovation Teams}. In \bibinfo{booktitle}{\emph{Designing Interactive Systems Conference}} (2022-06-13). \bibinfo{publisher}{ACM}, \bibinfo{address}{Virtual Event Australia}, \bibinfo{pages}{387--399}.
\newblock
\showISBNx{978-1-4503-9358-4}
\urldef\tempurl%
\url{https://doi.org/10.1145/3532106.3534570}
\showDOI{\tempurl}


\bibitem[Evans(2003)]%
        {Evans2003Trend}
\bibfield{author}{\bibinfo{person}{Martyn Evans}.} \bibinfo{year}{2003}\natexlab{}.
\newblock \showarticletitle{Trend Forecasting for Design Futures}. In \bibinfo{booktitle}{\emph{European Academy of Design Conference}} (2003).
\newblock


\bibitem[Feng et~al\mbox{.}(2021)]%
        {Feng2021Design}
\bibfield{author}{\bibinfo{person}{Yuanyuan Feng}, \bibinfo{person}{Yaxing Yao}, {and} \bibinfo{person}{Norman Sadeh}.} \bibinfo{year}{2021}\natexlab{}.
\newblock \showarticletitle{A Design Space for Privacy Choices: Towards Meaningful Privacy Control in the Internet of Things}.
\newblock \bibinfo{journal}{\emph{Proceedings of the 2021 CHI Conference on Human Factors in Computing Systems}}, \bibinfo{pages}{1--16}.
\newblock
\showISBNx{978-1-4503-8096-6}
\urldef\tempurl%
\url{https://doi.org/10.1145/3411764.3445148}
\showDOI{\tempurl}


\bibitem[Fiesler(2019)]%
        {Fiesler2019Ethical}
\bibfield{author}{\bibinfo{person}{Casey Fiesler}.} \bibinfo{year}{2019}\natexlab{}.
\newblock \showarticletitle{Ethical Considerations for Research Involving (Speculative) Public Data}.
\newblock \bibinfo{journal}{\emph{Proceedings of the ACM on Human-Computer Interaction}} \bibinfo{volume}{3}, \bibinfo{number}{GROUP} (\bibinfo{date}{5 12} \bibinfo{year}{2019}), \bibinfo{pages}{1--13}.
\newblock
\showISSN{2573-0142}
\urldef\tempurl%
\url{https://doi.org/10.1145/3370271}
\showDOI{\tempurl}


\bibitem[Fiesler(2021)]%
        {Fiesler2021Ethical}
\bibfield{author}{\bibinfo{person}{Casey Fiesler}.} \bibinfo{year}{2021}\natexlab{}.
\newblock \showarticletitle{Ethical Speculation in the Computing Classroom}. In \bibinfo{booktitle}{\emph{2021 Conference on Research in Equitable and Sustained Participation in Engineering, Computing, and Technology (RESPECT)}} (2021-05-23). \bibinfo{publisher}{IEEE}, \bibinfo{address}{Philadelphia, PA, USA}, \bibinfo{pages}{1--1}.
\newblock
\showISBNx{978-1-66544-905-2}
\urldef\tempurl%
\url{https://doi.org/10.1109/RESPECT51740.2021.9620641}
\showDOI{\tempurl}


\bibitem[Flanagan and Nissenbaum(2014)]%
        {Flanagan2014Groundwork}
\bibfield{author}{\bibinfo{person}{Mary Flanagan} {and} \bibinfo{person}{Helen Nissenbaum}.} \bibinfo{year}{2014}\natexlab{}.
\newblock \showarticletitle{Groundwork for Values in Games}.
\newblock In \bibinfo{booktitle}{\emph{Values at Play in Digital Games}}. \bibinfo{publisher}{MIT Press}, \bibinfo{address}{Cambridge, Massachusetts}.
\newblock


\bibitem[for Long-Term~Cybersecurity(2016)]%
        {CLTC2016Cybersecurity}
\bibfield{author}{\bibinfo{person}{Center for Long-Term~Cybersecurity}.} \bibinfo{year}{2016}\natexlab{}.
\newblock \bibinfo{booktitle}{\emph{Cybersecurity futures 2020}}.
\newblock \bibinfo{type}{{T}echnical {R}eport}. \bibinfo{pages}{128} pages.
\newblock
\urldef\tempurl%
\url{https://cltc.berkeley.edu/2016/04/28/cybersecurity-futures-2020/}
\showURL{%
\tempurl}


\bibitem[Freed et~al\mbox{.}(2018)]%
        {Freed2018"A}
\bibfield{author}{\bibinfo{person}{Diana Freed}, \bibinfo{person}{Jackeline Palmer}, \bibinfo{person}{Diana Minchala}, \bibinfo{person}{Karen Levy}, \bibinfo{person}{Thomas Ristenpart}, {and} \bibinfo{person}{Nicola Dell}.} \bibinfo{year}{2018}\natexlab{}.
\newblock \showarticletitle{"A Stalker's Paradise": How Intimate Partner Abusers Exploit Technology}. In \bibinfo{booktitle}{\emph{Proceedings of the 2018 CHI Conference on Human Factors in Computing Systems - CHI '18}} (2018). \bibinfo{publisher}{ACM Press}, \bibinfo{address}{New York, New York, USA}, \bibinfo{pages}{1--13}.
\newblock
\showISBNx{978-1-4503-5620-6}
\urldef\tempurl%
\url{https://doi.org/10.1145/3173574.3174241}
\showDOI{\tempurl}


\bibitem[Friedman et~al\mbox{.}(2017)]%
        {Friedman2017Survey}
\bibfield{author}{\bibinfo{person}{Batya Friedman}, \bibinfo{person}{David~G. Hendry}, {and} \bibinfo{person}{Alan Borning}.} \bibinfo{year}{2017}\natexlab{}.
\newblock \showarticletitle{A Survey of Value Sensitive Design Methods}.
\newblock \bibinfo{journal}{\emph{Foundations and Trends® in Human–Computer Interaction}} \bibinfo{volume}{11}, \bibinfo{number}{2} (\bibinfo{year}{2017}), \bibinfo{pages}{63--125}.
\newblock
\showISSN{1551-3955}
\urldef\tempurl%
\url{https://doi.org/10.1561/1100000015}
\showDOI{\tempurl}


\bibitem[Garg and Cui(2022)]%
        {Garg2022Social}
\bibfield{author}{\bibinfo{person}{Radhika Garg} {and} \bibinfo{person}{Hua Cui}.} \bibinfo{year}{2022}\natexlab{}.
\newblock \showarticletitle{Social Contexts, Agency, and Conflicts: Exploring Critical Aspects of Design for Future Smart Home Technologies}.
\newblock \bibinfo{journal}{\emph{ACM Transactions on Computer-Human Interaction}} \bibinfo{volume}{29}, \bibinfo{number}{2} (\bibinfo{date}{30 4} \bibinfo{year}{2022}), \bibinfo{pages}{1--30}.
\newblock
\showISSN{1073-0516, 1557-7325}
\urldef\tempurl%
\url{https://doi.org/10.1145/3485058}
\showDOI{\tempurl}


\bibitem[Gaver and Martin(2000)]%
        {Gaver2000Alternatives}
\bibfield{author}{\bibinfo{person}{Bill Gaver} {and} \bibinfo{person}{Heather Martin}.} \bibinfo{year}{2000}\natexlab{}.
\newblock \showarticletitle{Alternatives: Exploring Information Appliances through Conceptual Design Proposals}. In \bibinfo{booktitle}{\emph{Proceedings of the SIGCHI Conference on Human Factors in Computing Systems}} (The Hague, The Netherlands) \emph{(\bibinfo{series}{CHI '00})}. \bibinfo{publisher}{Association for Computing Machinery}, \bibinfo{address}{New York, NY, USA}, \bibinfo{pages}{209–216}.
\newblock
\showISBNx{1581132166}
\urldef\tempurl%
\url{https://doi.org/10.1145/332040.332433}
\showDOI{\tempurl}


\bibitem[Gaver(2011)]%
        {Gaver2011Making}
\bibfield{author}{\bibinfo{person}{William Gaver}.} \bibinfo{year}{2011}\natexlab{}.
\newblock \showarticletitle{Making spaces: how design workbooks work}. In \bibinfo{booktitle}{\emph{Proceedings of the SIGCHI Conference on Human Factors in Computing Systems (CHI '11)}} (2011). \bibinfo{pages}{1551--1560}.
\newblock
\showISBNx{978-1-4503-0267-8}
\urldef\tempurl%
\url{https://doi.org/10.1145/1978942.1979169}
\showDOI{\tempurl}


\bibitem[Giaccardi et~al\mbox{.}(2016)]%
        {Giaccardi2016ThingEthnography}
\bibfield{author}{\bibinfo{person}{Elisa Giaccardi}, \bibinfo{person}{Nazli Cila}, \bibinfo{person}{Chris Speed}, {and} \bibinfo{person}{Melissa Caldwell}.} \bibinfo{year}{2016}\natexlab{}.
\newblock \showarticletitle{Thing Ethnography: Doing Design Research with Non-Humans}. In \bibinfo{booktitle}{\emph{Proceedings of the 2016 ACM Conference on Designing Interactive Systems}} (Brisbane, QLD, Australia) \emph{(\bibinfo{series}{DIS '16})}. \bibinfo{publisher}{Association for Computing Machinery}, \bibinfo{address}{New York, NY, USA}, \bibinfo{pages}{377–387}.
\newblock
\showISBNx{9781450340311}
\urldef\tempurl%
\url{https://doi.org/10.1145/2901790.2901905}
\showDOI{\tempurl}


\bibitem[Harrington and Dillahunt(2021)]%
        {Harrington2021Eliciting}
\bibfield{author}{\bibinfo{person}{Christina Harrington} {and} \bibinfo{person}{Tawanna~R. Dillahunt}.} \bibinfo{year}{2021}\natexlab{}.
\newblock \showarticletitle{Eliciting Tech Futures Among Black Young Adults: A Case Study of Remote Speculative Co-Design}. In \bibinfo{booktitle}{\emph{Proceedings of the 2021 CHI Conference on Human Factors in Computing Systems}} (2021-05-06). \bibinfo{publisher}{ACM}, \bibinfo{address}{New York, NY, USA}, \bibinfo{pages}{1--15}.
\newblock
\showISBNx{978-1-4503-8096-6}
\urldef\tempurl%
\url{https://doi.org/10.1145/3411764.3445723}
\showDOI{\tempurl}


\bibitem[Hutchinson et~al\mbox{.}(2003)]%
        {Hutchinson2003Technology}
\bibfield{author}{\bibinfo{person}{Hilary Hutchinson}, \bibinfo{person}{Heiko Hansen}, \bibinfo{person}{Nicolas Roussel}, \bibinfo{person}{Björn Eiderbäck}, \bibinfo{person}{Wendy Mackay}, \bibinfo{person}{Bosse Westerlund}, \bibinfo{person}{Benjamin~B Bederson}, \bibinfo{person}{Allison Druin}, \bibinfo{person}{Catherine Plaisant}, \bibinfo{person}{Michel Beaudouin-Lafon}, \bibinfo{person}{Stéphane Conversy}, {and} \bibinfo{person}{Helen Evans}.} \bibinfo{year}{2003}\natexlab{}.
\newblock \showarticletitle{Technology probes}. In \bibinfo{booktitle}{\emph{Proceedings of the conference on Human factors in computing systems (CHI '03)}} (2003). \bibinfo{publisher}{ACM Press}, \bibinfo{address}{New York, New York, USA}, \bibinfo{pages}{17--24}.
\newblock
\showISBNx{1-58113-630-7}
\urldef\tempurl%
\url{https://doi.org/10.1145/642611.642616}
\showDOI{\tempurl}


\bibitem[Jacobs(2016)]%
        {Jacobs2016Attending}
\bibfield{author}{\bibinfo{person}{Alan Jacobs}.} \bibinfo{year}{2016}\natexlab{}.
\newblock \showarticletitle{Attending to Technology - Theses for Disputation}.
\newblock \bibinfo{journal}{\emph{The New Atlantis}} (\bibinfo{year}{2016}).
\newblock
\urldef\tempurl%
\url{https://www.thenewatlantis.com/publications/attending-to-technology-theses-for-disputation}
\showURL{%
\tempurl}


\bibitem[JafariNaimi et~al\mbox{.}(2015)]%
        {JafariNaimi2015Values}
\bibfield{author}{\bibinfo{person}{Nassim JafariNaimi}, \bibinfo{person}{Lisa Nathan}, {and} \bibinfo{person}{Ian Hargraves}.} \bibinfo{year}{2015}\natexlab{}.
\newblock \showarticletitle{Values as Hypotheses: Design, Inquiry, and the Service of Values}.
\newblock \bibinfo{journal}{\emph{Design Issues}} \bibinfo{volume}{31}, \bibinfo{number}{4} (\bibinfo{date}{10} \bibinfo{year}{2015}), \bibinfo{pages}{91--104}.
\newblock
\showISSN{0747-9360}
\urldef\tempurl%
\url{https://doi.org/10.1162/DESI_a_00354}
\showDOI{\tempurl}


\bibitem[Jenkins(2017)]%
        {Jenkins2017Living}
\bibfield{author}{\bibinfo{person}{Tom Jenkins}.} \bibinfo{year}{2017}\natexlab{}.
\newblock \showarticletitle{Living Apart, Together: Cohousing as a Site for ICT Design}. In \bibinfo{booktitle}{\emph{Proceedings of the 2017 Conference on Designing Interactive Systems}} (2017-06-10). \bibinfo{publisher}{ACM}, \bibinfo{address}{Edinburgh United Kingdom}, \bibinfo{pages}{1039--1051}.
\newblock
\showISBNx{978-1-4503-4922-2}
\urldef\tempurl%
\url{https://doi.org/10.1145/3064663.3064751}
\showDOI{\tempurl}


\bibitem[Key et~al\mbox{.}(2021)]%
        {Key2021Proceed}
\bibfield{author}{\bibinfo{person}{Cayla Key}, \bibinfo{person}{Fiona Browne}, \bibinfo{person}{Nick Taylor}, {and} \bibinfo{person}{Jon Rogers}.} \bibinfo{year}{2021}\natexlab{}.
\newblock \showarticletitle{Proceed with Care: Reimagining Home IoT Through a Care Perspective}. In \bibinfo{booktitle}{\emph{Proceedings of the 2021 CHI Conference on Human Factors in Computing Systems}} (Yokohama, Japan) \emph{(\bibinfo{series}{CHI '21})}. \bibinfo{publisher}{Association for Computing Machinery}, \bibinfo{address}{New York, NY, USA}, Article \bibinfo{articleno}{166}, \bibinfo{numpages}{15}~pages.
\newblock
\showISBNx{9781450380966}
\urldef\tempurl%
\url{https://doi.org/10.1145/3411764.3445602}
\showDOI{\tempurl}


\bibitem[Knowles et~al\mbox{.}(2019)]%
        {Knowles2019Scenario-Based}
\bibfield{author}{\bibinfo{person}{Bran Knowles}, \bibinfo{person}{Sophie Beck}, \bibinfo{person}{Joe Finney}, \bibinfo{person}{James Devine}, {and} \bibinfo{person}{Joseph Lindley}.} \bibinfo{year}{2019}\natexlab{}.
\newblock \showarticletitle{A Scenario-Based Methodology for Exploring Risks: Children and Programmable IoT}. In \bibinfo{booktitle}{\emph{Proceedings of the 2019 on Designing Interactive Systems Conference}} (2019-06-18). \bibinfo{publisher}{ACM}, \bibinfo{address}{New York, NY, USA}, \bibinfo{pages}{751--761}.
\newblock
\showISBNx{978-1-4503-5850-7}
\urldef\tempurl%
\url{https://doi.org/10.1145/3322276.3322315}
\showDOI{\tempurl}


\bibitem[Koshy et~al\mbox{.}(2021)]%
        {Koshy2021WeJustUse}
\bibfield{author}{\bibinfo{person}{Vinay Koshy}, \bibinfo{person}{Joon Sung~Sung Park}, \bibinfo{person}{Ti-Chung Cheng}, {and} \bibinfo{person}{Karrie Karahalios}.} \bibinfo{year}{2021}\natexlab{}.
\newblock \showarticletitle{“We Just Use What They Give Us”: Understanding Passenger User Perspectives in Smart Homes}. In \bibinfo{booktitle}{\emph{Proceedings of the 2021 CHI Conference on Human Factors in Computing Systems}} (Yokohama, Japan) \emph{(\bibinfo{series}{CHI '21})}. \bibinfo{publisher}{Association for Computing Machinery}, \bibinfo{address}{New York, NY, USA}, Article \bibinfo{articleno}{41}, \bibinfo{numpages}{14}~pages.
\newblock
\showISBNx{9781450380966}
\urldef\tempurl%
\url{https://doi.org/10.1145/3411764.3445598}
\showDOI{\tempurl}


\bibitem[Kozubaev et~al\mbox{.}(2020)]%
        {Kozubaev2020Expanding}
\bibfield{author}{\bibinfo{person}{Sandjar Kozubaev}, \bibinfo{person}{Chris Elsden}, \bibinfo{person}{Noura Howell}, \bibinfo{person}{Marie Louise~Juul Søndergaard}, \bibinfo{person}{Nick Merrill}, \bibinfo{person}{Britta Schulte}, {and} \bibinfo{person}{Richmond~Y Wong}.} \bibinfo{year}{2020}\natexlab{}.
\newblock \showarticletitle{Expanding Modes of Reflection in Design Futuring}. In \bibinfo{booktitle}{\emph{Proceedings of the 2020 CHI Conference on Human Factors in Computing Systems}} (2020-04-21). \bibinfo{publisher}{ACM}, \bibinfo{address}{New York, NY, USA}, \bibinfo{pages}{1--15}.
\newblock
\showISBNx{978-1-4503-6708-0}
\urldef\tempurl%
\url{https://doi.org/10.1145/3313831.3376526}
\showDOI{\tempurl}


\bibitem[Kozubaev et~al\mbox{.}(2019)]%
        {Kozubaev2019Spaces}
\bibfield{author}{\bibinfo{person}{Sandjar Kozubaev}, \bibinfo{person}{Fernando Rochaix}, \bibinfo{person}{Carl DiSalvo}, {and} \bibinfo{person}{Christopher~A. Le~Dantec}.} \bibinfo{year}{2019}\natexlab{}.
\newblock \showarticletitle{Spaces and Traces: Implications of Smart Technology in Public Housing}. In \bibinfo{booktitle}{\emph{Proceedings of the 2019 CHI Conference on Human Factors in Computing Systems - CHI '19}} (2019). \bibinfo{publisher}{ACM Press}, \bibinfo{address}{New York, New York, USA}, \bibinfo{pages}{1--13}.
\newblock
\showISBNx{978-1-4503-5970-2}
\urldef\tempurl%
\url{https://doi.org/10.1145/3290605.3300669}
\showDOI{\tempurl}


\bibitem[Kurze et~al\mbox{.}(2020)]%
        {Kurze2020Guess}
\bibfield{author}{\bibinfo{person}{Albrecht Kurze}, \bibinfo{person}{Andreas Bischof}, \bibinfo{person}{Sören Totzauer}, \bibinfo{person}{Michael Storz}, \bibinfo{person}{Maximilian Eibl}, \bibinfo{person}{Margot Brereton}, {and} \bibinfo{person}{Arne Berger}.} \bibinfo{year}{2020}\natexlab{}.
\newblock \showarticletitle{Guess the Data: Data Work to Understand How People Make Sense of and Use Simple Sensor Data from Homes}. In \bibinfo{booktitle}{\emph{Proceedings of the 2020 CHI Conference on Human Factors in Computing Systems}} (2020-04-21). \bibinfo{publisher}{ACM}, \bibinfo{address}{Honolulu HI USA}, \bibinfo{pages}{1--12}.
\newblock
\showISBNx{978-1-4503-6708-0}
\urldef\tempurl%
\url{https://doi.org/10.1145/3313831.3376273}
\showDOI{\tempurl}


\bibitem[Lau et~al\mbox{.}(2018)]%
        {Lau2018Alexa}
\bibfield{author}{\bibinfo{person}{Josephine Lau}, \bibinfo{person}{Benjamin Zimmerman}, {and} \bibinfo{person}{Florian Schaub}.} \bibinfo{year}{2018}\natexlab{}.
\newblock \showarticletitle{Alexa, Are You Listening?: Privacy Perceptions, Concerns and Privacy-seeking Behaviors with Smart Speakers}.
\newblock \bibinfo{journal}{\emph{Proceedings of the ACM on Human-Computer Interaction}} \bibinfo{volume}{2}, \bibinfo{number}{CSCW} (\bibinfo{date}{11} \bibinfo{year}{2018}), \bibinfo{pages}{1--31}.
\newblock
\showISSN{2573-0142}
\urldef\tempurl%
\url{https://doi.org/10.1145/3274371}
\showDOI{\tempurl}


\bibitem[Le~Dantec et~al\mbox{.}(2009)]%
        {LeDantec2009Values}
\bibfield{author}{\bibinfo{person}{Christopher~A. Le~Dantec}, \bibinfo{person}{Erika~Shehan Poole}, {and} \bibinfo{person}{Susan~P. Wyche}.} \bibinfo{year}{2009}\natexlab{}.
\newblock \showarticletitle{Values as lived experience: Evolving value sensitive design in support of value discovery}. In \bibinfo{booktitle}{\emph{Proceedings of the 27th international conference on Human factors in computing systems - CHI 09}} (2009). \bibinfo{publisher}{ACM Press}, \bibinfo{address}{New York, New York, USA}, \bibinfo{pages}{1141}.
\newblock
\showISBNx{978-1-60558-246-7}
\urldef\tempurl%
\url{https://doi.org/10.1145/1518701.1518875}
\showDOI{\tempurl}


\bibitem[Lee and Kobsa(2020)]%
        {Lee2020Confident}
\bibfield{author}{\bibinfo{person}{Hosub Lee} {and} \bibinfo{person}{Alfred Kobsa}.} \bibinfo{year}{2020}\natexlab{}.
\newblock \showarticletitle{Confident Privacy Decision-Making in IoT Environments}.
\newblock \bibinfo{journal}{\emph{ACM Transactions on Computer-Human Interaction}} \bibinfo{volume}{27}, \bibinfo{number}{1} (\bibinfo{date}{23 1} \bibinfo{year}{2020}), \bibinfo{pages}{1--39}.
\newblock
\showISSN{1073-0516}
\urldef\tempurl%
\url{https://doi.org/10.1145/3364223}
\showDOI{\tempurl}


\bibitem[Lindley et~al\mbox{.}(2020a)]%
        {Lindley2020DesignResearch}
\bibfield{author}{\bibinfo{person}{Joseph Lindley}, \bibinfo{person}{Haider~Ali Akmal}, {and} \bibinfo{person}{Paul Coulton}.} \bibinfo{year}{2020}\natexlab{a}.
\newblock \showarticletitle{Design Research and Object-Oriented Ontology}.
\newblock \bibinfo{journal}{\emph{Open Philosophy}} \bibinfo{volume}{3}, \bibinfo{number}{1} (\bibinfo{date}{Jan.} \bibinfo{year}{2020}), \bibinfo{pages}{11--41}.
\newblock
\urldef\tempurl%
\url{https://doi.org/10.1515/opphil-2020-0002}
\showDOI{\tempurl}


\bibitem[Lindley et~al\mbox{.}(2020b)]%
        {Lindley2020Researching}
\bibfield{author}{\bibinfo{person}{Joseph Lindley}, \bibinfo{person}{Haider~Ali Akmal}, \bibinfo{person}{Franziska Pilling}, {and} \bibinfo{person}{Paul Coulton}.} \bibinfo{year}{2020}\natexlab{b}.
\newblock \showarticletitle{Researching AI Legibility through Design}. In \bibinfo{booktitle}{\emph{Proceedings of the 2020 CHI Conference on Human Factors in Computing Systems}} (Honolulu, HI, USA) \emph{(\bibinfo{series}{CHI '20})}. \bibinfo{publisher}{Association for Computing Machinery}, \bibinfo{address}{New York, NY, USA}, \bibinfo{pages}{1–13}.
\newblock
\showISBNx{9781450367080}
\urldef\tempurl%
\url{https://doi.org/10.1145/3313831.3376792}
\showDOI{\tempurl}


\bibitem[Liu et~al\mbox{.}(2018)]%
        {Liu2018Design}
\bibfield{author}{\bibinfo{person}{Jen Liu}, \bibinfo{person}{Daragh Byrne}, {and} \bibinfo{person}{Laura Devendorf}.} \bibinfo{year}{2018}\natexlab{}.
\newblock \showarticletitle{Design for Collaborative Survival: An Inquiry into Human-Fungi Relationships}. In \bibinfo{booktitle}{\emph{Proceedings of the 2018 CHI Conference on Human Factors in Computing Systems}} (Montreal QC, Canada) \emph{(\bibinfo{series}{CHI '18})}. \bibinfo{publisher}{Association for Computing Machinery}, \bibinfo{address}{New York, NY, USA}, \bibinfo{pages}{1–13}.
\newblock
\showISBNx{9781450356206}
\urldef\tempurl%
\url{https://doi.org/10.1145/3173574.3173614}
\showDOI{\tempurl}


\bibitem[Lockton et~al\mbox{.}(2019)]%
        {Lockton2019New}
\bibfield{author}{\bibinfo{person}{Dan Lockton}, \bibinfo{person}{Devika Singh}, \bibinfo{person}{Saloni Sabnis}, \bibinfo{person}{Michelle Chou}, \bibinfo{person}{Sarah Foley}, {and} \bibinfo{person}{Alejandro Pantoja}.} \bibinfo{year}{2019}\natexlab{}.
\newblock \showarticletitle{New Metaphors: A workshop method for generating ideas and reframing problems in design and beyond}. In \bibinfo{booktitle}{\emph{Proceedings of the 2019 on Creativity and Cognition}} (2019-06-13). \bibinfo{publisher}{ACM}, \bibinfo{address}{New York, NY, USA}, \bibinfo{pages}{319--332}.
\newblock
\showISBNx{978-1-4503-5917-7}
\urldef\tempurl%
\url{https://doi.org/10.1145/3325480.3326570}
\showDOI{\tempurl}


\bibitem[Marky et~al\mbox{.}(2020)]%
        {Marky2020You}
\bibfield{author}{\bibinfo{person}{Karola Marky}, \bibinfo{person}{Sarah Prange}, \bibinfo{person}{Florian Krell}, \bibinfo{person}{Max Mühlhäuser}, {and} \bibinfo{person}{Florian Alt}.} \bibinfo{year}{2020}\natexlab{}.
\newblock \showarticletitle{“You just can’t know about everything”: Privacy Perceptions of Smart Home Visitors}. In \bibinfo{booktitle}{\emph{19th International Conference on Mobile and Ubiquitous Multimedia}} (2020-11-22). \bibinfo{publisher}{ACM}, \bibinfo{address}{Essen Germany}, \bibinfo{pages}{83--95}.
\newblock
\showISBNx{978-1-4503-8870-2}
\urldef\tempurl%
\url{https://doi.org/10.1145/3428361.3428464}
\showDOI{\tempurl}


\bibitem[Marky et~al\mbox{.}(2021)]%
        {Marky2021Roles}
\bibfield{author}{\bibinfo{person}{Karola Marky}, \bibinfo{person}{Sarah Prange}, \bibinfo{person}{Max Mühlhäuser}, {and} \bibinfo{person}{Florian Alt}.} \bibinfo{year}{2021}\natexlab{}.
\newblock \showarticletitle{Roles Matter! Understanding Differences in the Privacy Mental Models of Smart Home Visitors and Residents}. In \bibinfo{booktitle}{\emph{20th International Conference on Mobile and Ubiquitous Multimedia}} (2021-05-12). \bibinfo{publisher}{ACM}, \bibinfo{address}{Leuven Belgium}, \bibinfo{pages}{108--122}.
\newblock
\showISBNx{978-1-4503-8643-2}
\urldef\tempurl%
\url{https://doi.org/10.1145/3490632.3490664}
\showDOI{\tempurl}


\bibitem[Martin(2009)]%
        {Martin2009Design}
\bibfield{author}{\bibinfo{person}{Roger Martin}.} \bibinfo{year}{2009}\natexlab{}.
\newblock \bibinfo{booktitle}{\emph{The Design of Business: Why Design Thinking is the Next Competitive Edge}}.
\newblock \bibinfo{publisher}{Harvard Business Press}, \bibinfo{address}{Cambridge, Massachusetts}.
\newblock


\bibitem[Maz{\'e} et~al\mbox{.}(2019)]%
        {maze2019politics}
\bibfield{author}{\bibinfo{person}{Ramia Maz{\'e}} {et~al\mbox{.}}} \bibinfo{year}{2019}\natexlab{}.
\newblock \showarticletitle{Politics of designing visions of the future}.
\newblock \bibinfo{journal}{\emph{Journal of Futures Studies}} \bibinfo{volume}{23}, \bibinfo{number}{3} (\bibinfo{year}{2019}), \bibinfo{pages}{23--38}.
\newblock
\urldef\tempurl%
\url{https://doi.org/10.6531/JFS.201903_23(3).0003}
\showDOI{\tempurl}


\bibitem[McDonald et~al\mbox{.}(2020)]%
        {McDonald2020Privacy}
\bibfield{author}{\bibinfo{person}{Nora McDonald}, \bibinfo{person}{Karla Badillo-Urquiola}, \bibinfo{person}{Morgan~G. Ames}, \bibinfo{person}{Nicola Dell}, \bibinfo{person}{Elizabeth Keneski}, \bibinfo{person}{Manya Sleeper}, {and} \bibinfo{person}{Pamela~J. Wisniewski}.} \bibinfo{year}{2020}\natexlab{}.
\newblock \showarticletitle{Privacy and Power: Acknowledging the Importance of Privacy Research and Design for Vulnerable Populations}. In \bibinfo{booktitle}{\emph{Extended Abstracts of the 2020 CHI Conference on Human Factors in Computing Systems}} (2020-04-25). \bibinfo{publisher}{ACM}, \bibinfo{address}{New York, NY, USA}, \bibinfo{pages}{1--8}.
\newblock
\showISBNx{978-1-4503-6819-3}
\urldef\tempurl%
\url{https://doi.org/10.1145/3334480.3375174}
\showDOI{\tempurl}


\bibitem[McDonald and Forte(2020)]%
        {McDonald2020Politics}
\bibfield{author}{\bibinfo{person}{Nora McDonald} {and} \bibinfo{person}{Andrea Forte}.} \bibinfo{year}{2020}\natexlab{}.
\newblock \showarticletitle{The Politics of Privacy Theories: Moving from Norms to Vulnerabilities}. In \bibinfo{booktitle}{\emph{Proceedings of the 2020 CHI Conference on Human Factors in Computing Systems}} (2020-04-21). \bibinfo{publisher}{ACM}, \bibinfo{address}{New York, NY, USA}, \bibinfo{pages}{1--14}.
\newblock
\showISBNx{978-1-4503-6708-0}
\urldef\tempurl%
\url{https://doi.org/10.1145/3313831.3376167}
\showDOI{\tempurl}


\bibitem[Merrill(2020)]%
        {Merrill2020Security}
\bibfield{author}{\bibinfo{person}{Nick Merrill}.} \bibinfo{year}{2020}\natexlab{}.
\newblock \showarticletitle{Security Fictions: Bridging Speculative Design and Computer Security}. In \bibinfo{booktitle}{\emph{Proceedings of the 2020 ACM Designing Interactive Systems Conference}} (2020-07-03). \bibinfo{publisher}{ACM}, \bibinfo{address}{New York, NY, USA}, \bibinfo{pages}{1727--1735}.
\newblock
\showISBNx{978-1-4503-6974-9}
\urldef\tempurl%
\url{https://doi.org/10.1145/3357236.3395451}
\showDOI{\tempurl}


\bibitem[Mulligan and King(2011)]%
        {Mulligan2011Bridging}
\bibfield{author}{\bibinfo{person}{Deirdre~K. Mulligan} {and} \bibinfo{person}{Jennifer King}.} \bibinfo{year}{2011}\natexlab{}.
\newblock \showarticletitle{Bridging the gap between privacy and design}.
\newblock \bibinfo{journal}{\emph{University of Pennsylvania Journal of Constitutional Law}} (\bibinfo{year}{2011}), \bibinfo{pages}{989--1034}.
\newblock


\bibitem[Mulligan et~al\mbox{.}(2016)]%
        {Mulligan2016Privacy}
\bibfield{author}{\bibinfo{person}{Deirdre~K. Mulligan}, \bibinfo{person}{Colin Koopman}, {and} \bibinfo{person}{Nick Doty}.} \bibinfo{year}{2016}\natexlab{}.
\newblock \showarticletitle{Privacy is an essentially contested concept: a multi-dimensional analytic for mapping privacy}.
\newblock \bibinfo{journal}{\emph{Philosophical Transactions of the Royal Society A: Mathematical, Physical and Engineering Sciences}} \bibinfo{volume}{374}, \bibinfo{number}{2083} (\bibinfo{date}{28 12} \bibinfo{year}{2016}), \bibinfo{pages}{1--17}.
\newblock
\showISSN{1364-503X}
\urldef\tempurl%
\url{https://doi.org/10.1098/rsta.2016.0118}
\showDOI{\tempurl}


\bibitem[Mulligan et~al\mbox{.}(2019)]%
        {Mulligan2019This}
\bibfield{author}{\bibinfo{person}{Deirdre~K. Mulligan}, \bibinfo{person}{Joshua~A. Kroll}, \bibinfo{person}{Nitin Kohli}, {and} \bibinfo{person}{Richmond~Y. Wong}.} \bibinfo{year}{2019}\natexlab{}.
\newblock \showarticletitle{This Thing Called Fairness: Disciplinary confusion realizing a value in technology}.
\newblock \bibinfo{journal}{\emph{Proceedings of the ACM on Human-Computer Interaction}} \bibinfo{volume}{3}, \bibinfo{number}{CSCW} (\bibinfo{date}{7 11} \bibinfo{year}{2019}), \bibinfo{pages}{1--36}.
\newblock
\showISSN{25730142}
\urldef\tempurl%
\url{https://doi.org/10.1145/3359221}
\showDOI{\tempurl}


\bibitem[Mulligan and Nissenbaum(2020)]%
        {Mulligan2020Concept}
\bibfield{author}{\bibinfo{person}{Deirdre~K. Mulligan} {and} \bibinfo{person}{Helen Nissenbaum}.} \bibinfo{year}{2020}\natexlab{}.
\newblock \showarticletitle{The Concept of Handoff as a Model for Ethical Analysis and Design}.
\newblock In \bibinfo{booktitle}{\emph{The Oxford Handbook of Ethics of AI}}, \bibfield{editor}{\bibinfo{person}{Markus~D. Dubber}, \bibinfo{person}{Frank Pasquale}, {and} \bibinfo{person}{Sunit Das}} (Eds.). \bibinfo{publisher}{Oxford University Press}, \bibinfo{pages}{231--251}.
\newblock
\showISBNx{978-0-19-006739-7}
\urldef\tempurl%
\url{https://doi.org/10.1093/oxfordhb/9780190067397.013.15}
\showDOI{\tempurl}


\bibitem[Naeini et~al\mbox{.}(2017)]%
        {Naeini2017Privacy}
\bibfield{author}{\bibinfo{person}{Pardis~Emami Naeini}, \bibinfo{person}{Sruti Bhagavatula}, \bibinfo{person}{Hana Habib}, \bibinfo{person}{Martin Degeling}, \bibinfo{person}{Lujo Bauer}, \bibinfo{person}{Lorrie Cranor}, {and} \bibinfo{person}{Norman Sadeh}.} \bibinfo{year}{2017}\natexlab{}.
\newblock \showarticletitle{Privacy Expectations and Preferences in an IoT World}. In \bibinfo{booktitle}{\emph{Thirteenth Symposium on Usable Privacy and Security (SOUPS 2017)}} (2017). \bibinfo{pages}{399--412}.
\newblock
\showISBNx{978-1-931971-39-3}


\bibitem[N\"{a}gele et~al\mbox{.}(2018)]%
        {Nagele2018PDFi}
\bibfield{author}{\bibinfo{person}{Larissa~Vivian N\"{a}gele}, \bibinfo{person}{Merja Ry\"{o}ppy}, {and} \bibinfo{person}{Danielle Wilde}.} \bibinfo{year}{2018}\natexlab{}.
\newblock \showarticletitle{PDFi: Participatory Design Fiction with Vulnerable Users}. In \bibinfo{booktitle}{\emph{Proceedings of the 10th Nordic Conference on Human-Computer Interaction}} (Oslo, Norway) \emph{(\bibinfo{series}{NordiCHI '18})}. \bibinfo{publisher}{Association for Computing Machinery}, \bibinfo{address}{New York, NY, USA}, \bibinfo{pages}{819–831}.
\newblock
\showISBNx{9781450364379}
\urldef\tempurl%
\url{https://doi.org/10.1145/3240167.3240272}
\showDOI{\tempurl}


\bibitem[Nathan et~al\mbox{.}(2008)]%
        {Nathan2008Envisioning}
\bibfield{author}{\bibinfo{person}{Lisa~P. Nathan}, \bibinfo{person}{Batya Friedman}, \bibinfo{person}{Predrag Klasnja}, \bibinfo{person}{Shaun~K. Kane}, {and} \bibinfo{person}{Jessica~K. Miller}.} \bibinfo{year}{2008}\natexlab{}.
\newblock \showarticletitle{Envisioning systemic effects on persons and society throughout interactive system design}. In \bibinfo{booktitle}{\emph{Proceedings of the 7th ACM conference on Designing interactive systems - DIS '08}} (2008). \bibinfo{publisher}{ACM Press}, \bibinfo{address}{New York, New York, USA}, \bibinfo{pages}{1--10}.
\newblock
\showISBNx{978-1-60558-002-9}
\urldef\tempurl%
\url{https://doi.org/10.1145/1394445.1394446}
\showDOI{\tempurl}


\bibitem[Nissenbaum(2009)]%
        {Nissenbaum2009Privacy}
\bibfield{author}{\bibinfo{person}{Helen Nissenbaum}.} \bibinfo{year}{2009}\natexlab{}.
\newblock \bibinfo{booktitle}{\emph{Privacy in Context: Technology, Policy, and the Integrity of Social Life}}.
\newblock \bibinfo{publisher}{Stanford University Press}, \bibinfo{address}{Stanford, California}.
\newblock
\showISBNx{978-0-8047-5237-4}


\bibitem[Noortman et~al\mbox{.}(2019)]%
        {Noortman2019HawkEye}
\bibfield{author}{\bibinfo{person}{Renee Noortman}, \bibinfo{person}{Britta~F. Schulte}, \bibinfo{person}{Paul Marshall}, \bibinfo{person}{Saskia Bakker}, {and} \bibinfo{person}{Anna~L. Cox}.} \bibinfo{year}{2019}\natexlab{}.
\newblock \showarticletitle{HawkEye - Deploying a Design Fiction Probe}. In \bibinfo{booktitle}{\emph{Proceedings of the 2019 CHI Conference on Human Factors in Computing Systems - CHI '19}} (2019). \bibinfo{publisher}{ACM Press}, \bibinfo{address}{New York, New York, USA}, \bibinfo{pages}{1--14}.
\newblock
\showISBNx{978-1-4503-5970-2}
\urldef\tempurl%
\url{https://doi.org/10.1145/3290605.3300652}
\showDOI{\tempurl}


\bibitem[Odom et~al\mbox{.}(2019)]%
        {Odom2019Diversifying}
\bibfield{author}{\bibinfo{person}{William Odom}, \bibinfo{person}{Sumeet Anand}, \bibinfo{person}{Doenja Oogjes}, {and} \bibinfo{person}{Jo Shin}.} \bibinfo{year}{2019}\natexlab{}.
\newblock \showarticletitle{Diversifying the Domestic: A Design Inquiry into Collective and Mobile Living}. In \bibinfo{booktitle}{\emph{Proceedings of the 2019 on Designing Interactive Systems Conference - DIS '19}} (2019). \bibinfo{publisher}{ACM Press}, \bibinfo{address}{New York, New York, USA}, \bibinfo{pages}{1377--1390}.
\newblock
\showISBNx{978-1-4503-5850-7}
\urldef\tempurl%
\url{https://doi.org/10.1145/3322276.3323687}
\showDOI{\tempurl}


\bibitem[Oogjes et~al\mbox{.}(2018)]%
        {Oogjes2018Designing}
\bibfield{author}{\bibinfo{person}{Doenja Oogjes}, \bibinfo{person}{William Odom}, {and} \bibinfo{person}{Pete Fung}.} \bibinfo{year}{2018}\natexlab{}.
\newblock \showarticletitle{Designing for an other Home: Expanding and Speculating on Different Forms of Domestic Life}. In \bibinfo{booktitle}{\emph{Proceedings of the 2018 on Designing Interactive Systems Conference 2018 - DIS '18}} (2018). \bibinfo{publisher}{ACM Press}, \bibinfo{address}{New York, New York, USA}, \bibinfo{pages}{313--326}.
\newblock
\showISBNx{978-1-4503-5198-0}
\urldef\tempurl%
\url{https://doi.org/10.1145/3196709.3196810}
\showDOI{\tempurl}


\bibitem[Pierce(2019a)]%
        {Pierce2019Lamps}
\bibfield{author}{\bibinfo{person}{James Pierce}.} \bibinfo{year}{2019}\natexlab{a}.
\newblock \showarticletitle{Lamps, Curtains, Robots: 3 scenarios for the future of the smart home}. In \bibinfo{booktitle}{\emph{Proceedings of the 2019 on Creativity and Cognition - C\&C '19}} (2019). \bibinfo{publisher}{ACM Press}, \bibinfo{address}{New York, New York, USA}, \bibinfo{pages}{423--424}.
\newblock
\showISBNx{978-1-4503-5917-7}
\urldef\tempurl%
\url{https://doi.org/10.1145/3325480.3329181}
\showDOI{\tempurl}


\bibitem[Pierce(2019b)]%
        {Pierce2019Smart}
\bibfield{author}{\bibinfo{person}{James Pierce}.} \bibinfo{year}{2019}\natexlab{b}.
\newblock \showarticletitle{Smart Home Security Cameras and Shifting Lines of Creepiness: A Design-Led Inquiry}. In \bibinfo{booktitle}{\emph{Proceedings of the 2019 CHI Conference on Human Factors in Computing Systems}} (2019-05-02). \bibinfo{publisher}{ACM}, \bibinfo{address}{Glasgow Scotland Uk}, \bibinfo{pages}{1--14}.
\newblock
\showISBNx{978-1-4503-5970-2}
\urldef\tempurl%
\url{https://doi.org/10.1145/3290605.3300275}
\showDOI{\tempurl}


\bibitem[Pierce(2021)]%
        {Pierce2021eccentric}
\bibfield{author}{\bibinfo{person}{James Pierce}.} \bibinfo{year}{2021}\natexlab{}.
\newblock \showarticletitle{Eccentric Sensing Devices: Using Conceptual Design Notes to Understand Design Opportunities, Limitations, and Concerns Connected to Digital Sensing}. In \bibinfo{booktitle}{\emph{Creativity and Cognition}} (Virtual Event, Italy) \emph{(\bibinfo{series}{C\&C '21})}. \bibinfo{publisher}{Association for Computing Machinery}, \bibinfo{address}{New York, NY, USA}, Article \bibinfo{articleno}{37}, \bibinfo{numpages}{14}~pages.
\newblock
\showISBNx{9781450383769}
\urldef\tempurl%
\url{https://doi.org/10.1145/3450741.3466775}
\showDOI{\tempurl}


\bibitem[Pierce and DiSalvo(2018)]%
        {Pierce2018AddressingNetwork}
\bibfield{author}{\bibinfo{person}{James Pierce} {and} \bibinfo{person}{Carl DiSalvo}.} \bibinfo{year}{2018}\natexlab{}.
\newblock \showarticletitle{Addressing Network Anxieties with Alternative Design Metaphors}. In \bibinfo{booktitle}{\emph{Proceedings of the 2018 CHI Conference on Human Factors in Computing Systems}} (Montreal QC, Canada) \emph{(\bibinfo{series}{CHI '18})}. \bibinfo{publisher}{Association for Computing Machinery}, \bibinfo{address}{New York, NY, USA}, \bibinfo{pages}{1–13}.
\newblock
\showISBNx{9781450356206}
\urldef\tempurl%
\url{https://doi.org/10.1145/3173574.3174123}
\showDOI{\tempurl}


\bibitem[Pierce et~al\mbox{.}(2018a)]%
        {Pierce2018Differential}
\bibfield{author}{\bibinfo{person}{James Pierce}, \bibinfo{person}{Sarah Fox}, \bibinfo{person}{Nick Merrill}, {and} \bibinfo{person}{Richmond Wong}.} \bibinfo{year}{2018}\natexlab{a}.
\newblock \showarticletitle{Differential Vulnerabilities and a Diversity of Tactics: What Toolkits Teach Us about Cybersecurity}.
\newblock \bibinfo{journal}{\emph{Proceedings of the ACM on Human-Computer Interaction}} \bibinfo{volume}{2}, \bibinfo{number}{CSCW} (\bibinfo{date}{1 11} \bibinfo{year}{2018}), \bibinfo{pages}{1--24}.
\newblock
\showISSN{25730142}
\urldef\tempurl%
\url{https://doi.org/10.1145/3274408}
\showDOI{\tempurl}


\bibitem[Pierce et~al\mbox{.}(2018b)]%
        {Pierce2018Interface}
\bibfield{author}{\bibinfo{person}{James Pierce}, \bibinfo{person}{Sarah Fox}, \bibinfo{person}{Nick Merrill}, \bibinfo{person}{Richmond Wong}, {and} \bibinfo{person}{Carl DiSalvo}.} \bibinfo{year}{2018}\natexlab{b}.
\newblock \showarticletitle{An Interface without A User: An Exploratory Design Study of Online Privacy Policies and Digital Legalese}. In \bibinfo{booktitle}{\emph{Proceedings of the 2018 Designing Interactive Systems Conference}} (Hong Kong, China) \emph{(\bibinfo{series}{DIS '18})}. \bibinfo{publisher}{Association for Computing Machinery}, \bibinfo{address}{New York, NY, USA}, \bibinfo{pages}{1345–1358}.
\newblock
\showISBNx{9781450351980}
\urldef\tempurl%
\url{https://doi.org/10.1145/3196709.3196818}
\showDOI{\tempurl}


\bibitem[Pierce et~al\mbox{.}(2022)]%
        {Pierce2022Addressing}
\bibfield{author}{\bibinfo{person}{James Pierce}, \bibinfo{person}{Claire Weizenegger}, \bibinfo{person}{Parag Nandi}, \bibinfo{person}{Isha Agarwal}, \bibinfo{person}{Gwenna Gram}, \bibinfo{person}{Jade Hurrle}, \bibinfo{person}{Hannah Liao}, \bibinfo{person}{Betty Lo}, \bibinfo{person}{Aaron Park}, \bibinfo{person}{Aivy Phan}, \bibinfo{person}{Mark Shumskiy}, {and} \bibinfo{person}{Grace Sturlaugson}.} \bibinfo{year}{2022}\natexlab{}.
\newblock \showarticletitle{Addressing Adjacent Actor Privacy: Designing for Bystanders, Co-Users, and Surveilled Subjects of Smart Home Cameras}. In \bibinfo{booktitle}{\emph{Designing Interactive Systems Conference}} (2022-06-13). \bibinfo{publisher}{ACM}, \bibinfo{address}{Virtual Event Australia}, \bibinfo{pages}{26--40}.
\newblock
\showISBNx{978-1-4503-9358-4}
\urldef\tempurl%
\url{https://doi.org/10.1145/3532106.3535195}
\showDOI{\tempurl}


\bibitem[Pierce et~al\mbox{.}(2020)]%
        {Pierce2020Sensor}
\bibfield{author}{\bibinfo{person}{James Pierce}, \bibinfo{person}{Richmond~Y. Wong}, {and} \bibinfo{person}{Nick Merrill}.} \bibinfo{year}{2020}\natexlab{}.
\newblock \showarticletitle{Sensor Illumination: Exploring Design Qualities and Ethical Implications of Smart Cameras and Image/Video Analytics}. In \bibinfo{booktitle}{\emph{Proceedings of the 2020 CHI Conference on Human Factors in Computing Systems (CHI ’20)}} (2020). \bibinfo{pages}{1--19}.
\newblock
\showISBNx{978-1-4503-6708-0}
\urldef\tempurl%
\url{https://doi.org/10.1145/3313831.3376347}
\showDOI{\tempurl}


\bibitem[Rogers et~al\mbox{.}(2019)]%
        {Rogers2019OurFriends}
\bibfield{author}{\bibinfo{person}{Jon Rogers}, \bibinfo{person}{Loraine Clarke}, \bibinfo{person}{Martin Skelly}, \bibinfo{person}{Nick Taylor}, \bibinfo{person}{Pete Thomas}, \bibinfo{person}{Michelle Thorne}, \bibinfo{person}{Solana Larsen}, \bibinfo{person}{Katarzyna Odrozek}, \bibinfo{person}{Julia Kloiber}, \bibinfo{person}{Peter Bihr}, \bibinfo{person}{Anab Jain}, \bibinfo{person}{Jon Arden}, {and} \bibinfo{person}{Max von Grafenstein}.} \bibinfo{year}{2019}\natexlab{}.
\newblock \showarticletitle{Our Friends Electric: Reflections on Advocacy and Design Research for the Voice Enabled Internet}. In \bibinfo{booktitle}{\emph{Proceedings of the 2019 CHI Conference on Human Factors in Computing Systems}} (Glasgow, Scotland Uk) \emph{(\bibinfo{series}{CHI '19})}. \bibinfo{publisher}{Association for Computing Machinery}, \bibinfo{address}{New York, NY, USA}, \bibinfo{pages}{1–13}.
\newblock
\showISBNx{9781450359702}
\urldef\tempurl%
\url{https://doi.org/10.1145/3290605.3300344}
\showDOI{\tempurl}


\bibitem[Saldaña(2013)]%
        {Saldana2013Coding}
\bibfield{author}{\bibinfo{person}{Johnny Saldaña}.} \bibinfo{year}{2013}\natexlab{}.
\newblock \bibinfo{booktitle}{\emph{The Coding Manual for Qualitative Researchers}}.
\newblock \bibinfo{publisher}{Sage}, \bibinfo{address}{Los Angeles}.
\newblock


\bibitem[Shilton(2018)]%
        {Shilton2018Values}
\bibfield{author}{\bibinfo{person}{Katie Shilton}.} \bibinfo{year}{2018}\natexlab{}.
\newblock \showarticletitle{Values and Ethics in Human-Computer Interaction}.
\newblock \bibinfo{journal}{\emph{Foundations and Trends® in Human–Computer Interaction}} \bibinfo{volume}{12}, \bibinfo{number}{2} (\bibinfo{year}{2018}), \bibinfo{pages}{107--171}.
\newblock
\showISSN{1551-3955}
\urldef\tempurl%
\url{https://doi.org/10.1561/1100000073}
\showDOI{\tempurl}


\bibitem[Shilton et~al\mbox{.}(2020)]%
        {Shilton2020Role-Playing}
\bibfield{author}{\bibinfo{person}{Katie Shilton}, \bibinfo{person}{Donal Heidenblad}, \bibinfo{person}{Adam Porter}, \bibinfo{person}{Susan Winter}, {and} \bibinfo{person}{Mary Kendig}.} \bibinfo{year}{2020}\natexlab{}.
\newblock \showarticletitle{Role-Playing Computer Ethics: Designing and Evaluating the Privacy by Design (PbD) Simulation}.
\newblock \bibinfo{journal}{\emph{Science and Engineering Ethics}} (\bibinfo{date}{1 7} \bibinfo{year}{2020}).
\newblock
\showISSN{1353-3452}
\urldef\tempurl%
\url{https://doi.org/10.1007/s11948-020-00250-0}
\showDOI{\tempurl}


\bibitem[Solove(2002)]%
        {Solove2002Conceptualizing}
\bibfield{author}{\bibinfo{person}{Daniel~J. Solove}.} \bibinfo{year}{2002}\natexlab{}.
\newblock \showarticletitle{Conceptualizing privacy}.
\newblock \bibinfo{journal}{\emph{California Law Review}}  \bibinfo{volume}{90} (\bibinfo{year}{2002}), \bibinfo{pages}{1087--1155}.
\newblock
\showISSN{00081221}
\urldef\tempurl%
\url{https://doi.org/10.1145/1929609.1929610}
\showDOI{\tempurl}


\bibitem[Solove(2003)]%
        {Solove2003Taxonomy}
\bibfield{author}{\bibinfo{person}{Daniel~J. Solove}.} \bibinfo{year}{2003}\natexlab{}.
\newblock \showarticletitle{A Taxonomy of Privacy}.
\newblock \bibinfo{journal}{\emph{University of Pennsylvania Law Review}} \bibinfo{volume}{154}, \bibinfo{number}{3} (\bibinfo{year}{2003}), \bibinfo{pages}{477--560}.
\newblock


\bibitem[Solove(2008)]%
        {Solove2008Privacy:}
\bibfield{author}{\bibinfo{person}{Daniel~J Solove}.} \bibinfo{year}{2008}\natexlab{}.
\newblock \showarticletitle{Privacy: A Concept in Disarray}.
\newblock In \bibinfo{booktitle}{\emph{Understanding privacy}}. \bibinfo{publisher}{Harvard University Press}, \bibinfo{address}{Cambridge, Massachusetts}.
\newblock


\bibitem[Stark and Levy(2018)]%
        {Stark2018surveillant}
\bibfield{author}{\bibinfo{person}{Luke Stark} {and} \bibinfo{person}{Karen Levy}.} \bibinfo{year}{2018}\natexlab{}.
\newblock \showarticletitle{The surveillant consumer}.
\newblock \bibinfo{journal}{\emph{Media, Culture \& Society}} \bibinfo{volume}{40}, \bibinfo{number}{8} (\bibinfo{date}{25 11} \bibinfo{year}{2018}), \bibinfo{pages}{1202--1220}.
\newblock
\showISSN{0163-4437}
\urldef\tempurl%
\url{https://doi.org/10.1177/0163443718781985}
\showDOI{\tempurl}


\bibitem[Strengers et~al\mbox{.}(2019)]%
        {Strengers2019Protection}
\bibfield{author}{\bibinfo{person}{Yolande Strengers}, \bibinfo{person}{Jenny Kennedy}, \bibinfo{person}{Paula Arcari}, \bibinfo{person}{Larissa Nicholls}, {and} \bibinfo{person}{Melissa Gregg}.} \bibinfo{year}{2019}\natexlab{}.
\newblock \showarticletitle{Protection, Productivity and Pleasure in the Smart Home: Emerging Expectations and Gendered Insights from Australian Early Adopters}. In \bibinfo{booktitle}{\emph{Proceedings of the 2019 CHI Conference on Human Factors in Computing Systems}} (2019-05-02). \bibinfo{publisher}{ACM}, \bibinfo{address}{Glasgow Scotland Uk}, \bibinfo{pages}{1--13}.
\newblock
\showISBNx{978-1-4503-5970-2}
\urldef\tempurl%
\url{https://doi.org/10.1145/3290605.3300875}
\showDOI{\tempurl}


\bibitem[Swauger(2020)]%
        {Swauger2020Software}
\bibfield{author}{\bibinfo{person}{Shea Swauger}.} \bibinfo{year}{2020}\natexlab{}.
\newblock \showarticletitle{Software that monitors students during tests perpetuates inequality and violates their privacy}.
\newblock \bibinfo{journal}{\emph{MIT Technology Review}} (\bibinfo{date}{7 8} \bibinfo{year}{2020}).
\newblock
\urldef\tempurl%
\url{https://www.technologyreview.com/2020/08/07/1006132/software-algorithms-proctoring-online-tests-ai-ethics/}
\showURL{%
\tempurl}


\bibitem[Tabassum et~al\mbox{.}(2020)]%
        {Tabassum2020Smart}
\bibfield{author}{\bibinfo{person}{Madiha Tabassum}, \bibinfo{person}{Jess Kropczynski}, \bibinfo{person}{Pamela Wisniewski}, {and} \bibinfo{person}{Heather~Richter Lipford}.} \bibinfo{year}{2020}\natexlab{}.
\newblock \showarticletitle{Smart Home Beyond the Home: A Case for Community-Based Access Control}. In \bibinfo{booktitle}{\emph{Proceedings of the 2020 CHI Conference on Human Factors in Computing Systems}} (2020-04-21). \bibinfo{publisher}{ACM}, \bibinfo{address}{Honolulu HI USA}, \bibinfo{pages}{1--12}.
\newblock
\showISBNx{978-1-4503-6708-0}
\urldef\tempurl%
\url{https://doi.org/10.1145/3313831.3376255}
\showDOI{\tempurl}


\bibitem[Tan et~al\mbox{.}(2022a)]%
        {Tan2022Critical-Playful}
\bibfield{author}{\bibinfo{person}{Neilly~H. Tan}, \bibinfo{person}{Brian Kinnee}, \bibinfo{person}{Dana Langseth}, \bibinfo{person}{Sean A.~Munson}, {and} \bibinfo{person}{Audrey Desjardins}.} \bibinfo{year}{2022}\natexlab{a}.
\newblock \showarticletitle{Critical-Playful Speculations with Cameras in the Home}. In \bibinfo{booktitle}{\emph{CHI Conference on Human Factors in Computing Systems}} (2022-04-29). \bibinfo{publisher}{ACM}, \bibinfo{address}{New Orleans LA USA}, \bibinfo{pages}{1--22}.
\newblock
\showISBNx{978-1-4503-9157-3}
\urldef\tempurl%
\url{https://doi.org/10.1145/3491102.3502109}
\showDOI{\tempurl}


\bibitem[Tan et~al\mbox{.}(2022b)]%
        {Tan2022Monitoring}
\bibfield{author}{\bibinfo{person}{Neilly~H. Tan}, \bibinfo{person}{Richmond~Y. Wong}, \bibinfo{person}{Audrey Desjardins}, \bibinfo{person}{Sean~A. Munson}, {and} \bibinfo{person}{James Pierce}.} \bibinfo{year}{2022}\natexlab{b}.
\newblock \showarticletitle{Monitoring Pets, Deterring Intruders, and Casually Spying on Neighbors: Everyday Uses of Smart Home Cameras}. In \bibinfo{booktitle}{\emph{CHI Conference on Human Factors in Computing Systems}} (2022-04-29). \bibinfo{publisher}{ACM}, \bibinfo{address}{New Orleans LA USA}, \bibinfo{pages}{1--25}.
\newblock
\showISBNx{978-1-4503-9157-3}
\urldef\tempurl%
\url{https://doi.org/10.1145/3491102.3517617}
\showDOI{\tempurl}


\bibitem[Thakkar et~al\mbox{.}(2022)]%
        {Thakkar2022It}
\bibfield{author}{\bibinfo{person}{Parth~Kirankumar Thakkar}, \bibinfo{person}{Shijing He}, \bibinfo{person}{Shiyu Xu}, \bibinfo{person}{Danny~Yuxing Huang}, {and} \bibinfo{person}{Yaxing Yao}.} \bibinfo{year}{2022}\natexlab{}.
\newblock \showarticletitle{“It would probably turn into a social faux-pas”: Users’ and Bystanders’ Preferences of Privacy Awareness Mechanisms in Smart Homes}. In \bibinfo{booktitle}{\emph{CHI Conference on Human Factors in Computing Systems}} (2022-04-29). \bibinfo{publisher}{ACM}, \bibinfo{address}{New Orleans LA USA}, \bibinfo{pages}{1--13}.
\newblock
\showISBNx{978-1-4503-9157-3}
\urldef\tempurl%
\url{https://doi.org/10.1145/3491102.3502137}
\showDOI{\tempurl}


\bibitem[Tran~O'Leary et~al\mbox{.}(2019)]%
        {TranOLeary2019Who}
\bibfield{author}{\bibinfo{person}{Jasper Tran~O'Leary}, \bibinfo{person}{Sara Zewde}, \bibinfo{person}{Jennifer Mankoff}, {and} \bibinfo{person}{Daniela~K Rosner}.} \bibinfo{year}{2019}\natexlab{}.
\newblock \showarticletitle{Who Gets to Future? Race, Representation, and Design Methods in Africatown}. In \bibinfo{booktitle}{\emph{Proceedings of the 2019 CHI Conference on Human Factors in Computing Systems - CHI '19}} (2019). \bibinfo{publisher}{ACM Press}, \bibinfo{address}{New York, New York, USA}, \bibinfo{pages}{1--13}.
\newblock
\showISBNx{978-1-4503-5970-2}
\urldef\tempurl%
\url{https://doi.org/10.1145/3290605.3300791}
\showDOI{\tempurl}


\bibitem[Ur et~al\mbox{.}(2014)]%
        {Ur2014Intruders}
\bibfield{author}{\bibinfo{person}{Blase Ur}, \bibinfo{person}{Jaeyeon Jung}, {and} \bibinfo{person}{Stuart Schechter}.} \bibinfo{year}{2014}\natexlab{}.
\newblock \showarticletitle{Intruders versus intrusiveness: Teens’ and Parents’ Perspectives on Home-Entryway Surveillance}. In \bibinfo{booktitle}{\emph{Proceedings of the 2014 ACM International Joint Conference on Pervasive and Ubiquitous Computing}} (2014-09-13). \bibinfo{publisher}{ACM}, \bibinfo{address}{New York, NY, USA}, \bibinfo{pages}{129--139}.
\newblock
\showISBNx{978-1-4503-2968-2}
\urldef\tempurl%
\url{https://doi.org/10.1145/2632048.2632107}
\showDOI{\tempurl}


\bibitem[Vaara and Whittington(2012)]%
        {Vaara2012Strategy-as-Practice:}
\bibfield{author}{\bibinfo{person}{Eero Vaara} {and} \bibinfo{person}{Richard Whittington}.} \bibinfo{year}{2012}\natexlab{}.
\newblock \showarticletitle{Strategy-as-Practice: Taking Social Practices Seriously}.
\newblock \bibinfo{journal}{\emph{Academy of Management Annals}} \bibinfo{volume}{6}, \bibinfo{number}{1} (\bibinfo{date}{6} \bibinfo{year}{2012}), \bibinfo{pages}{285--336}.
\newblock
\showISSN{1941-6520}
\urldef\tempurl%
\url{https://doi.org/10.5465/19416520.2012.672039}
\showDOI{\tempurl}


\bibitem[van~der Heijden(2005)]%
        {Heijden2005Scenarios:}
\bibfield{author}{\bibinfo{person}{Kees van~der Heijden}.} \bibinfo{year}{2005}\natexlab{}.
\newblock \bibinfo{booktitle}{\emph{Scenarios: The Art of Strategic Conversation}}.
\newblock \bibinfo{publisher}{John Wiley \& Sons, Inc.}
\newblock


\bibitem[Vervoort et~al\mbox{.}(2015)]%
        {Vervoort2015Scenarios}
\bibfield{author}{\bibinfo{person}{Joost~M. Vervoort}, \bibinfo{person}{Roy Bendor}, \bibinfo{person}{Aisling Kelliher}, \bibinfo{person}{Oscar Strik}, {and} \bibinfo{person}{Ariella~E.R. Helfgott}.} \bibinfo{year}{2015}\natexlab{}.
\newblock \showarticletitle{Scenarios and the art of worldmaking}.
\newblock \bibinfo{journal}{\emph{Futures}}  \bibinfo{volume}{74} (\bibinfo{date}{11} \bibinfo{year}{2015}), \bibinfo{pages}{62--70}.
\newblock
\showISSN{00163287}
\urldef\tempurl%
\url{https://doi.org/10.1016/j.futures.2015.08.009}
\showDOI{\tempurl}


\bibitem[Wack(1985)]%
        {Wack1985Scenarios:}
\bibfield{author}{\bibinfo{person}{Pierre Wack}.} \bibinfo{year}{1985}\natexlab{}.
\newblock \showarticletitle{Scenarios: uncharted waters ahead}.
\newblock \bibinfo{journal}{\emph{Harvard Business Review}} \bibinfo{volume}{63}, \bibinfo{number}{5} (\bibinfo{year}{1985}), \bibinfo{pages}{73--89}.
\newblock


\bibitem[Wakkary(2021)]%
        {wakkary2021things}
\bibfield{author}{\bibinfo{person}{Ron Wakkary}.} \bibinfo{year}{2021}\natexlab{}.
\newblock \bibinfo{booktitle}{\emph{Things We Could Design: For More Than Human-Centered Worlds}}.
\newblock \bibinfo{publisher}{MIT Press}.
\newblock


\bibitem[Weber(1996)]%
        {Weber1996Counterfactuals}
\bibfield{author}{\bibinfo{person}{Steven Weber}.} \bibinfo{year}{1996}\natexlab{}.
\newblock \showarticletitle{Counterfactuals, past and future}.
\newblock In \bibinfo{booktitle}{\emph{Counterfactual Thought Experiments in World Politics: Logical, Methodological, and Psychological Perspectives}}, \bibfield{editor}{\bibinfo{person}{Phlip~E. Tetlock} {and} \bibinfo{person}{Aaron Belkin}} (Eds.). \bibinfo{publisher}{Princeton University Press}, \bibinfo{address}{Princeton, New Jersey}.
\newblock


\bibitem[Wong et~al\mbox{.}(2020a)]%
        {Wong2020Beyond}
\bibfield{author}{\bibinfo{person}{Richmond~Y Wong}, \bibinfo{person}{Karen Boyd}, \bibinfo{person}{Jake Metcalf}, {and} \bibinfo{person}{Katie Shilton}.} \bibinfo{year}{2020}\natexlab{a}.
\newblock \showarticletitle{Beyond Checklist Approaches to Ethics in Design}. In \bibinfo{booktitle}{\emph{Conference Companion Publication of the 2020 on Computer Supported Cooperative Work and Social Computing}} (2020-10-17). \bibinfo{publisher}{ACM}, \bibinfo{address}{New York, NY, USA}, \bibinfo{pages}{511--517}.
\newblock
\showISBNx{978-1-4503-8059-1}
\urldef\tempurl%
\url{https://doi.org/10.1145/3406865.3418590}
\showDOI{\tempurl}


\bibitem[Wong and Khovanskaya(2018)]%
        {Wong2018Speculative}
\bibfield{author}{\bibinfo{person}{Richmond~Y. Wong} {and} \bibinfo{person}{Vera Khovanskaya}.} \bibinfo{year}{2018}\natexlab{}.
\newblock \showarticletitle{Speculative Design in HCI: From Corporate Imaginations to Critical Orientations}.
\newblock In \bibinfo{booktitle}{\emph{New Directions in 3rd Wave HCI}}, \bibfield{editor}{\bibinfo{person}{Michael Filimowicz}} (Ed.). \bibinfo{publisher}{Springer}, \bibinfo{address}{Cham, Switzerland}, \bibinfo{pages}{175--202}.
\newblock
\urldef\tempurl%
\url{https://doi.org/10.1007/978-3-319-73374-6_10}
\showDOI{\tempurl}


\bibitem[Wong et~al\mbox{.}(2020b)]%
        {Wong2020Infrastructural}
\bibfield{author}{\bibinfo{person}{Richmond~Y Wong}, \bibinfo{person}{Vera Khovanskaya}, \bibinfo{person}{Sarah~E Fox}, \bibinfo{person}{Nick Merrill}, {and} \bibinfo{person}{Phoebe Sengers}.} \bibinfo{year}{2020}\natexlab{b}.
\newblock \showarticletitle{Infrastructural Speculations: Tactics for Designing and Interrogating Lifeworlds}. In \bibinfo{booktitle}{\emph{Proceedings of the 2020 CHI Conference on Human Factors in Computing Systems}} (2020-04-21). \bibinfo{publisher}{ACM}, \bibinfo{address}{New York, NY, USA}, \bibinfo{pages}{1--15}.
\newblock
\showISBNx{978-1-4503-6708-0}
\urldef\tempurl%
\url{https://doi.org/10.1145/3313831.3376515}
\showDOI{\tempurl}


\bibitem[Wong and Mulligan(2019)]%
        {Wong2019Bringing}
\bibfield{author}{\bibinfo{person}{Richmond~Y. Wong} {and} \bibinfo{person}{Deirdre~K. Mulligan}.} \bibinfo{year}{2019}\natexlab{}.
\newblock \showarticletitle{Bringing Design to the Privacy Table: Broadening "Design" in "Privacy by Design" Through the Lens of HCI}. In \bibinfo{booktitle}{\emph{CHI Conference on Human Factors in Computing Systems (CHI 2019)}} (2019).
\newblock
\showISBNx{978-1-4503-5970-2}
\urldef\tempurl%
\url{https://doi.org/10.1145/3290605.3300492}
\showDOI{\tempurl}


\bibitem[Wong et~al\mbox{.}(2017)]%
        {Wong2017Eliciting}
\bibfield{author}{\bibinfo{person}{Richmond~Y. Wong}, \bibinfo{person}{Deirdre~K Mulligan}, \bibinfo{person}{Ellen Van~Wyk}, \bibinfo{person}{James Pierce}, {and} \bibinfo{person}{John Chuang}.} \bibinfo{year}{2017}\natexlab{}.
\newblock \showarticletitle{Eliciting Values Reflections by Engaging Privacy Futures Using Design Workbooks}.
\newblock \bibinfo{journal}{\emph{Proceedings of the ACM on Human Computer Interaction}} \bibinfo{volume}{1}, \bibinfo{number}{CSCW} (\bibinfo{year}{2017}).
\newblock
\urldef\tempurl%
\url{https://doi.org/10.1145/3134746}
\showDOI{\tempurl}


\bibitem[Wong and Nguyen(2021)]%
        {Wong2021Timelines:}
\bibfield{author}{\bibinfo{person}{Richmond~Y. Wong} {and} \bibinfo{person}{Tonya Nguyen}.} \bibinfo{year}{2021}\natexlab{}.
\newblock \showarticletitle{Timelines: A World-Building Activity for Values Advocacy}. In \bibinfo{booktitle}{\emph{Proceedings of the 2021 CHI Conference on Human Factors in Computing Systems}} (2021-05-06). \bibinfo{publisher}{ACM}, \bibinfo{address}{New York, NY, USA}, \bibinfo{pages}{1--15}.
\newblock
\showISBNx{978-1-4503-8096-6}
\urldef\tempurl%
\url{https://doi.org/10.1145/3411764.3445447}
\showDOI{\tempurl}


\bibitem[Wu et~al\mbox{.}(2023)]%
        {Wu2023doStreamers}
\bibfield{author}{\bibinfo{person}{Yanlai Wu}, \bibinfo{person}{Xinning Gui}, \bibinfo{person}{Pamela~J. Wisniewski}, {and} \bibinfo{person}{Yao Li}.} \bibinfo{year}{2023}\natexlab{}.
\newblock \showarticletitle{Do Streamers Care about Bystanders' Privacy? An Examination of Live Streamers' Considerations and Strategies for Bystanders' Privacy Management}.
\newblock \bibinfo{journal}{\emph{Proc. ACM Hum.-Comput. Interact.}} \bibinfo{volume}{7}, \bibinfo{number}{CSCW1}, Article \bibinfo{articleno}{127} (\bibinfo{date}{apr} \bibinfo{year}{2023}), \bibinfo{numpages}{29}~pages.
\newblock
\urldef\tempurl%
\url{https://doi.org/10.1145/3579603}
\showDOI{\tempurl}


\bibitem[Wyche(2021)]%
        {Wyche2021Benefits}
\bibfield{author}{\bibinfo{person}{Susan Wyche}.} \bibinfo{year}{2021}\natexlab{}.
\newblock \showarticletitle{The Benefits of Using Design Workbooks with Speculative Design Proposals in Information Communication Technology for Development (ICTD)}. In \bibinfo{booktitle}{\emph{Designing Interactive Systems Conference 2021}} (2021-06-28). \bibinfo{publisher}{ACM}, \bibinfo{address}{New York, NY, USA}, \bibinfo{pages}{1861--1874}.
\newblock
\showISBNx{978-1-4503-8476-6}
\urldef\tempurl%
\url{https://doi.org/10.1145/3461778.3462140}
\showDOI{\tempurl}


\bibitem[Yao et~al\mbox{.}(2019)]%
        {Yao2019Privacy}
\bibfield{author}{\bibinfo{person}{Yaxing Yao}, \bibinfo{person}{Justin~Reed Basdeo}, \bibinfo{person}{Oriana~Rosata Mcdonough}, {and} \bibinfo{person}{Yang Wang}.} \bibinfo{year}{2019}\natexlab{}.
\newblock \showarticletitle{Privacy Perceptions and Designs of Bystanders in Smart Homes}.
\newblock \bibinfo{journal}{\emph{Proceedings of the ACM on Human-Computer Interaction}} \bibinfo{volume}{3}, \bibinfo{number}{CSCW} (\bibinfo{date}{7 11} \bibinfo{year}{2019}), \bibinfo{pages}{1--24}.
\newblock
\showISSN{2573-0142}
\urldef\tempurl%
\url{https://doi.org/10.1145/3359161}
\showDOI{\tempurl}


\bibitem[Zeng et~al\mbox{.}(2017)]%
        {Zeng2017End}
\bibfield{author}{\bibinfo{person}{Eric Zeng}, \bibinfo{person}{Shrirang Mare}, {and} \bibinfo{person}{Franziska Roesner}.} \bibinfo{year}{2017}\natexlab{}.
\newblock \showarticletitle{End User Security and Privacy Concerns with Smart Homes}. In \bibinfo{booktitle}{\emph{Thirteenth Symposium on Usable Privacy and Security (SOUPS 2017)}} (2017). \bibinfo{pages}{65–80}.
\newblock
\showISBNx{978-1-931971-39-3}
\urldef\tempurl%
\url{https://www.usenix.org/conference/soups2017/technical-sessions/presentation/zeng}
\showURL{%
\tempurl}


\bibitem[Zeng and Roesner(2019)]%
        {Zeng2019Understanding}
\bibfield{author}{\bibinfo{person}{Eric Zeng} {and} \bibinfo{person}{Franziska Roesner}.} \bibinfo{year}{2019}\natexlab{}.
\newblock \showarticletitle{Understanding and Improving Security and Privacy in Multi-User Smart Homes: A Design Exploration and In-Home User Study}. In \bibinfo{booktitle}{\emph{28th USENIX Security Symposium (USENIX Security 19)}} (2019). \bibinfo{publisher}{USENIX Association}, \bibinfo{pages}{159---176}.
\newblock
\urldef\tempurl%
\url{https://www.usenix.org/conference/usenixsecurity19/presentation/zeng}
\showURL{%
\tempurl}


\end{thebibliography}

\appendix
\section{Design Process Details and Images of the Scenarios}
\label{sec:appendix}
Developing the scenarios was an emergent iterative process. To develop scenarios for the design workbook, we identified current trends in smart home technologies synthesized from prior research and news reporting. In parallel, we generated a wide range of hypothetical scenarios. Throughout we tended to focus on crafting scenarios that involve a future version of a smart home security camera because this technology is currently popular but still emerging as a product category, it raises myriad social and ethical concerns, and it involves both camera and microphone sensors. 

\begin{figure}[h]
  \centering
  \includegraphics[width=\linewidth]{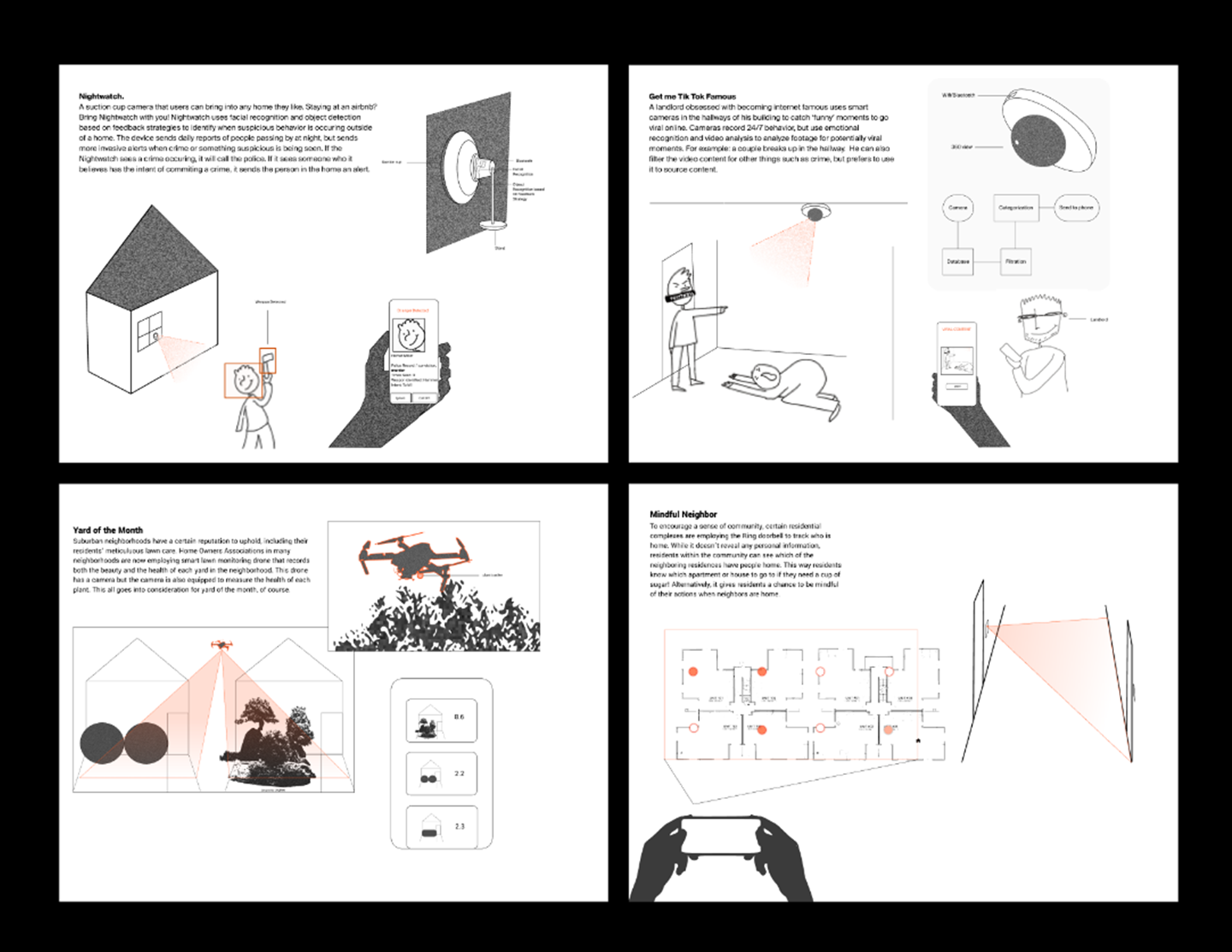}
  \caption{Earlier in our process, we explored many different themes and scenarios ranging from counter-surveillance tactics to futuristic new smart home technologies.}
  \Description{ 4 slides showing design explorations. While the words are too small to make out, they include sketches of homes with cameras, drones, IoT devices, and wearable devices.}
   \label{fig:B1process}
\end{figure}

\begin{figure}[h]
  \centering
  \includegraphics[width=\linewidth]{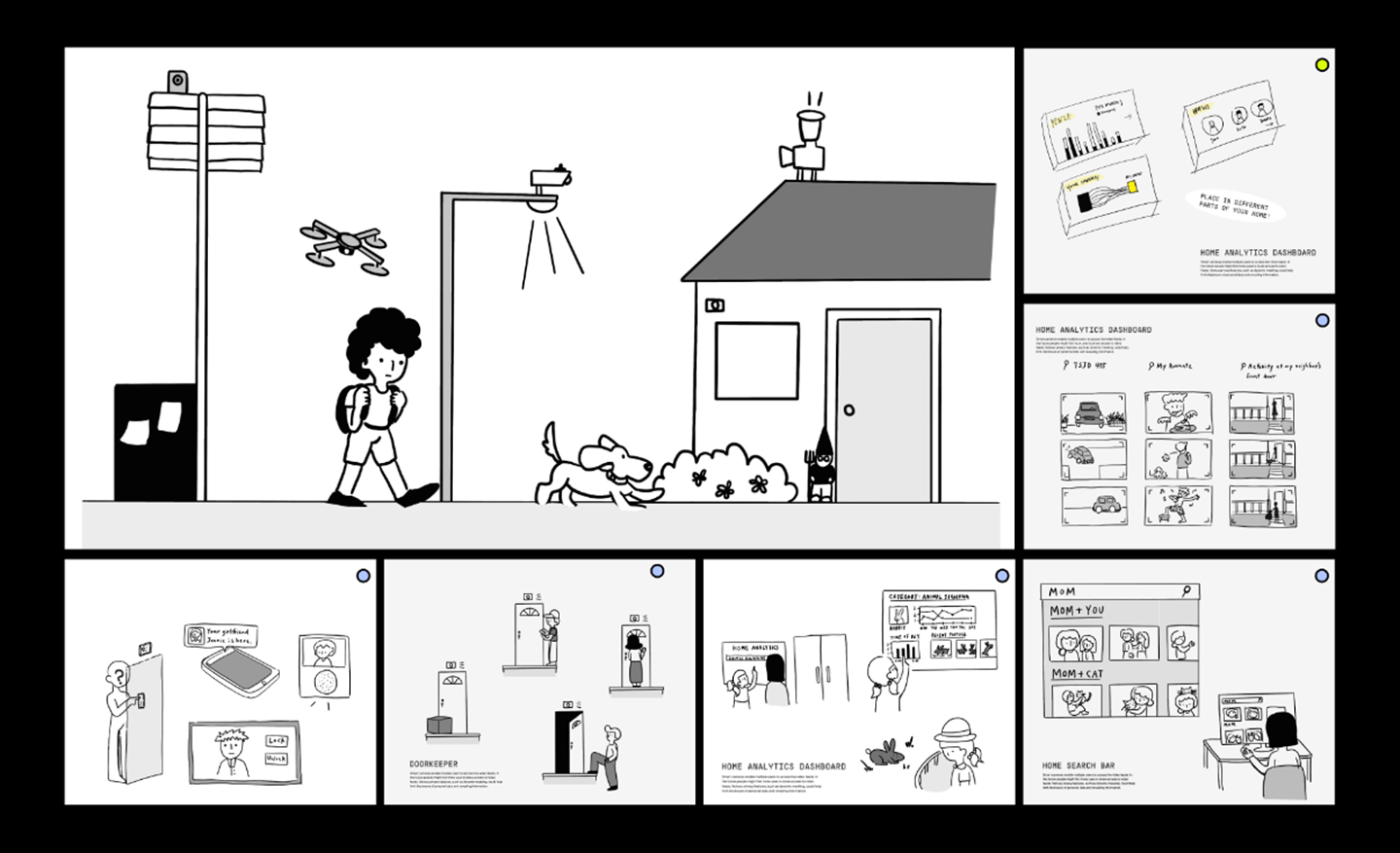}
  \caption{ We developed scenarios for many other themes that we ultimately did not include in this study because they exceeded our scope. The images above are examples of semi-refined scenarios for a theme addressing the “sensorification of everyday life” and ways that advanced data analytics may enable new forms of everyday monitoring, search, and curation of content.}
  \Description{ Screenshot of 7 slides showing early sketches of scenarios (showing smart home cameras used in different settings)
}
   \label{fig:B2process}
\end{figure}

During the initial brainstorming process, we focused on three elements to generate ideas for scenarios which we anticipated would generally be legible, comprehensible, relatable, engaging in a manner that would lead to discussion, reflection, criticism, and imagination (Figure \ref{fig:B1process}). This compositional framework consists of three basic elements: 

\begin{itemize}
    \item \textbf{Hooks} - \textit{entry points} that engage viewers, readers, users, participants.
    \item \textbf{Anchors} - \textit{grounding points} to real-world events.
    \item \textbf{Orbits} - Tacitly embedded themes or ideas to prompt reflection, discussion, and inspiration. 
\end{itemize}

Roughly, we constructed scenarios by combining general themes, or orbits, we wished to explore and communicate (e.g., parents surveilling kids) with specific real-world anchors (e.g., the Amazon drone smart home camera \cite{Bohn2020Ring}, helicopter parenting \cite{Barrett2015Fuhu's}, and emergent technologies for analyzing sociality, test-taking, and eye contact among kids \cite{Swauger2020Software}). We also designed each scenario with an eye toward one or a few well-designed hooks--carefully iterated details designed to draw the viewer in. In some cases we built scenarios around a what we thought might be a clever hook, like a “drone parent” as an extension and play on “helicopter parents”. In other cases we instead relied on cliches--such as a child sneaking of the window at night, or an overprotective dad--to engage the viewer in a specific narrative snapshot while gesturing toward other possibilities. 

We typically began the design of scenarios focusing on text and writing. We first composed lists of scenarios that consisted of a short descriptive title followed by a few sentences. From here, we selected some to iterate and added very crude sketches. We then continued to select and refine. The next stage was creating illustrations. One group of authors created rough illustrations while another focused on refining the text. 
As we explored these iterations, we continued to reflect upon themes that the scenarios explored. Example of these early themes include (1) the sensorification of daily life, (2) perceptually powerful devices, (3) behind-the-scenes actors and data misuse, (4) beyond safety and security: social, reflective, and aesthetic applications, (5) environmental sensor pollution, and (6) social tensions and asymmetric power relations. (Figure \ref{fig:B2process})

Ultimately we decided to focus on the theme of everyday social tensions and asymmetric power relations connected to smart cameras. We chose this theme for several reasons. First, it captured social and ethical issues at stake with smart devices. Smart home cameras are an exemplary site for investigating the topics of privacy, surveillance, and other technology ethics issues. Second, the scenarios in this category appeared well-suited for eliciting discussion and provoking responses based on our own internal discussions among authors. Third, practically we needed to select a small subset of scenarios that we could reasonably discuss within a 60-90 minute interview. Fourth, several authors have conducted prior research and had expertise in this space.

\subsection{Images of the Workbook}
\label{sec:appendixAimages}
\begin{figure}[p]
  \centering
  \includegraphics[width=0.9\linewidth]{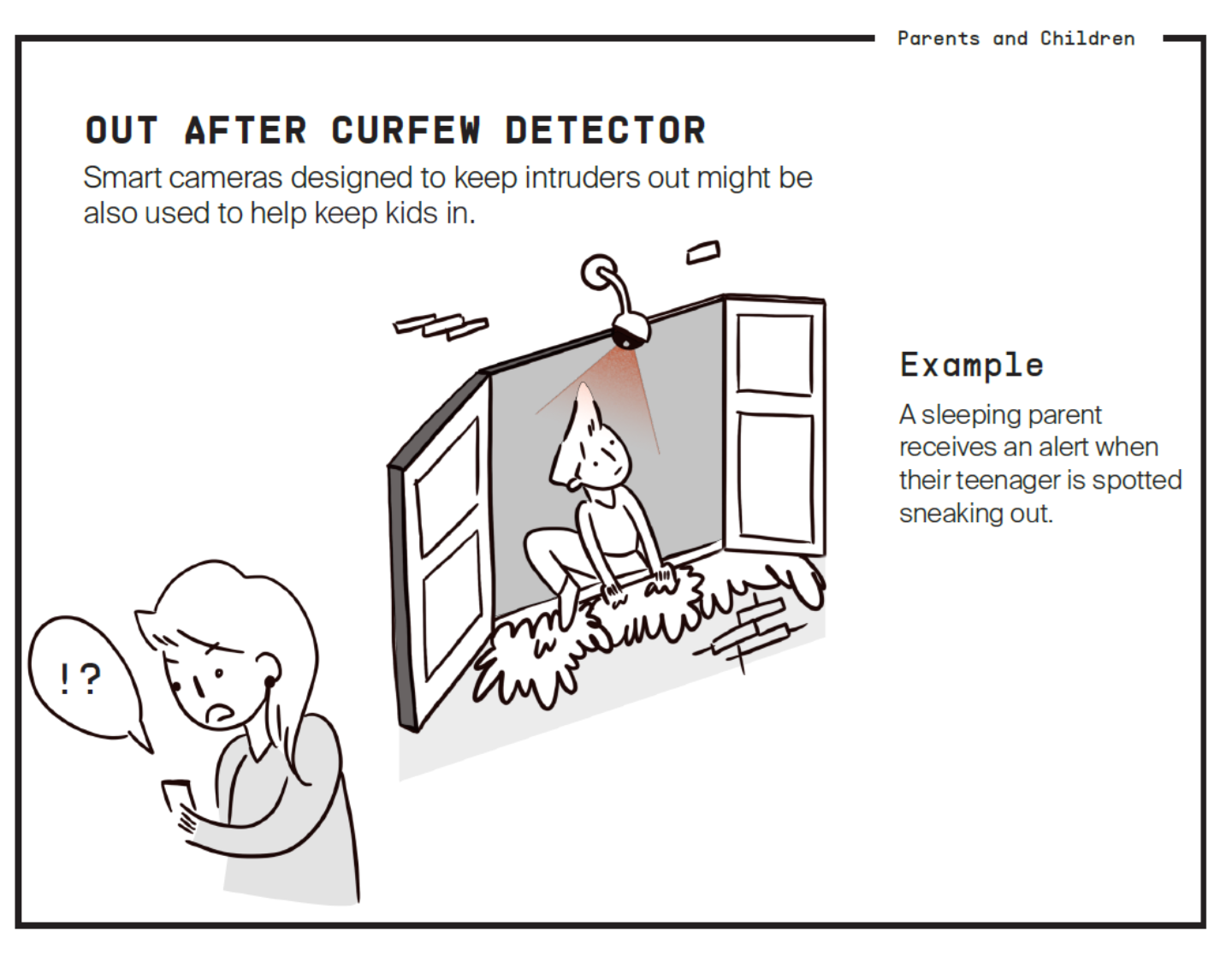}
  \caption{Out After Curfew Detector Scenario.}
  \Description{Drawing of Out After Curfew Detector. Smart cameras designed to keep intruders out might be also used to help keep kids in.}
   \label{fig:A1curfew}
\end{figure}

\begin{figure}[p]
  \centering
  \includegraphics[width=0.9\linewidth]{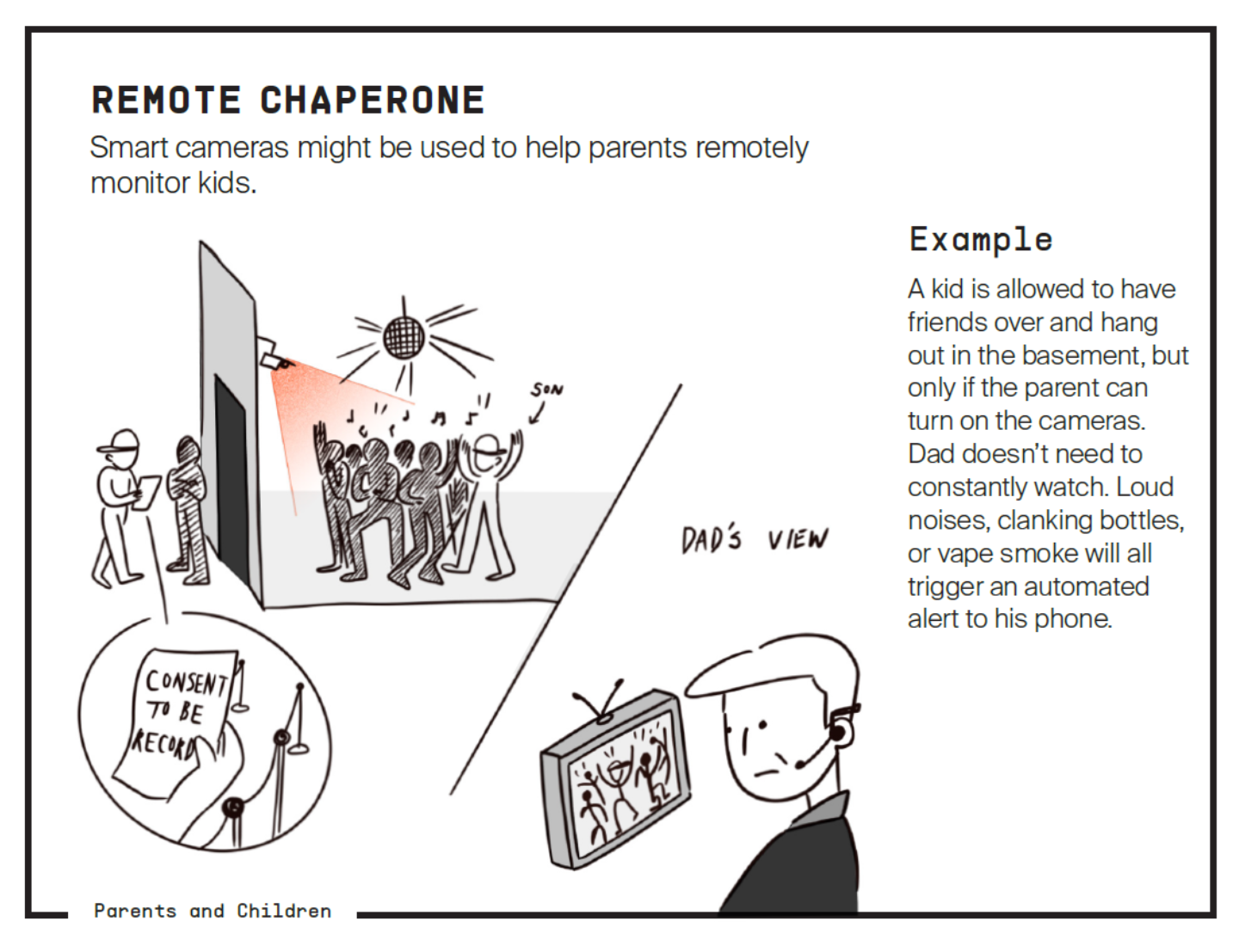}
  \caption{Remote Chaperone scenario.}
  \Description{Drawing of Remote Chaperone. Smart cameras might be used to help parents remotely monitor kids.}
   \label{fig:A2chaperone}
\end{figure}

\begin{figure}[p]
  \centering
  \includegraphics[width=0.9\linewidth]{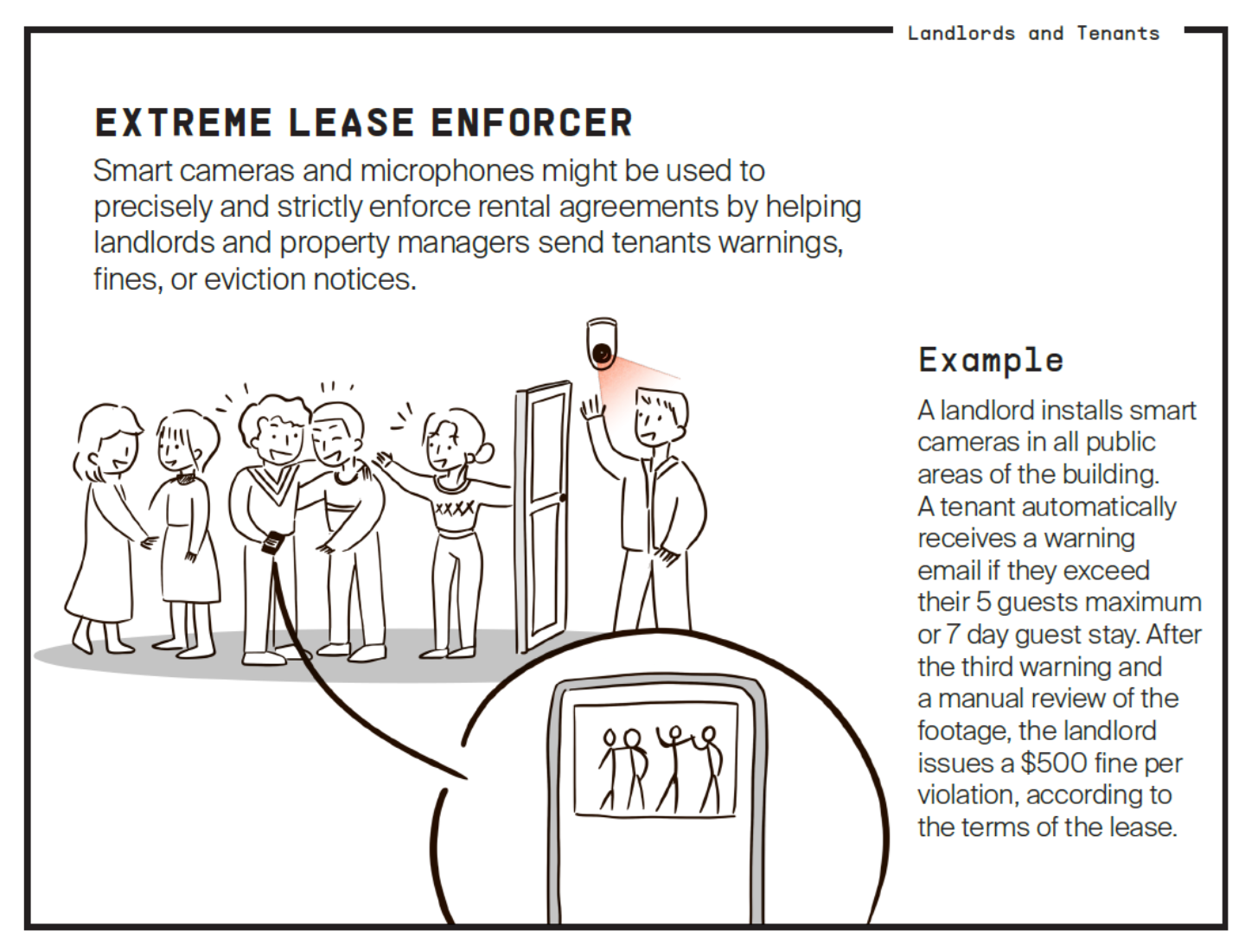}
  \caption{Extreme Lease Enforcer Scenario.}
  \Description{Drawing of Extreme Lease Enforcer. Smart cameras and microphones might be used to precisely and strictly enforce rental agreements by helping landlords and property managers send warnings, fines, or eviction notices}
   \label{fig:A3extremelease}
\end{figure}

\begin{figure}[p]
  \centering
  \includegraphics[width=0.9\linewidth]{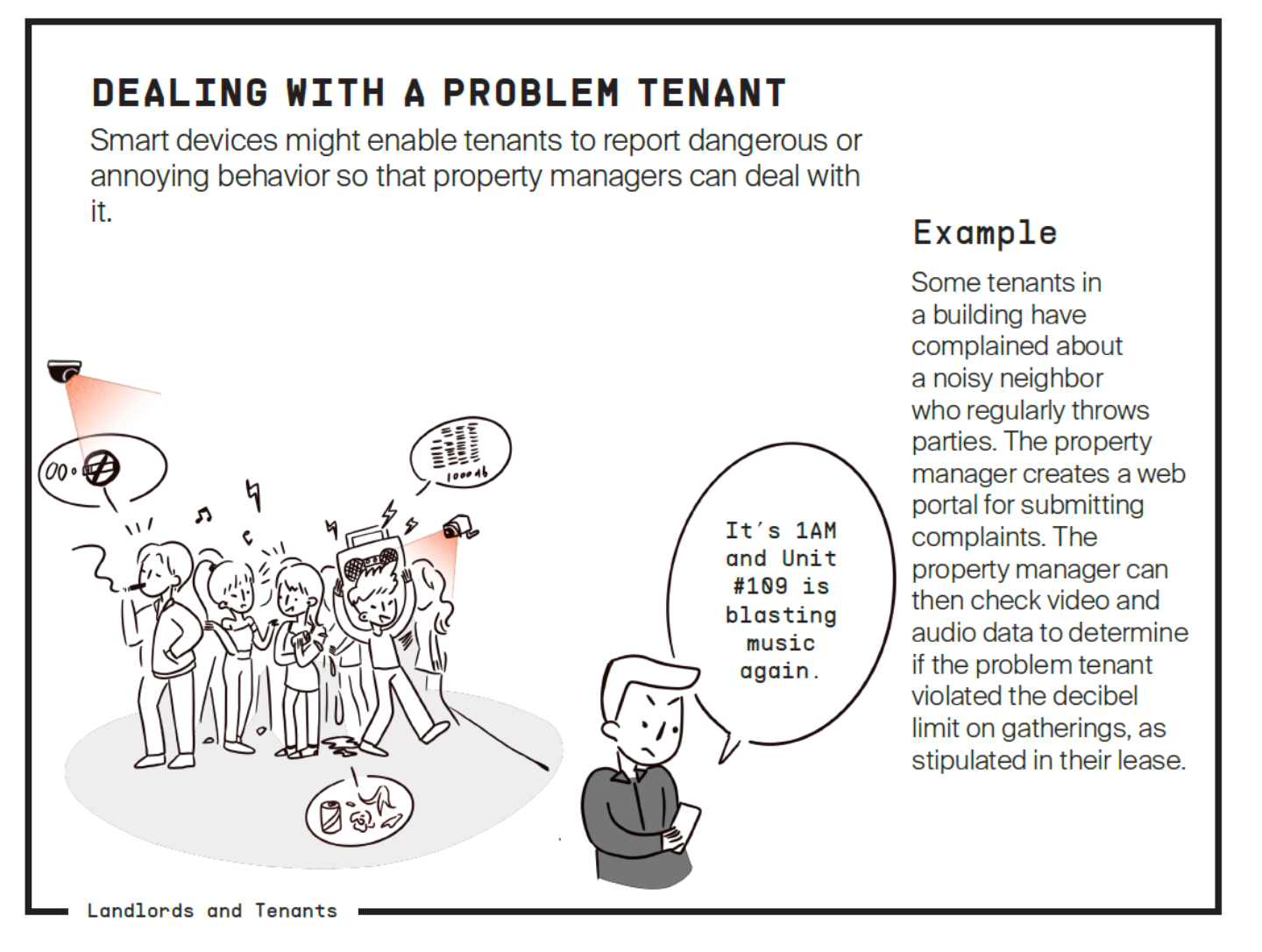}
  \caption{Dealing with a Problem Tenant scenario.}
  \Description{Drawing of Dealing with a Problem Tenant. Smart devices might enable tenants to report dangerous or annoying behavior so that property managers can deal with it.}
   \label{fig:A4problemtenant}
\end{figure}

\begin{figure}[p]
  \centering
  \includegraphics[width=0.9\linewidth]{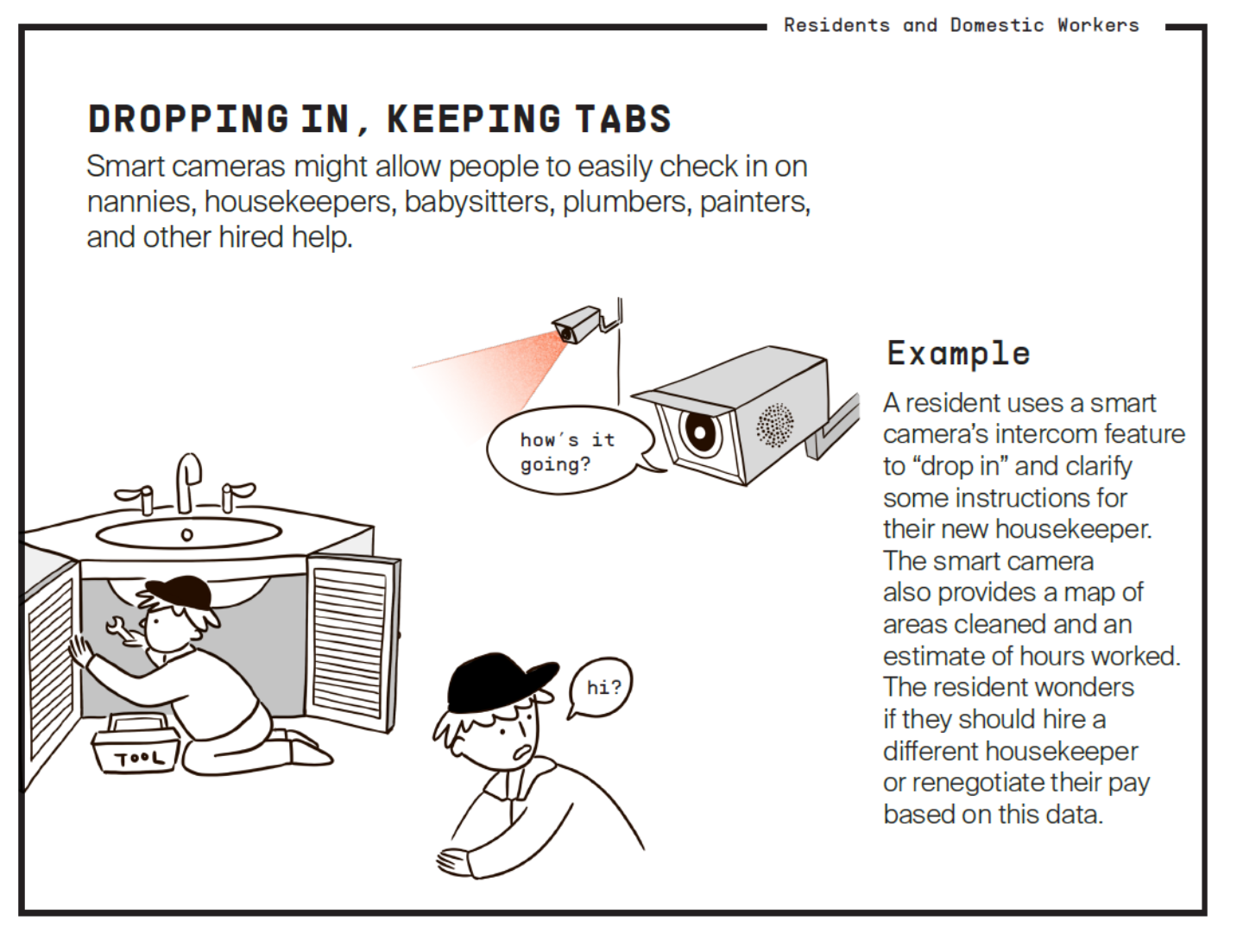}
  \caption{Dropping In, Keeping Tabs scenario.}
  \Description{Drawing of Dropping in, Keeping Tabs. Smart cameras might allow people to easily check in on nannies, housekeepers, babysitters, plumbers, painters, and other hired help. }
   \label{fig:A5droppingin}
\end{figure}

\begin{figure}[p]
  \centering
  \includegraphics[width=0.9\linewidth]{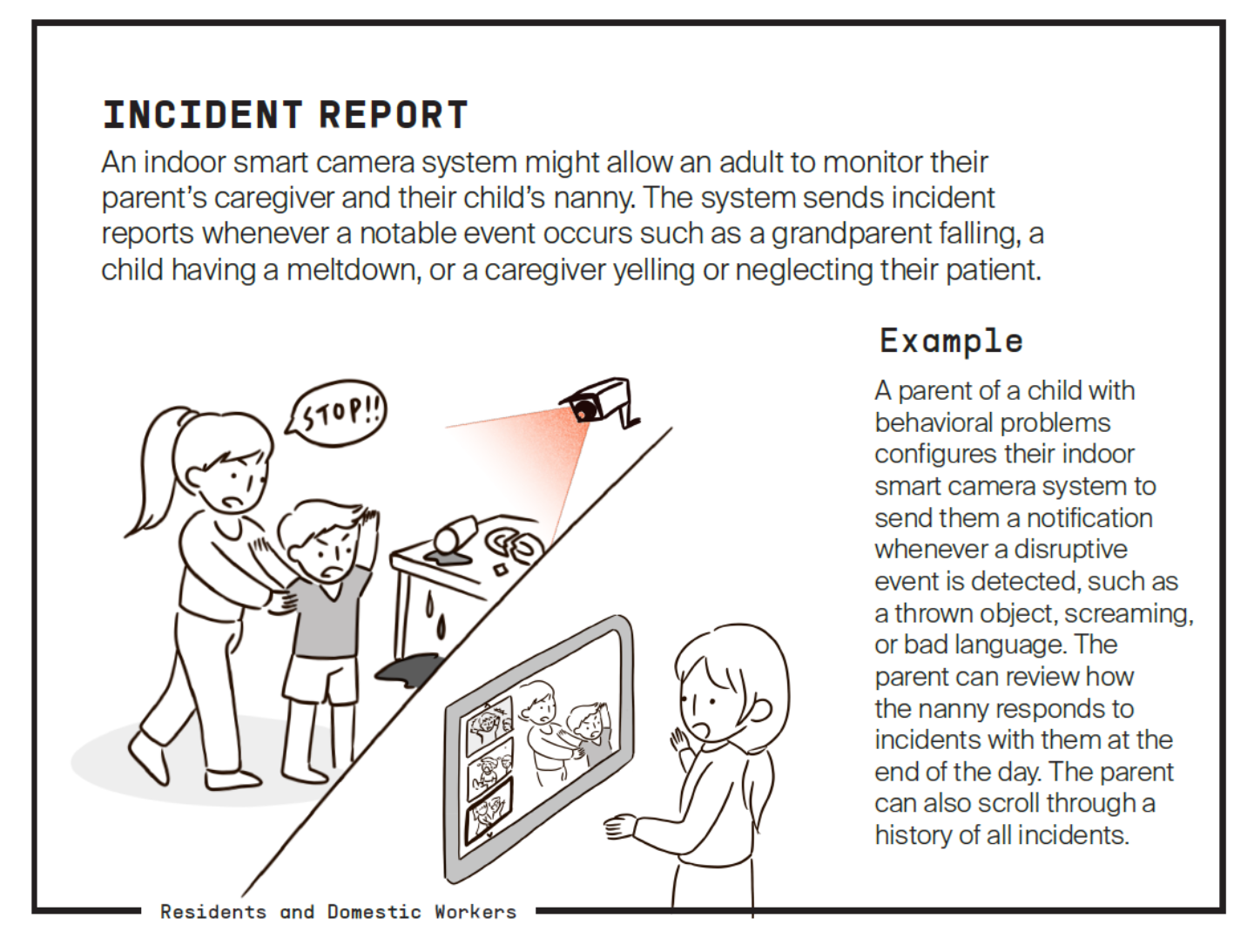}
  \caption{Incident Report scenario.}
  \Description{Drawing of Incident Report. An indoor smart camera system might allow an adult to monitor their parent’s caregiver and their child’s nanny. The system sends incident reports whenever a notable event occurs such as a grandparent falling, a child having a meltdown, or a caregiver yelling or neglecting their patient. 3. Mood Check. Smart camera systems might also monitor the internal emotional states of people }
   \label{fig:A6incident}
\end{figure}

Figures \ref{fig:A1curfew}-\ref{fig:A6incident} depict screenshots of additional scenarios from the workbook.

\end{document}